\documentclass[11pt,a4paper,superscriptaddress]{revtex4-1}
\usepackage{graphicx,color,hyperref}
\usepackage[top=30mm,bottom=30mm,left=30mm,right=30mm]{geometry}
\usepackage{placeins}
\usepackage{caption}
\usepackage{float}
\usepackage{multirow}
\usepackage{dcolumn}
\usepackage{amssymb}
\usepackage[normalem]{ulem}
\usepackage{subcaption}
\usepackage{indentfirst}

\newcommand\onepot{\ensuremath{7.8\times 10^{21}\;\mbox{protons-on-target}} }
\newcommand\twopot{\ensuremath{20\times 10^{21}\;\mbox{protons-on-target}} }
\newcommand\nupot{\ensuremath{6.6\times 10^{20}\;\mbox{protons-on-target}} }

\newcommand\onepott{\ensuremath{7.8\times 10^{21}~\mbox{POT}} }
\newcommand\twopott{\ensuremath{20\times 10^{21}~\mbox{POT}} }

\graphicspath{{/figure}}
\begin{document}
\title{Proposal for an Extended Run of T2K to $20\times10^{21}$ POT}

\newcommand{\INSTFE}{\affiliation{Boston University, Department of Physics, Boston, Massachusetts, U.S.A.}}
\newcommand{\INSTD}{\affiliation{University of British Columbia, Department of Physics and Astronomy, Vancouver, British Columbia, Canada}}
\newcommand{\INSTGA}{\affiliation{University of California, Irvine, Department of Physics and Astronomy, Irvine, California, U.S.A.}}
\newcommand{\INSTI}{\affiliation{IRFU, CEA Saclay, Gif-sur-Yvette, France}}
\newcommand{\INSTGB}{\affiliation{University of Colorado at Boulder, Department of Physics, Boulder, Colorado, U.S.A.}}
\newcommand{\INSTFG}{\affiliation{Colorado State University, Department of Physics, Fort Collins, Colorado, U.S.A.}}
\newcommand{\INSTFH}{\affiliation{Duke University, Department of Physics, Durham, North Carolina, U.S.A.}}
\newcommand{\INSTBA}{\affiliation{Ecole Polytechnique, IN2P3-CNRS, Laboratoire Leprince-Ringuet, Palaiseau, France }}
\newcommand{\INSTEG}{\affiliation{University of Geneva, Section de Physique, DPNC, Geneva, Switzerland}}
\newcommand{\INSTDG}{\affiliation{H. Niewodniczanski Institute of Nuclear Physics PAN, Cracow, Poland}}
\newcommand{\INSTCB}{\affiliation{High Energy Accelerator Research Organization (KEK), Tsukuba, Ibaraki, Japan}}
\newcommand{\INSTED}{\affiliation{Institut de Fisica d'Altes Energies (IFAE), The Barcelona Institute of Science and Technology, Campus UAB, Bellaterra (Barcelona) Spain}}
\newcommand{\INSTEC}{\affiliation{IFIC (CSIC \& University of Valencia), Valencia, Spain}}
\newcommand{\INSTEI}{\affiliation{Imperial College London, Department of Physics, London, United Kingdom}}
\newcommand{\INSTGF}{\affiliation{INFN Sezione di Bari and Universit\`a e Politecnico di Bari, Dipartimento Interuniversitario di Fisica, Bari, Italy}}
\newcommand{\INSTBE}{\affiliation{INFN Sezione di Napoli and Universit\`a di Napoli, Dipartimento di Fisica, Napoli, Italy}}
\newcommand{\INSTBF}{\affiliation{INFN Sezione di Padova and Universit\`a di Padova, Dipartimento di Fisica, Padova, Italy}}
\newcommand{\INSTBD}{\affiliation{INFN Sezione di Roma and Universit\`a di Roma ``La Sapienza'', Roma, Italy}}
\newcommand{\INSTEB}{\affiliation{Institute for Nuclear Research of the Russian Academy of Sciences, Moscow, Russia}}
\newcommand{\INSTHA}{\affiliation{Kavli Institute for the Physics and Mathematics of the Universe (WPI), The University of Tokyo Institutes for Advanced Study, University of Tokyo, Kashiwa, Chiba, Japan}}
\newcommand{\INSTCC}{\affiliation{Kobe University, Kobe, Japan}}
\newcommand{\INSTCD}{\affiliation{Kyoto University, Department of Physics, Kyoto, Japan}}
\newcommand{\INSTEJ}{\affiliation{Lancaster University, Physics Department, Lancaster, United Kingdom}}
\newcommand{\INSTFC}{\affiliation{University of Liverpool, Department of Physics, Liverpool, United Kingdom}}
\newcommand{\INSTFI}{\affiliation{Louisiana State University, Department of Physics and Astronomy, Baton Rouge, Louisiana, U.S.A.}}
\newcommand{\INSTHB}{\affiliation{Michigan State University, Department of Physics and Astronomy,  East Lansing, Michigan, U.S.A.}}
\newcommand{\INSTCE}{\affiliation{Miyagi University of Education, Department of Physics, Sendai, Japan}}
\newcommand{\INSTDF}{\affiliation{National Centre for Nuclear Research, Warsaw, Poland}}
\newcommand{\INSTFJ}{\affiliation{State University of New York at Stony Brook, Department of Physics and Astronomy, Stony Brook, New York, U.S.A.}}
\newcommand{\INSTGJ}{\affiliation{Okayama University, Department of Physics, Okayama, Japan}}
\newcommand{\INSTCF}{\affiliation{Osaka City University, Department of Physics, Osaka, Japan}}
\newcommand{\INSTGG}{\affiliation{Oxford University, Department of Physics, Oxford, United Kingdom}}
\newcommand{\INSTBB}{\affiliation{UPMC, Universit\'e Paris Diderot, CNRS/IN2P3, Laboratoire de Physique Nucl\'eaire et de Hautes Energies (LPNHE), Paris, France}}
\newcommand{\INSTGC}{\affiliation{University of Pittsburgh, Department of Physics and Astronomy, Pittsburgh, Pennsylvania, U.S.A.}}
\newcommand{\INSTFA}{\affiliation{Queen Mary University of London, School of Physics and Astronomy, London, United Kingdom}}
\newcommand{\INSTE}{\affiliation{University of Regina, Department of Physics, Regina, Saskatchewan, Canada}}
\newcommand{\INSTGD}{\affiliation{University of Rochester, Department of Physics and Astronomy, Rochester, New York, U.S.A.}}
\newcommand{\INSTHC}{\affiliation{Royal Holloway University of London, Department of Physics, Egham, Surrey, United Kingdom}}
\newcommand{\INSTBC}{\affiliation{RWTH Aachen University, III. Physikalisches Institut, Aachen, Germany}}
\newcommand{\INSTDA}{\affiliation{University Autonoma Madrid, Department of Theoretical Physics, Madrid, Spain}}
\newcommand{\INSTFB}{\affiliation{University of Sheffield, Department of Physics and Astronomy, Sheffield, United Kingdom}}
\newcommand{\INSTDI}{\affiliation{University of Silesia, Institute of Physics, Katowice, Poland}}
\newcommand{\INSTEH}{\affiliation{STFC, Rutherford Appleton Laboratory, Harwell Oxford,  and  Daresbury Laboratory, Warrington, United Kingdom}}
\newcommand{\INSTCH}{\affiliation{University of Tokyo, Department of Physics, Tokyo, Japan}}
\newcommand{\INSTBJ}{\affiliation{University of Tokyo, Institute for Cosmic Ray Research, Kamioka Observatory, Kamioka, Japan}}
\newcommand{\INSTCG}{\affiliation{University of Tokyo, Institute for Cosmic Ray Research, Research Center for Cosmic Neutrinos, Kashiwa, Japan}}
\newcommand{\INSTGI}{\affiliation{Tokyo Metropolitan University, Department of Physics, Tokyo, Japan}}
\newcommand{\INSTF}{\affiliation{University of Toronto, Department of Physics, Toronto, Ontario, Canada}}
\newcommand{\INSTB}{\affiliation{TRIUMF, Vancouver, British Columbia, Canada}}
\newcommand{\INSTDJ}{\affiliation{University of Warsaw, Faculty of Physics, Warsaw, Poland}}
\newcommand{\INSTDH}{\affiliation{Warsaw University of Technology, Institute of Radioelectronics, Warsaw, Poland}}
\newcommand{\INSTFD}{\affiliation{University of Warwick, Department of Physics, Coventry, United Kingdom}}
\newcommand{\INSTGE}{\affiliation{University of Washington, Department of Physics, Seattle, Washington, U.S.A.}}
\newcommand{\INSTGH}{\affiliation{University of Winnipeg, Department of Physics, Winnipeg, Manitoba, Canada}}
\newcommand{\INSTEA}{\affiliation{Wroclaw University, Faculty of Physics and Astronomy, Wroclaw, Poland}}
\newcommand{\INSTH}{\affiliation{York University, Department of Physics and Astronomy, Toronto, Ontario, Canada}}

\INSTFE
\INSTGA
\INSTI
\INSTGB
\INSTFG
\INSTFH
\INSTBA
\INSTEG
\INSTDG
\INSTCB
\INSTED
\INSTEC
\INSTEI
\INSTGF
\INSTBE
\INSTBF
\INSTBD
\INSTEB
\INSTHA
\INSTCC
\INSTCD
\INSTEJ
\INSTFC
\INSTFI
\INSTHB
\INSTCE
\INSTDF
\INSTFJ
\INSTGJ
\INSTCF
\INSTGG
\INSTBB
\INSTGC
\INSTFA
\INSTE
\INSTGD
\INSTHC
\INSTBC
\INSTDA
\INSTFB
\INSTDI
\INSTEH
\INSTCH
\INSTBJ
\INSTCG
\INSTGI
\INSTF
\INSTB
\INSTDJ
\INSTDH
\INSTFD
\INSTGE
\INSTGH
\INSTEA
\INSTH

\author{K.\,Abe}\INSTBJ
\author{H.\,Aihara}\INSTCH\INSTHA
\author{A.\,Ajmi}\INSTBF
\author{J.\,Amey}\INSTEI
\author{C.\,Andreopoulos}\INSTEH\INSTFC
\author{M.\,Antonova}\INSTEB
\author{S.\,Aoki}\INSTCC
\author{A.\,Atherton}\INSTEH
\author{S.\,Ban}\INSTCD
\author{F.C.T.\,Barbato}\INSTBE
% ...

\collaboration{The T2K Collaboration}\noaffiliation

\date{\today}
\maketitle
\begin{center}
  {\Large\bf Abstract}
\end{center}

Recent measurements by the T2K neutrino oscillation experiment
indicate that CP violation in neutrino mixing may be observed in the future
 by long-baseline neutrino oscillation experiments. 
We propose 
an extension to the currently approved T2K running from $\onepot$ to $\twopot$,
aiming at initial observation of CP violation with 3$\,\sigma$ or higher significance for the case of maximum CP violation.
The program also contains a measurement of mixing parameters, $\theta_{23}$ and $\Delta m^2_{32}$,
  with a precision of 1.7$^\circ$ or better and 1\%, respectively.
  With accelerator and beamline 
upgrades, as well as analysis improvements, this program would occur
before the next generation of long-baseline neutrino oscillation experiments that are expected 
to start operation in 2026.

\thispagestyle{empty}
\newpage

\renewcommand{\thepage}{\roman{page}}
\setcounter{page}{1}
\noindent{\Large\bf Executive Summary}
\vskip 0.5 cm
The discovery of $\nu_\mu\to \nu_e$ oscillations by T2K\cite{Abe:2013hdq} 
has opened the possibility of observing CP-violation (CPV) in the lepton sector,
 which would be a crucial input towards understanding the
matter-antimatter asymmetry of the universe. In neutrino oscillations, CPV 
can arise from $\delta_{CP}$,  an  irreducible CP-odd phase
in the lepton mixing matrix.
It can be measured at 
accelerator-based long basline neutrino oscillation experiments by comparing the $\nu_\mu\to\nu_e$ 
and $\bar{\nu}_\mu\to\bar{\nu}_e$ oscillation probabilities or by comparing 
these probabilities with $\bar{\nu}_e$ disappearance probabilities measured by reactors-based experiments. 
While the current significance is marginal, 
measurements by the T2K experiment
with $\nupot$ (POT) hint at maximum  CP violation with 
$\delta_{CP}\sim-\frac{\pi}{2}$ and normal mass hierarchy\cite{Abe:2015awa}.
In such case T2K could observe CPV with 90\% C.L. sensitivity with the 
currently approved exposure of $\onepott$\cite{Abe:2014tzr} expected 
around 2020.
Future proposed projects such as Hyper-Kamiokande\cite{Abe:2015zbg} 
and DUNE\cite{dune} aim to achieve $>3\;\sigma$ sensitivity to CPV across a 
wide range of  $\delta_{CP}$ on the time scale of 2026 and beyond.
By increasing the beam power and extending T2K data-taking to 2026, 
when Hyper-Kamiokande and DUNE are expected to start, sensitivity to CPV can be 
significantly improved with the additional statistics. This would also have 
the benefit of establishing higher beam power for the next generation of 
measurements at Hyper-Kamiokande from the start.

The T2K collaboration proposes to extend the run from $\onepott$ to $\twopott$
in a five or six year period after the currently approved running
to explore CP violation with sensitivity greater than $3\,\sigma$
if $\delta_{CP}\sim-\frac{\pi}{2}$ and the mass hierarchy is normal.
We refer to this extended running as ``T2K Phase II'', hereafter abbreviated as T2K-II in this document.

\begin{figure}[bhtp]
\centering
\includegraphics[width=0.8\textwidth]{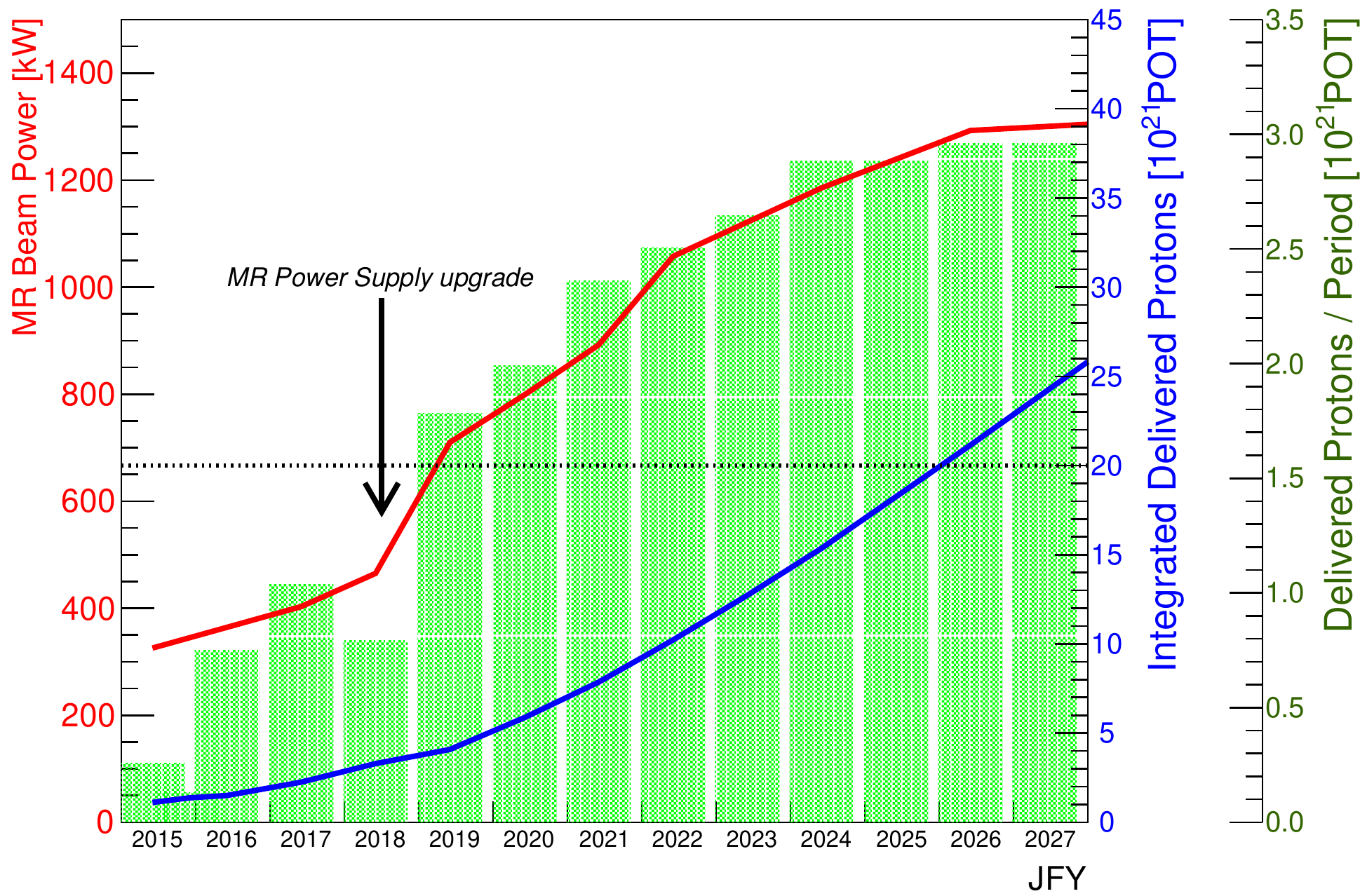}
\caption[POT]{Anticipated MR beam power and POT accumulation as function of Japanese Fiscal Year (JFY) which starts 1 April of the corresponding calendar year.
\label{fig:POT}}
\end{figure}
Since the start of its operation,
the J-PARC MR beam power has steadily increased.
In May 2016, 420~kW beam with 2.2$\times10^{14}$ protons-per-pulse (ppp)
delivered with a 2.48 second period was successfully provided to the T2K neutrino beamline.
In order to achieve the design power of 750~kW, J-PARC plans 
to reduce the repetition cycle of the MR to 1.3 seconds
with an upgrade to the power supplies for the main magnets,
RF cavities, and some injection and extraction devices by January 2019. 
Studies to increase the ppp are also in progress, with
 $2.73\times 10^{14}$ ppp equivalent beam
with acceptable beam loss already demonstrated in a test operation
with two bunches.
Based on these developments,
MR beam power prospects were updated and presented 
in the accelerator report at the past PAC in July 2015\cite{pac16}
and anticipated beam power of 1.3~MW with 3.2$\times$10$^{14}$ ppp
and a repetition cycle of 1.16 seconds are presented
at international workshops{\cite{jparcnuws,nuinfra}}.
Figure \ref{fig:POT}  shows our projected data accumulation scenario where five months of
neutrino operation each year and running time efficiency of 90\%
are assumed.
In this scenario, we expect to accumulate $\twopott$ by JFY2026.
If six months operation each year is assumed, this goal can be accomplished by JFY2025.

The T2K collaboration is also working intensively to increase the
effective statistics and sensitivity of the experiment per POT.
Increasing the electromagnetic horn current from the present 250~kA
to the designed 320~kA will result in  10\% greater neutrino flux.
The current efficiency to select oscillated $\nu_e$ CC events 
at the far detector, Super-Kamiokande (SK), is 66\%.
The main inefficiency results from only selecting events with a single Cherenkov ring
from the outgoing lepton without additional rings or decay electrons that may
arise from pions produced in the interaction.
We expect to increase the efficiency to 70-80\%
by selecting additional events accompanied with a decay electron
and multi-ring events with an improved event reconstruction algorithm at SK.
This algorithm may also allow the  fiducial volume at SK to be increased by 10-15\%.
Taken together, the beamline upgrades and analysis improvements
can potentially increase the effective statistics
of T2K by up to 50\%.

The number of events expected at the far detector
for an exposure of \(20\times10^{21}\) POT with 50\% statistical
  improvement is given in Table \ref{tab:detevts} for  \(\delta_{CP} = 0\) or \(-\pi / 2\). 

\begin{table}[htpb]
\begin{center}
\caption[Number of Expected Events]{Number of events expected to be observed
        at the far detector for
\(10\times10^{21}\)~POT \(\nu\)- + \(10\times10^{21}\)~POT \(\bar{\nu}\)-mode
with a 50\% statistical improvement.
Assumed relevant oscillation parameters are:
\(\sin^22\theta_{13}=0.085\), \(\sin^2\theta_{23}=0.5\), 
\(\Delta m^2_{32}=2.5\times10^{-3}\) eV\(^2\), and normal mass hierarchy (NH).
\label{tab:detevts}}
\begin{tabular}{c | c | c | c | c | c | c | c   } \hline
& & & Signal & Signal & Beam CC & Beam CC & \\
& True \(\delta_{CP}\) & Total & \(\nu_{\mu} \rightarrow \nu_e\) & \(\bar{\nu}_{\mu} \rightarrow
\bar{\nu}_e\) & \(\nu_e + \bar{\nu}_e \) & \(\nu_{\mu} + \bar{\nu}_{\mu} \) & NC\\ \hline\hline
\(\nu\)-mode & 0  & 467.6 & 356.3 &  4.0 & 73.3 & 1.8 & 32.3 \\ \cline{2-8}
$\nu_e$ sample & \(-\pi/2\) & 558.7 & 448.6 &  2.8 & 73.3 & 1.8 & 32.3 \\ \hline \hline
\(\bar{\nu}\)-mode & 0          & 133.9 & 16.7 &  73.6 & 29.2 & 0.4 & 14.1 \\ \cline{2-8}
$\bar{\nu}_e$ sample & \(-\pi/2\) & 115.8 & 19.8 &  52.3 & 29.2 & 0.4 & 14.1 \\ \hline 
\end{tabular} 
\vskip 0.4cm
\begin{tabular}{  c | c | c | c | c | c | c   } \hline
& & Beam CC & Beam CC & Beam CC & \(\nu_{\mu} \rightarrow \nu_e +\) & \\
& Total & \(\nu_\mu\) & \(\bar{\nu}_\mu\) & \(\nu_e + \bar{\nu}_e \) & \(\bar{\nu}_{\mu} \rightarrow \bar{\nu}_e\) & NC\\ \hline\hline
\(\nu\)-mode $\nu_\mu$ sample & 2735.0 & 2393.0 &  158.2 & 1.6 & 7.2 & 175.0 \\ \hline \hline
\(\bar{\nu}\)-mode $\bar{\nu}_\mu$ sample & 1283.5 & 507.8 &  707.9 & 0.6 & 1.0 & 66.2 \\ \hline
\end{tabular}
\end{center}
\end{table}


In T2K, we have achieved 5.5\% to 6.8\% systematic error on the predicted number of events at the far detector and its influence on the oscillation measurement has been modest thus far.
With the much higher ultimate statistics at T2K-II, however, the physics reach will be significantly enhanced by reducing the systematic errors.
Considering the present understanding and projected
improvements, we consider that 4\% systematic error  is a reachable
and reasonable target for T2K-II.
In case some uncertainties prevent us from achieving this goal, we are preparing to pursue necessary actions.
For example,  we have been improving our model of the neutrino-nucleus interactions, which is a significant source of systematic error, 
with our near detector measurements  and measurements from other experiments
by working closely with theorists. In case these uncertainties are not resolved,
we are investigating possible near detector upgrades to
resolve uncertainties from  neutrino-nucleus interaction modelling.

\begin{figure} [thbp]
\begin{center} 
\begin{subfigure}[H]{7.2cm}
\includegraphics[width=7.2cm]{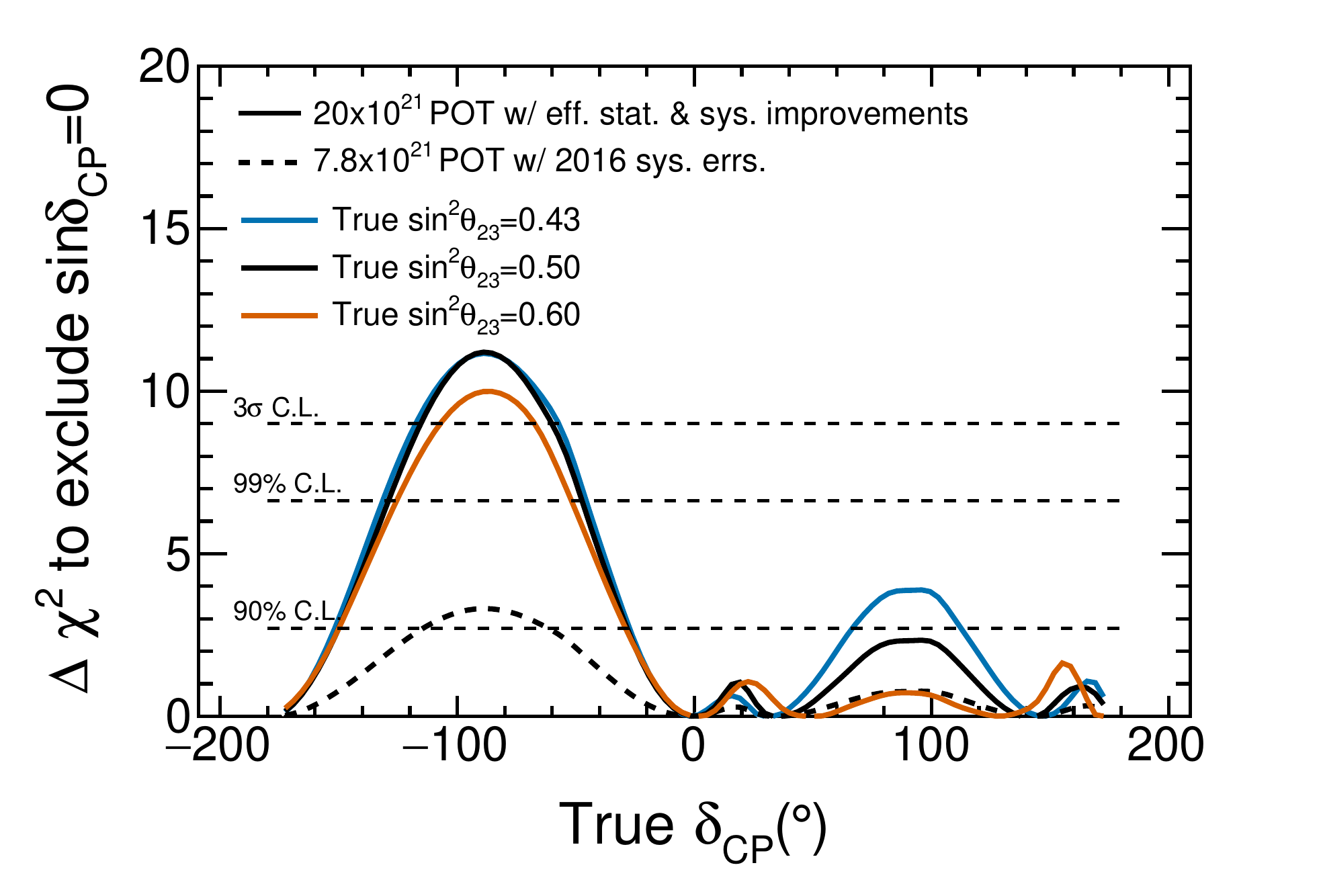}\caption{Assuming the MH is unknown.} \label{fig:CPVvsdCP_unknownMH}
\end{subfigure} \quad 
\begin{subfigure}[H]{7.2cm}
\includegraphics[width=7.2cm]{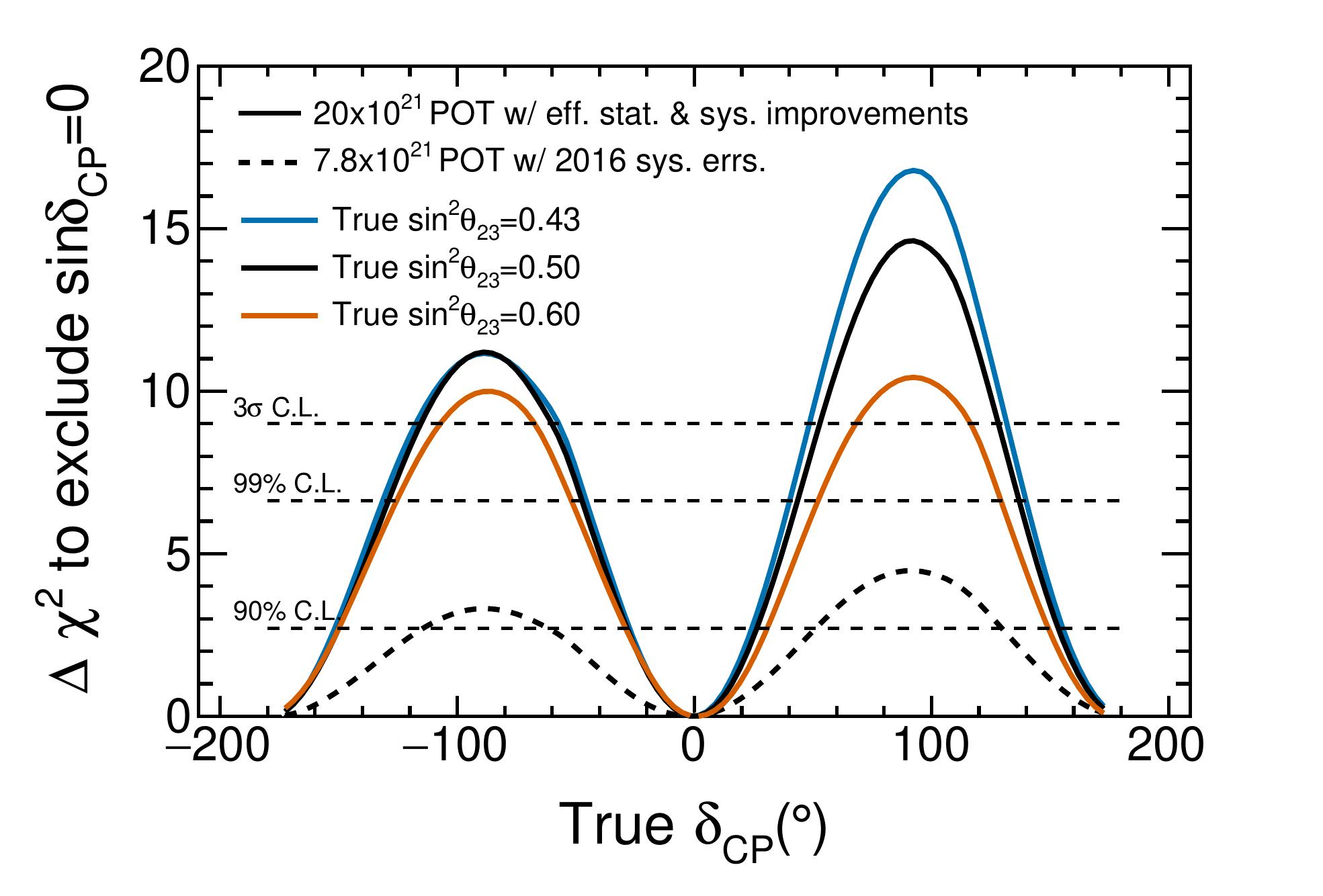}\caption{Assuming the MH is known -- measured by an outside experiment.} \label{fig:CPVvsdCP_knownMH}
\end{subfigure} \quad
\end{center}
\caption[CPV vs dCP]{Sensitivity to CP violation as a function of true
\(\delta_{CP}\) for the full T2K-II exposure of $20\times 10^{21}$ POT
with a 50\% improvement in the effective statistics, a reduction of the systematic uncertainties to 2/3 of their current size, and assuming that the  true MH is the normal MH. The left plot is with assumption of unknown mass hierarchy and the right is with known mass hierarchy.
Sensitivities at three different values of $\sin^2\theta_{23}$ (0.43, 0.5 and 0.6) are shown.
\label{fig:CPVvsdCP0}} \end{figure}

With these accelerator and beamline upgrades, as well as analysis improvements,
our sensitivity to CP violation is shown in Figure~\ref{fig:CPVvsdCP0}.
The sensitivity reaches 3 $\sigma$ or higher
for the oscillation parameter region favored by our latest result:
$\delta_\mathrm{CP}=-\frac{\pi}{2}$, $0.43<\sin^2\theta_{23}<0.6$, and normal mass hierarchy.
The fractional region for which \(\sin\delta_{CP}=0\) can be excluded at the 99\% (3\(\sigma\)) 
C.L.\ for \mbox{$\sin^2\theta_{23}=0.5$} case is 49\% (36\%) of possible true values of \(\delta_{CP}\)
assuming that the MH has been determined by an outside experiment. 

It was surprising that the flavor-mass mixing in the lepton sector is very different
from that in the quark sector.
The current measured value of $\theta_{23}$ is consistent
with maximum mixing: \mbox{45$^\circ$} with \mbox{3.2$^\circ$} uncertainty.
The precise determination of this value, whether the mixing is maximal or not,
would guide us to understand the origin of the flavor-mass mixing.
We expect that the precision of $\theta_{23}$ reaches 1.7$^\circ$ or better with this program.
The squared mass difference $\Delta m^2_{32}$ will be determined with $\sim 1\%$ precision.

Precise measurements of neutrino-nucleus interactions at the near detectors during T2K-II
would contribute critically to the reduction of systematic uncertainties arising from neutrino interaction modelling in future accelerator-based long baseline experiments.
T2K-II would also perform 
searches for physics beyond the standard model.
In particular, the combination of accelerator-based long baseline measurements
with $\nu_\mu/\bar{\nu}_\mu$ beams and reactor measurements with
$\bar{\nu}_e$ flux would give redundant constraints
on ($\Delta m^2_{32}, \sin^2\theta_{23}, \delta_{CP}$).
New physics could show up as an inconsistency in these measurements.

To realize these physics goals, especially the first observation of CP violation at the $3\;\sigma$ level by JFY2026,
we propose to extend the run from $\onepott$ to $\twopott$
with J-PARC Main Ring upgrades to operate at 1.3~MW
following the timeline 
shown in Figure \ref{fig:POT}.

\newpage

\tableofcontents

\newpage
\setcounter{page}{1}
\renewcommand{\thepage}{\arabic{page}}
\section{Introduction}
The T2K long baseline neutrino oscillation experiment sends a beam of muon neutrinos ($\nu_\mu$) produced at the J-PARC accelerator in Tokai to the Super-Kamiokande detector 295 km away to study neutrino oscillations arising from the mixing of neutrino flavor and  mass eigenstates. There, the depletion of muon neutrinos due to their conversion into other neutrino flavors ($\nu_e$ and $\nu_\tau$) can be precisely measured along with the appearance of $\nu_e$ interactions arising from the oscillation process. A muon antineutrino ($\bar{\nu}_\mu$) beam can also be produced to study the corresponding antineutrino processes. 

Since starting operations in 2010, the experiment has achieved a number of major milestones in the study of  neutrino oscillations:
\begin{itemize}
\item The observation of $\nu_\mu\to\nu_e$ oscillations in a series of analyses spanning 2011-2013\cite{Abe:2011sj,Abe:2013xua,Abe:2013hdq}.
  This was the first time neutrinos produced in one flavor has been explicitly observed interacting as another flavor
  due to the oscillation process. It  has opened the possibility for measuring CP violation arising from an irreducible CP-odd phase ($\delta_{CP}$) in the mixing as described later.
\item The most precise measurement of $\theta_{23}$ through $\nu_\mu$ disappearance, one of three mixing angles fundamental to neutrino mixing in 2014\cite{Abe:2014ugx}.
\item A joint analysis of the $\nu_\mu\to\nu_e$ appearance and $\nu_\mu$ disappearance channels to place the first significant constraints on $\delta_{CP}$ in 2015\cite{Abe:2015awa}. Combined with information from other experiments, the relatively large signal of  $\nu_\mu\to\nu_e$ observed at T2K may be a hint of large CP violating effects that  enhance this transition while suppressing the corresponding $\bar{\nu}_\mu\to\bar{\nu}_e$ process in antineutrinos.
\item The measurement of $\bar{\nu}_\mu$ disappearance with one year of antineutrino running in 2015\cite{Abe:2015ibe}
  with precision competitive to other experiments.
\end{itemize}
In addition to these  achievements, the collaboration has engaged in an extensive program of systematic error reduction through improving the modelling of the neutrino flux and developing near detector measurements to constrain the uncertainties resulting from backgrounds and the modelling of neutrino-nucleus interactions.  Dedicated programs of neutrino-nucleus interaction studies and searches for exotic neutrino properties are also under way. After accumulating our first substantial sample of antineutrino interactions, T2K is now strengthening what may be the first, albeit inconclusive, indications of CP violation in neutrinos by directly comparing the $\nu_\mu\to\nu_e$ transition to its antineutrino counterpart. This process has been expedited by continuous improvements in accelerator performance which now allow 420 kW operation. We  welcome the prospect of further increasing the beam power following the upgrade of the Main Ring power supplies in the next few years to the design power of 750 kW and beyond.

These developments raise the possibility that the  observation of CP violation in neutrino oscillations, recognized globally as one of the next major goals in particle physics, may be achieved at T2K with increased statistics in advance of the next generation of experiments that are expected to start  {\em circa} 2026. We propose to extend the T2K run beyond the currently approved $\onepott$ to $\twopott$, which will allow T2K to observe CP violation with $>3 \sigma$ significance if the neutrino oscillation parameters are close to their currently favored values. Accomplishing this by 2026  will require accelerator and beamline upgrades to handle even higher beam power, eventually reaching 1.3 MW, and improvements to the horn magnetic focussing devices to increase the neutrino flux. Also needed are  analysis improvements to increase the effective efficiency for identifying oscillated $\nu_e/\bar{\nu}_e$ events in the far detector and the reduction of systematic uncertainties.

In what follows, we briefly review the physics of neutrino oscillations  (Section~\ref{sec:nuosc})  and the current status of the T2K experimental apparatuses  (Section \ref{sec:t2know}) and analysis effort (Section \ref{sec:t2kana}). In Section \ref{sec:t2kupgrade} we describe the hardware and analysis improvements necessary to accomplish the physics goals described  in Section \ref{sec:physics}. We dconclude with a summary in Section \ref{sec:summary}. 

\newpage
\section{Neutrino Oscillations}
\label{sec:nuosc}
\subsection{Three Flavor Formalism}
Neutrino oscillations, the evolution of the flavor content of a neutrino as it propagates in space and time, result from the mixing of neutrino flavor and mass eigenstates\cite{Pontecorvo:1957cp,Pontecorvo:1957qd,Pontecorvo:1967fh,Maki:1962mu}. The discovery of neutrino oscillations in  atmospheric\cite{Fukuda:1998mi} and solar \cite{Hirata:1989zj,Ahmad:2001an,Ahmad:2002jz,Fukuda:1998fd} neutrinos established that neutrinos in fact have non-zero and non-degenerate masses, the only indication of phenomena beyond the Standard Model in particle physics, and that the mixing is large. Since then, a variety of experiments have studied several modes of oscillations and established an overall picture that is consistent with the three flavor framework,  though some possibility of phenomena beyond this framework have been reported. 

Recently, T2K\cite{Abe:2013hdq} and reactor experiments\cite{An:2016bvr,RENO:2015ksa,Abe:2014bwa} have established that the full three-flavor mixing needed to induce interference terms leading to $CP$-violating effects\cite{Kobayashi:1973fv} in neutrino oscillations is present, opening up the possibility to observe and study $CP$-violation in neutrino oscillations. This would have profound implications for particle physics as it would constitute the only observed source of $CP$-violation outside of quark mixing, and for cosmology,  new sources of $CP$-violation  are necessary to explain the observed matter-dominance of the universe. In {\em leptogenesis}, CP violation related to neutrinos in the early universe is responsible for this primordial asymmetry\cite{Fukugita:1986hr,Pascoli:2006ci}.

In the three-flavor framework, unitary neutrino mixing can be parametrized by three mixing angles ($\theta_{12}$, $\theta_{23}$, and $\theta_{13}$) and a $CP$-odd phase ($\delta_{CP}$), as follows:
$$
\tiny{
\left(
\begin{array}{c}
\nu_e \\
\nu_\mu \\
\nu_\tau
\end{array}
\right)
=  
\left(
\begin{array}{ccc}
1			&	0			&		0\\
0			&     \cos\theta_{23} 	&  \sin\theta_{23} \\
0			& -\sin\theta_{23} 	& \cos\theta_{23} 
\end{array}
\right)
\left(
\begin{array}{ccc}
\cos\theta_{13} &  0		& \sin\theta_{13}e^{-i\delta_{CP}} \\
0			& 1		& 0			\\
-\sin\theta_{13} e^{+i\delta_{CP}}	& 0		& \cos\theta_{13} 
\end{array}
\right)
\left(
\begin{array}{ccc}
\cos\theta_{12} 	& \sin\theta_{12} & 0 \\
-\sin\theta_{12} 	& \cos\theta_{12} & 0 \\
0		 	& 0			& 1 \\
\end{array}
\right)
\left(
\begin{array}{c}
\nu_1 \\
\nu_2 \\
\nu_3
\end{array}
\right)
}
$$
where the $\theta_{ij}$ parameterize  $2\times 2$ rotations of the $i$th and $j$th rows and columns  and $\nu_i$  are the mass eigenstates of the neutrino.  In general, the mixing angles $\theta_{ij}$ determine the amplitudes of the interfering oscillation terms, while the mass-splittings $\Delta m_{ij}^2 \equiv m_i^2 - m_j^2$ determine their frequency in terms of $L/E$, where $L$ is the distance between the production and detection of the neutrino and $E$ its energy.  The irreducible $CP$-odd phase $\delta_{CP}$ gives rise to asymmetries in the neutrino oscillations relative to the corresponding antineutrino process if $\sin\delta_{CP} \neq 0$. For neutrinos propagating through matter, coherent forward-scattering effects also induce differences in neutrino and antineutrino oscillations, and are also sensitive to the ordering of the mass eigenvalues\cite{Wolfenstein:1977ue,Mikheev:1986gs}. 

\subsection{Current Status of Parameters}
\begin{figure}[t]
\begin{subtable}[l]{0.52\textwidth}
\begin{tabular}{ll}\hline\hline
Mixing angles  		&    						\\ \hline
$\sin^2 \theta_{12}$ 	&  $0.304\pm 0014$			\\
$\sin^2 \theta_{23}$ 	&  $0.514[0.511]^{+0.055}_{-0.056}$			\\
$\sin^2 \theta_{13}$  &  $(2.19\pm 0.12)\times 10^{-2}$			\\ \hline
				&									\\ \hline\hline
Mass splittings		&									\\ \hline				
$\Delta m^2_{21}$ 	&  $(7.53 \pm 0.018)\times10^{-5} \mbox{eV}^2$		\\
$|\Delta m^2_{32}|$ 	&  $(2.44 [2.49]\pm 0.06)\times10^{-3}\mbox{eV}^2$	\\ \hline
\end{tabular}
\end{subtable}
~~~~~~~~
\begin{subfigure}[r]{0.38\textwidth}
 \includegraphics[width=1.00\textwidth]{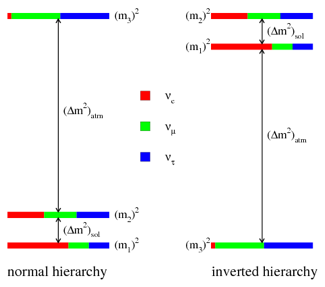} 
\end{subfigure}
\caption{Left: Current values of neutrino mixing and mass parameters.
  For $\sin^2\theta_{23}$ and $|\Delta m^2_{32}|$, the  pair of values (with one in brackets) indicate values extracted assuming the normal [inverted] hierarchy\cite{pdg}.
 Right: Representation of the two possible mass hierarchies with flavor content of each mass eigenstate.}
\label{fig:numassmix}
\end{figure}

A summary of the current status of the mixing parameters is shown in Figure \ref{fig:numassmix}. Measurements of $\theta_{12}$ and $\Delta m^2_{21}$ come from analysis of solar and long-baseline reactor data, while $\theta_{23}$ and $\Delta m^2_{32}$ are measured with long-baseline neutrino experiments (see below) and atmospheric neutrinos. Short baseline reactor experiments have recently provided precise measurements of $\theta_{13}$.
       
  It should be noted  that the flavor-mass mixing in the lepton sector
  is very different from that in the quark sector.
All elements of the mixing matrix are large, even the smallest, mixing angle $\theta_{13}$ is about $8^\circ$.
  The measured value of the mixing angle $\theta_{23}$ is consistent
with maximum mixing : \mbox{45$^\circ$} with \mbox{3.2$^\circ$} uncertainty.
The precise determination of this value, whether the mixing is maximal or not,
would guide us to understand the origin of the flavor-mass mixing.

\subsection{Neutrino Oscillations at T2K}
\begin{figure}[bhtp] 
  \centering
  \includegraphics[width=0.7\textwidth]{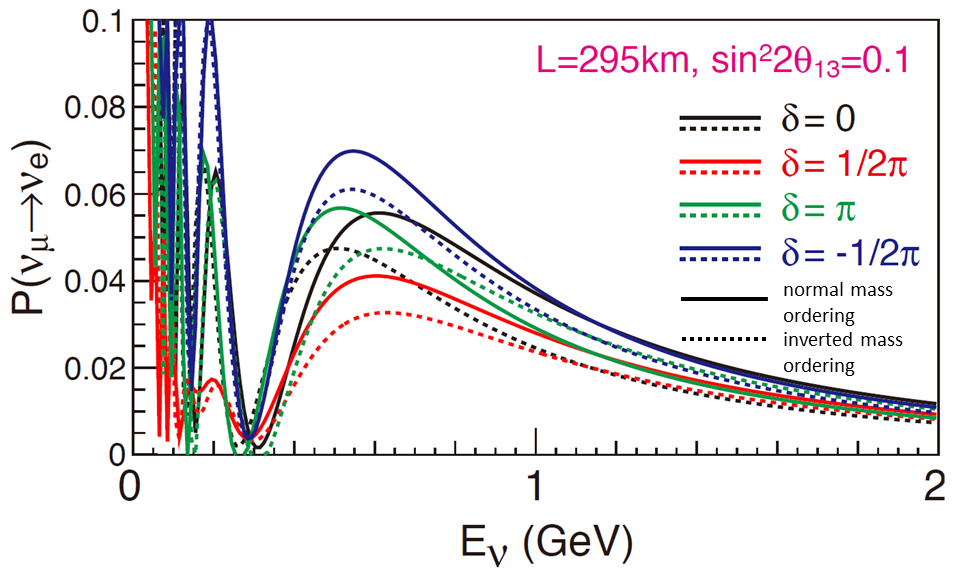} 
  \caption{$\nu_\mu\to\nu_e$ oscillation probability at T2K as a function of neutrino energy for various values of
    $\delta_\mathrm{CP}$ and mass hierarchies. $\sin^2\theta_{23}$ and $\sin^22\theta_{13}$ are fixed to 0.5 and 0.1.}
  \label{fig:oscprob}
\end{figure}

At T2K, (anti)neutrino oscillations are studied primarily through two channels with the following oscillation probabilities governed by the mixing parameters:
\begin{itemize}
\item {\bf $\nu_\mu\to\nu_{x\neq\mu}$ disappearance:}
The survival probability of muon neutrinos produced in the beam as they propagate to the far detector is given by:
\begin{equation}
P(\nu_\mu\to\nu_\mu) \approx 1- \left(\cos^4 2\theta_{13}\sin^2 2\theta_{23} +\sin^2 2\theta_{13} \sin^2\theta_{23}\right) \sin^2\Delta_{31} 
\end{equation}

Here, $\Delta_{31} \equiv \Delta m_{31}^2\frac{L}{4E}$.
While $\theta_{13}\neq 0$ introduces a dependence on $\sin^2\theta_{23}$ which is in principle sensitive to the ``octant'' of $\theta_{23}$ ({\em i.e.} if $\theta_{23}\neq \pi/4$, which side of $\pi/4$ it lies), an effective degeneracy still exists in that nearly identical oscillation probabilities result for pairs of $\theta_{23}$ values on either side of $\pi/4$ for currently allowed values of $\theta_{23}$\cite{parke}.  Note that the disappearance probability is maximized for $\theta_{23}\neq \pi/4$, resulting in distinct ``maximal mixing'' and ``maximal oscillation'' conditions. $\nu_\mu$ disappearance is the channel through which precise measurements of $\sin^22\theta_{23}$ and $\Delta m^2_{32}$ can be obtained.
\item {\bf $\nu_\mu\to\nu_e$ appearance}\cite{Freund:2001pn}
\begin{equation}
\begin{array}{llll}
P(\nu_\mu\to\nu_e)  \approx 	& \sin^2 2\theta_{13} 	& \times \sin^2\theta_{23} 								& \times \frac{\sin^2[(1-x)\Delta_{31}]}{(1-x)^2} \\
 						& -\alpha \sin\delta_{CP}  	& \times \sin 2\theta_{12} \sin 2\theta_{13} \sin 2\theta_{23} 	
											& \times \sin\Delta_{31} \frac{\sin[x\Delta_{31}]}{x} \frac{\sin[(1-x)\Delta_{31}]}{1-x} \\
 						& +\alpha \cos\delta_{CP}	& \times \sin 2\theta_{12} \sin 2\theta_{13} \sin 2\theta_{23} 	
											& \times \cos\Delta_{31} \frac{\sin[x\Delta_{31}]}{x} \frac{\sin[(1-x)\Delta_{31}]}{1-x} \\
						& + \mathcal{O}(\alpha^2)	&												&	
\end{array}
\label{eq:nueapp}
\end{equation}
Here, the terms with  $x=\frac{2\sqrt{2}G_F N_e E}{\Delta m^2_{31}}$ accounts for matter effects which alter this oscillation probability depending on the mass hierarchy ({\em i.e.} the sign of $\Delta m^2_{31}$) and switches sign depending on whether we consider neutrino or antineutrino oscillations.  The expression results from an  expansion in $\alpha\equiv \Delta m_{21}^2/\Delta m_{31}^2 \sim 1/30$ that separates oscillations driven by the ``solar'' ($\Delta m_{21}^2$) and ``atmospheric'' ($\Delta m_{31}^2$) mass splittings.
The second term proportional to $\sin\delta$ is $CP$-odd, switching signs when considering the antineutrino channel and changes
the oscillation probability by $\pm27\%$ at most,
while the third term proportional to $\cos\delta_{CP}$ is $CP$-even. 
Figure~\ref{fig:oscprob} gives oscillation probabilities for various values of $\delta_\mathrm{CP}$ and mass hierarchies.

The oscillation probability depends on all three mixing angles, including the $\theta_{23}$ octant, and the mass hierarchy.
With $\theta_{13}$ and $\theta_{12}$ determined precisely by reactor and solar neutrino experiments,
and the matter effect relatively small ($\sim\pm10\%$ for $L=295$ km, $E=0.6$ GeV),
the probability is sensitive to $\sin^2\theta_{23}$ and to $\delta_{CP}$.
Since $\sin^22\theta_{23}$ is most precisely measured by $\nu_\mu$ disappearance measurements at T2K,
this naturally leads to a joint analysis of both modes across neutrino and antineutrino channels.
\end{itemize}

While the expressions shown result on approximations based on the relative sizes of the mixing angles as well as the magnitude of solar and atmospheric terms at the T2K baseline and energy, exact expressions are used when performing oscillation analyses at T2K. 

\begin{figure}
  \centering
  \begin{subfigure}{0.48\textwidth}
    \includegraphics[width=\textwidth]{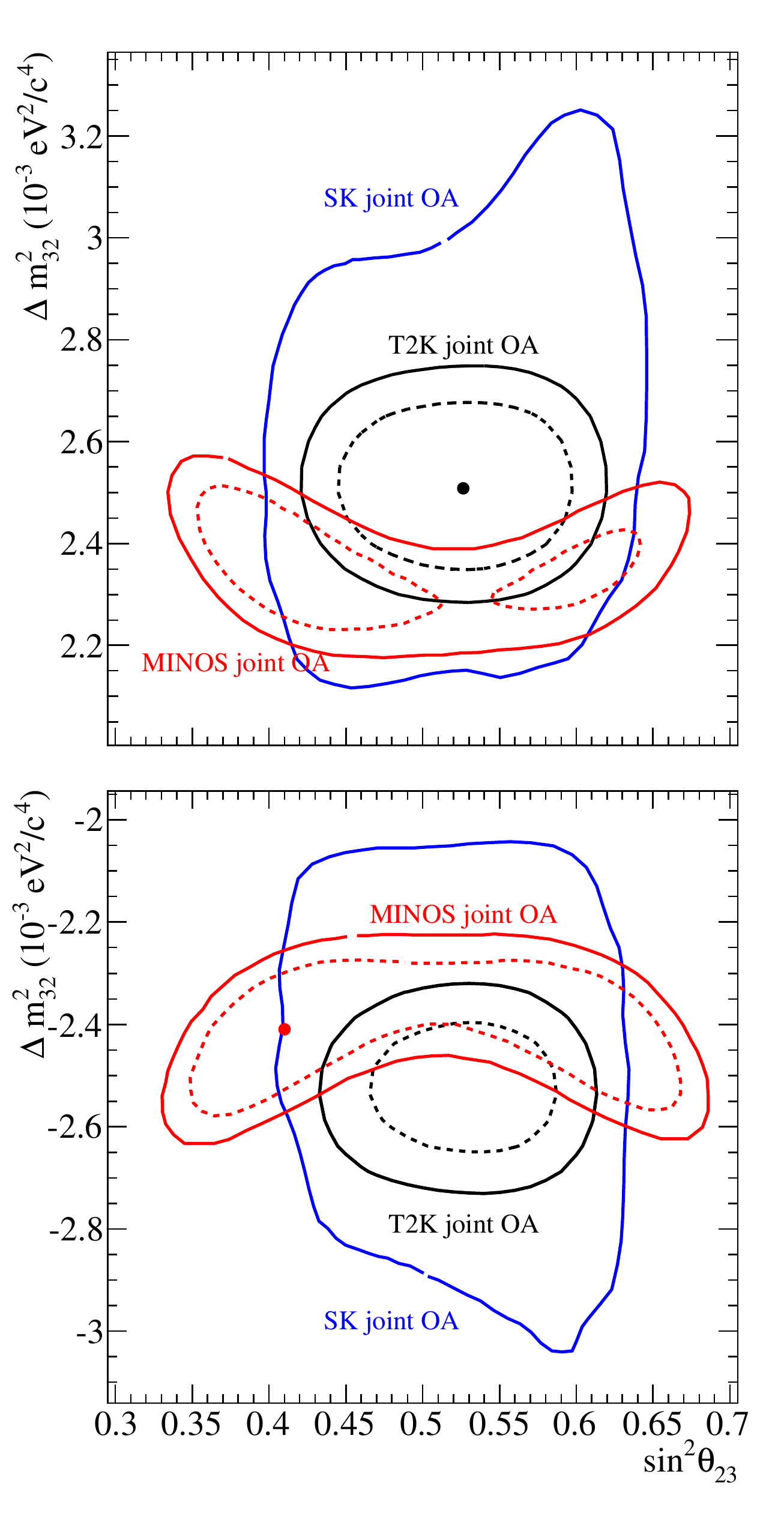}
    \caption{68\% (dashed) and 90\% (solid) C.L. regions for normal (top) and inverted (bottom) mass
      hierarchy in the $(\sin^2\theta_{23},\Delta m^2_{32})$ space.}
    \label{fig:th_23}
  \end{subfigure}
  \begin{subfigure}{0.48\textwidth}
    \includegraphics[width=\textwidth]{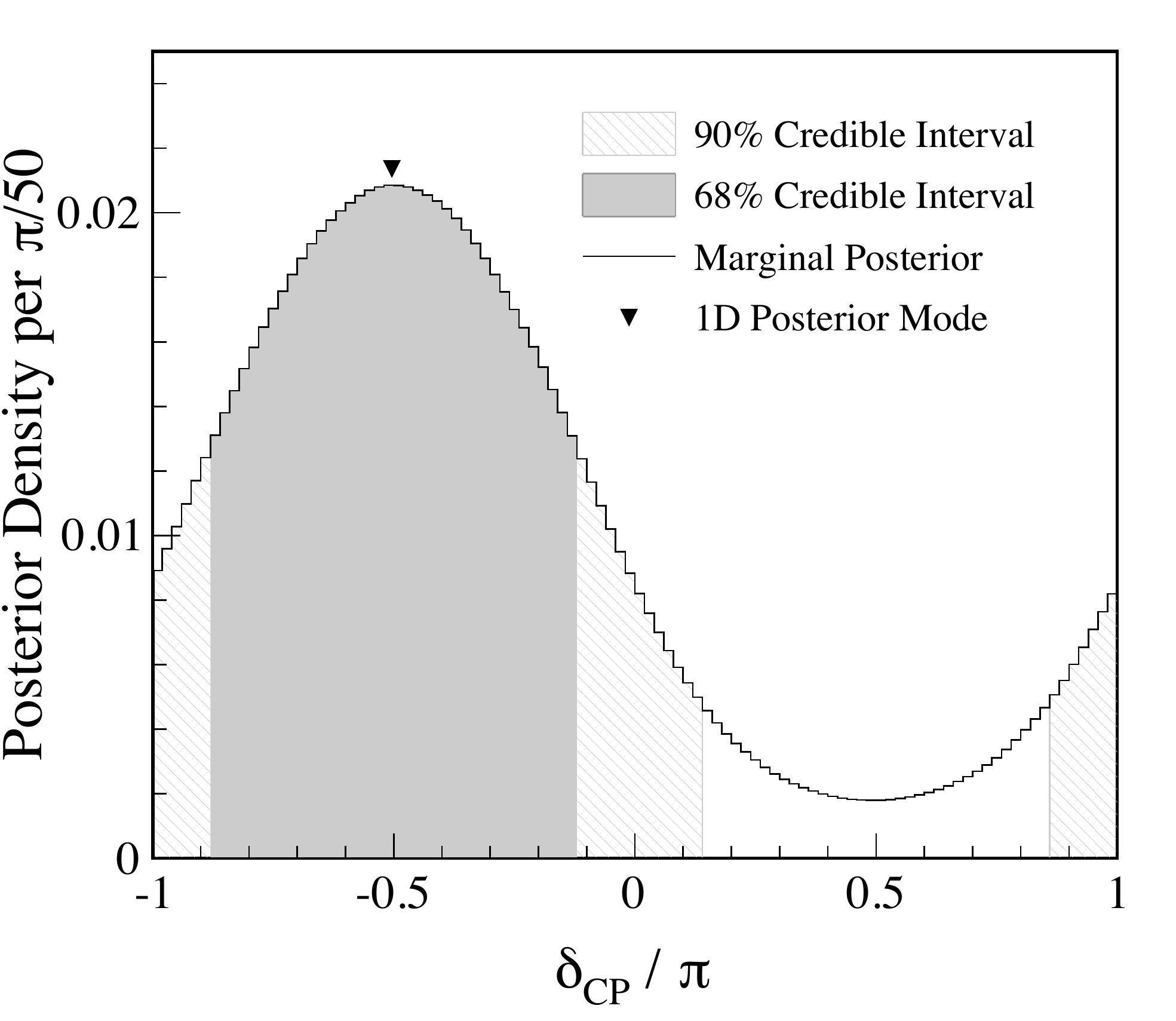}
    \caption{$\delta_{CP}$ constraints}
    \label{fig:dcp}
    \vskip 2cm
    \begin{tabular}{lcc|c}\hline\hline
      & NH		& IH 		& Sum \\ \hline
      $\sin^2\theta_{23} \leq 0.5$ 	& 0.179		& 0.078	& 0.257 \\
      $\sin^2\theta_{23} > 0.5$ 		& 0.505		& 0.238	& 0.743 \\ \hline
      Sum						& 0.684		& 0.316	& 1.000 \\ \hline
    \end{tabular}
    \caption{Bayesian posterior probabilities for combinations
      of neutrino mass hierarchy (normal (NH) or inverted(IH))
      \\ and $\theta_{23}$ octant.}
  \end{subfigure}
  \caption{Current T2K oscillation results obtained from Run1-4 $\nu-$mode data~\cite{Abe:2015awa}
  \label{fig:run1-4results}}
\end{figure}

\begin{figure}
  \centering
  \includegraphics[width=0.7\textwidth]{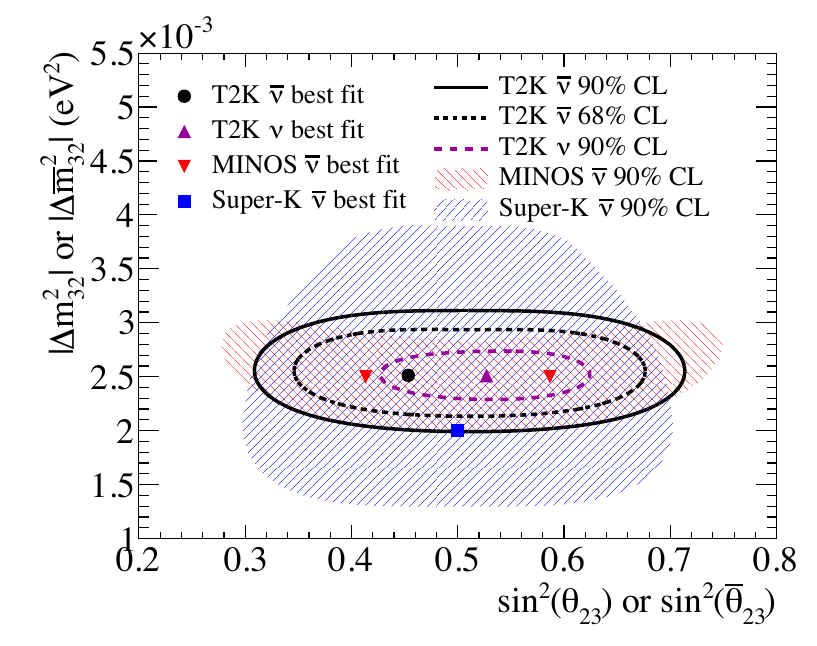}
  \caption{$\sin^2\bar{\theta}_{23}$ and $\Delta \bar{m}^2_{32}$ constraints using
    Run5 anti-neutrino data~\cite{Abe:2015ibe}}
  \label{fig:anuth_23}
\end{figure}

The current oscillation results from T2K are shown in Figures \ref{fig:run1-4results}\cite{Abe:2015awa}
and \ref{fig:anuth_23}~\cite{Abe:2015ibe}, which include world-leading measurements 
of the mixing angle $\sin^2\theta_{23}$ measured with neutrinos, the best constraint of $\delta_{CP}$ measured both with T2K  data only and in combination with reactor data, and competitive measurements of $\sin^2\theta_{23}$ measured with anti-neutrinos. According to Equation \ref{eq:nueapp}, normal hierarchy ($\Delta m^2_{31}>0$) and $\delta_{CP}\approx -\pi/2$ maximizes $P(\nu_\mu\to\nu_e)$ while minimizing $P(\bar{\nu}_\mu\to\bar{\nu}_e)$. The relatively large $\nu_e$ appearance signal observed at T2K in $\nu$-mode weakly favors these parameters. While the statistics are too small to make any conclusion, the initial search for $\bar{\nu}_e$ appearance at T2K is consistent with this picture (3 events observed with an expectation of 3.2 for $\delta_{CP}=-\pi/2$ and normal hierarchy).

\subsection{Outlook}
In the next several years, continued operation of T2K and NOvA with higher beam power is expected to improve the precision on $\theta_{23}$ and $\Delta m^2_{32}$ and the constraints on $\delta_{CP}$, reaching up to $90\%$ confidence level sensitivity for CPV with the currently approved exposures on the timescale of $\sim 2021$. Reactor experiments will increase precision on $\sin^22\theta_{13}$  and $\Delta m^2_{ee}$ to $\sim 3\%$.

There are several opportunities in the near future for determining the mass hierarchy at $>3\sigma$ level. Due to its longer baseline and higher energy, NOvA has  greater sensitivity to the mass hierarchy than T2K in its oscillation measurements. Combined with T2K, the mass hierarchy sensitivity reaches $3\;\sigma$ in certain cases, including the currently favored oscillation parameters values. Super-Kamiokande, INO, ORCA, and PINGU also have the opportunity to use matter effects with atmospheric neutrinos to resolve the neutrino mass hierarchy, with ORCA and PINGU expected to reach $>3\;\sigma$ sensitivity with 3-4 years of operation. These experiments also expect to achieve precise measurements of $\theta_{23}$ comparable to those of T2K and NOvA. Finally, JUNO and RENO-50 aim to establish the mass hierarchy by studying $\bar{\nu}_e$ disappearance at $\sim 60$ km, where oscillations induced by the solar and atmospheric splittings interfere and produce a shift in the observed energy spectrum that depends on the mass hierarchy. INO, ORCA, PINGU, and RENO-50 are currently seeking approval.

\newpage
\section{Overview of Current T2K Experimental Setup}
\label{sec:t2know}
\begin{figure}[htbp] 
  \centering
  \includegraphics[width=0.7\textwidth]{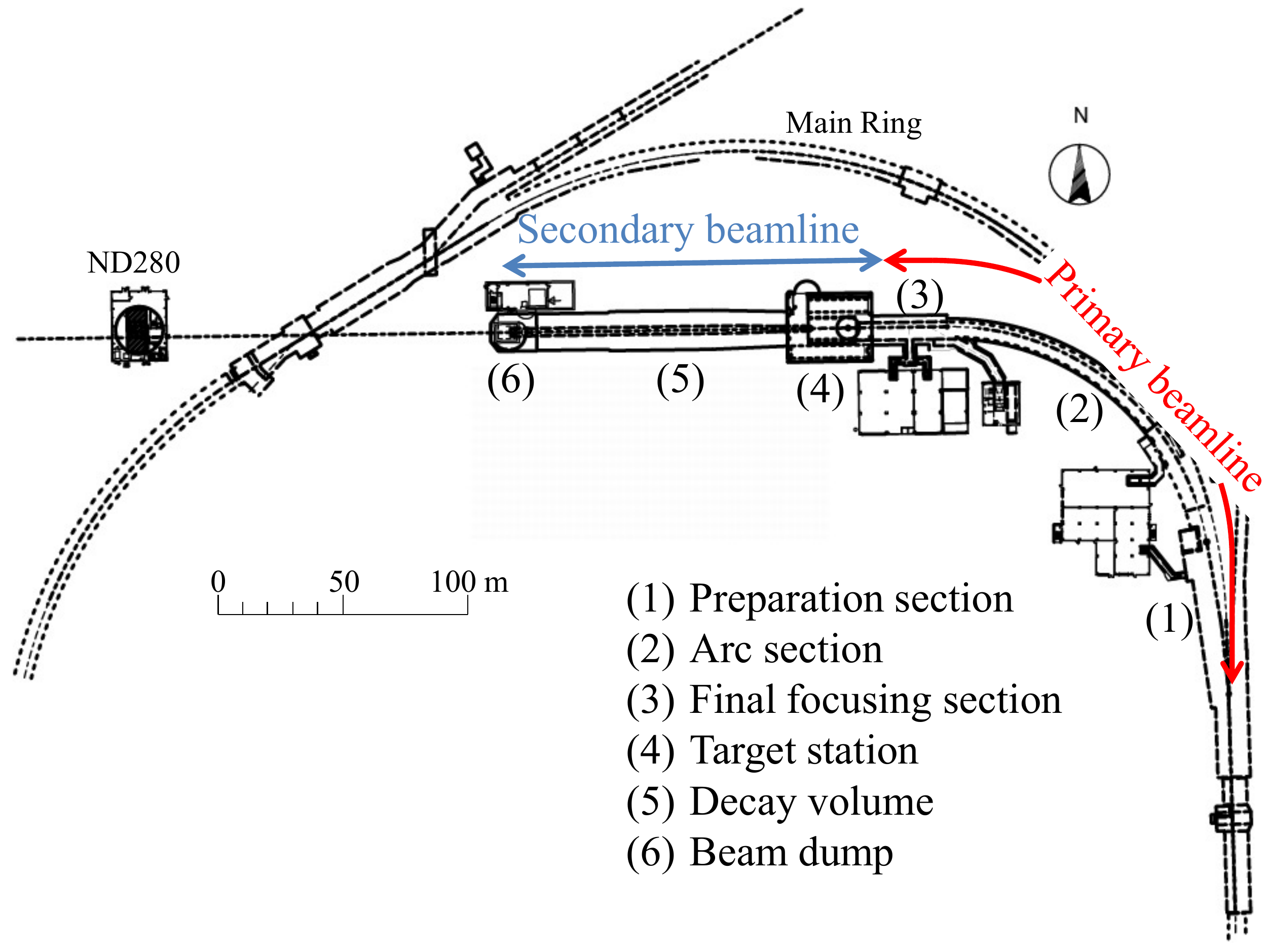} 
  \caption{Overview of the T2K neutrino beamline.}
  \label{fig:beamline}
\end{figure}

T2K uses 30 GeV protons from the J-PARC Main Ring (MR) to produce a beam of primarily muon (anti-)neutrinos whose center is directed 2.5 degrees off  the line of sight (off-axis) connecting 
the neutrino production target and the far detector, Super-Kamiokande (Super-K, SK), located 295~km away in Gifu prefecture.  The T2K neutrino beamline is shown in Figure \ref{fig:beamline}.
Magnetic horns surrounding and downstream of the production target focus charged pions along the beam axis, where their subsequent decays produce (anti-)neutrinos in the same direction.
Changing the polarity of the horn current enhances the resulting beam in either neutrinos ($\nu$-mode beam)
or antineutrinos ($\bar{\nu}$-mode beam).
In this configuration, the $\nu_{\mu} \rightarrow \nu_{e}$ oscillation probability is expected to be maximal for neutrinos with energies around $\sim$ 600 MeV, the peak energy of the neutrino energy spectrum at this off-axis angle.

Measurements at a complex of detectors (near detectors) located 280~m downstream of the target are used to provide constraints on the neutrino direction, flux, and interaction models 
before standard oscillation effects have distorted the neutrino spectrum. In addition, they make independent neutrino cross section measurements.

A detailed description of the T2K experiment can be found in Reference \cite{Abe:2011ks}. Here, we present a short summary of the current experimental apparatus.

\subsection{Primary and Secondary Beamlines}
\begin{figure}[htbp] 
  \centering
  \includegraphics[width=0.7\textwidth]{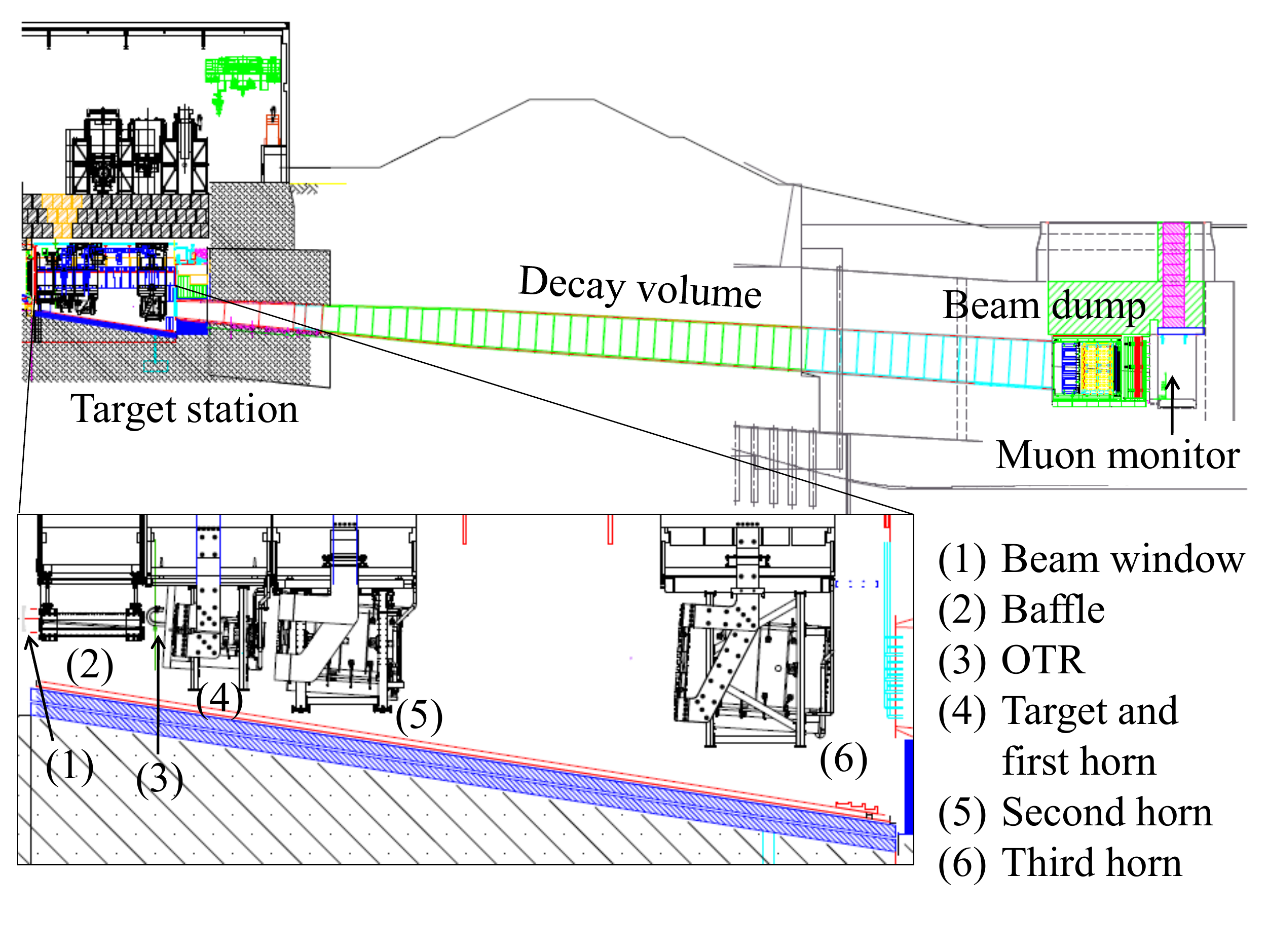} 
  \caption{Side view of the secondary beamline.}
  \label{fig:beamline_sec}
\end{figure}

The primary beamline transports the extracted protons toward the production target.
A series of normal-conducting and superconducting magnets are located along the beamline 
to focus the protons on the target with appropriate position, direction, and size. 
The intensity, profile, and position of the protons
are measured by various beam monitors to allow for precise control of the proton beam, and this information is also used in analysis to predict the expected neutrino flux.
The secondary beamline is shown in Figure~\ref{fig:beamline_sec}.

The production target is a graphite rod (90~cm in length and 2.6~cm in diameter)
enclosed by a titanium container and is designed to survive thermal shocks resulting from beam delivery with up to  $3.3\times 10^{14}$~protons/pulse (ppp). 
High speed helium gas flow of $\sim$200~m/s provides sufficient cooling for 750 kW beam operation. The three magnetic horns are designed for 320~kA pulsed current and maximize focussing of pions with low momentum and high emission angle. The aluminum conductors are cooled by sprayed water.

The decay volume is a 94~m-long tunnel with a vertically elongated rectangular cross-section allowing variation of the off-axis angle to SK from $2.0^\circ$ to  $2.5^{\circ}$. 
The beam dump is composed of graphite blocks with aluminum water cooling modules attached.
The target, magnetic horns, decay volume, and beam dump are enclosed by
a gigantic iron vessel, filled with 1~atm. helium gas 
to reduce pion absorption and  suppress tritium and nitrogen oxide production.
Water cooling channels are attached along the inner surface of the helium vessel. 
The helium vessel and beam dump, which are inaccessible due to the high radioactivity
after beam exposure, are designed to survive thermal stress from $3\sim 4$~MW beam. 
The muon monitor detects tertiary muons penetrating the beam dump
and monitors the direction, profile, and intensity of the muons to check the stability of the beamline, such as the primary proton beam optics, target, and horns.

\subsection{Near Detectors : INGRID and ND280}



\begin{figure} \centering
\includegraphics[width=0.7\textwidth]{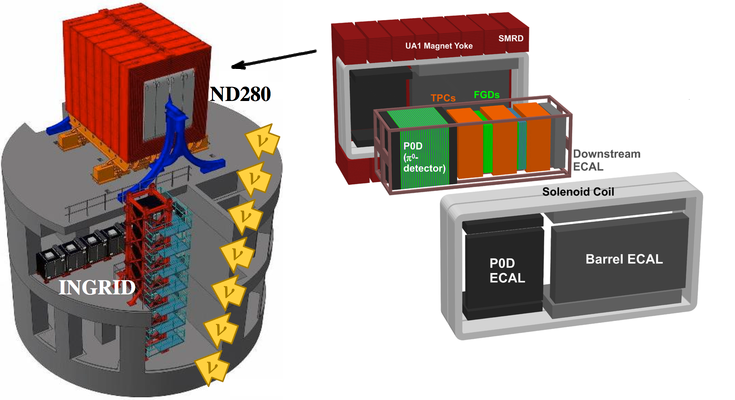}
\caption{ND280 detector suite.
\label{fig:ND280_layout}}
\end{figure}

The ND280 site, located 280 m from  the beam source, houses detectors that measure, monitor,
and constrain the beam flux before oscillation happens and neutrino-nucleus interactions.
Specifically the complex consists of an on-axis detector(INGRID) and off-axis detectors as shown in Figure~\ref{fig:ND280_layout}.
The Interactive Neutrino GRId Detector(INGRID) is composed of seven vertical and horizontal modules interleaved with planes of iron and segmented 
scintillator\cite{Abe:2011xv}. These tracker modules are arranged in a 10-m horizontal by 10-m vertical crossed array. This detector provides
high-statistics monitoring of the beam intensity, direction, profile, and stability using neutrino interactions.

The ND280 off-axis detector is a hybrid detector designed to provide constraints on the SK-directed neutrino flux, the neutrino interaction model, 
and the oscillation signal and backgrounds at the far detector. The reduced systematic errors improve the experimental sensitivity of T2K to both the appearance and disappearance oscillation signals. 
The off-axis detector is enclosed in a 0.2-T magnet which contains a sub-detector optimized to measure $\pi^{0}$ production on water (P$\O$D)\cite{Assylbekov:2011sh}, three time projection chambers (TPC1,2,3)\cite{Abgrall:2010hi} alternating with two one-ton fine grained scintillating bar detectors (FGD1,2)\cite{Amaudruz:2012agx} optimized to measure charged current interactions, and an electromagnetic calorimeter (ECal)\cite{Allan:2013ofa} that surrounds the TPC, FGD, and P$\O$D detectors.
A Side Muon Range Detector (SMRD)\cite{Aoki:2012mf}, built into slots in the
magnet flux return steel, detects muons that exit or stop in the magnet steel when the path length
exceeds the energy loss range.
The FGD1 is mainly made of plastic scintillator while FGD2 contains water layers.
The combination enables to measure the interaction on water.
Currently, detector-related systematic uncertainties of $\sim2\%$ have been
achieved in $\nu_\mu/\bar{\nu}_\mu$ charged-current samples selected in ND280.


\subsection{Far Detector : Super-Kamiokande}
The far detector, Super-Kamiokande, is a 50~kiloton cylindrical water Cherenkov detector instrumented with 11,129 20'' photomultiplier tubes (PMTs) viewing a 32~kton (22.5~kton fiducial) inner target volume \cite{Fukuda:2002uc}.
A 2~m thick cylindrical volume surrounding the target volume is instrumented with 1885 8'' PMTs and serves as an active and passive background veto.
In 2008 the detector front end electronics were upgraded to provide lossless acquisition of all channels in the detector ahead of the start of T2K beam running the 
following year \cite{Yamada:2010zzc}.
Cherenkov radiation produced in the inner volume projects onto the detector walls in ring-like patterns, whose number, topology, timing, and charge are used to 
infer the location, type, and kinematic properties of particles produced in interactions in the water. 
The reconstructed momentum and angular resolutions
for single electrons (muons) are estimated as $0.6\% + 2.6/\sqrt{P[\mbox{GeV}/c]}$ (  $1.7\% + 0.7/\sqrt{P[\mbox{GeV}/c]}$ ) and
$3.0^{\circ}$ ( $1.8^{\circ}$ ), respectively. 
Mistakenly identifying such an electron as a muon (or the reverse) is estimated to occur with probability 0.7\% (0.8\%). 
The efficiency for reconstructing delayed electrons from the decay of muons, an important discriminant in the selection of 
charged current quasi-elastic (CCQE) neutrino interactions, is 89.1\%.

Timing synchronization with the proton accelerator is an essential part of extracting beam-neutrino induced interactions from the rain of cosmic ray muons and 
atmospheric neutrinos passing through Super-K.
Two nearly identical timing systems, one each at the near and far detector complexes, are synchronized using a GPS-based method with better than 150~ns precision.
Trigger signals are generated at the accelerator and time stamped before distribution to Super-K via a virtual private network. 
At Super-Kamiokande the accelerator trigger is used as the center of a 1~ms timing window after correcting for the neutrino time-of-flight between the two 
sites. 
A software trigger is used to select interactions within this window with properties of potential interest to T2K analyses.

In 2015, the Super-K collaboration decided to proceed with an upgrade to dissolve Gadolinium Sulfate ($\mathrm{Gd_2(SO_4)_3}$) into the detector.
The very high cross section for neutron capture on gadolinium and the release of energetic gamma rays in the process will significantly enhance the neutron detection efficiency with the aim of detecting relic supernova neutrinos. 
A shutdown of Super-K is needed to enter the detector and repair leaks in the detector which is 
anticipated to coincide with a major maintenance period for the J-PARC accelerator complex such as the upgrade of the Main Ring power supplies. Following these repairs, a staged deployment of Gd with increasing concentration, starting at 0.002\% and eventually reaching 0.2\%, is planned. The precise schedule and logistics of the SK-Gd upgrade is under discussion between the SK and T2K collaborations.

The T2K-II proposal is officially supported by the Super-Kamiokande collaboration.
  

\newpage
\section{Overview of the Current T2K Measurements}
\label{sec:t2kana}

\begin{figure}[htbp] 
   \centering
   \includegraphics[width=0.7\textwidth]{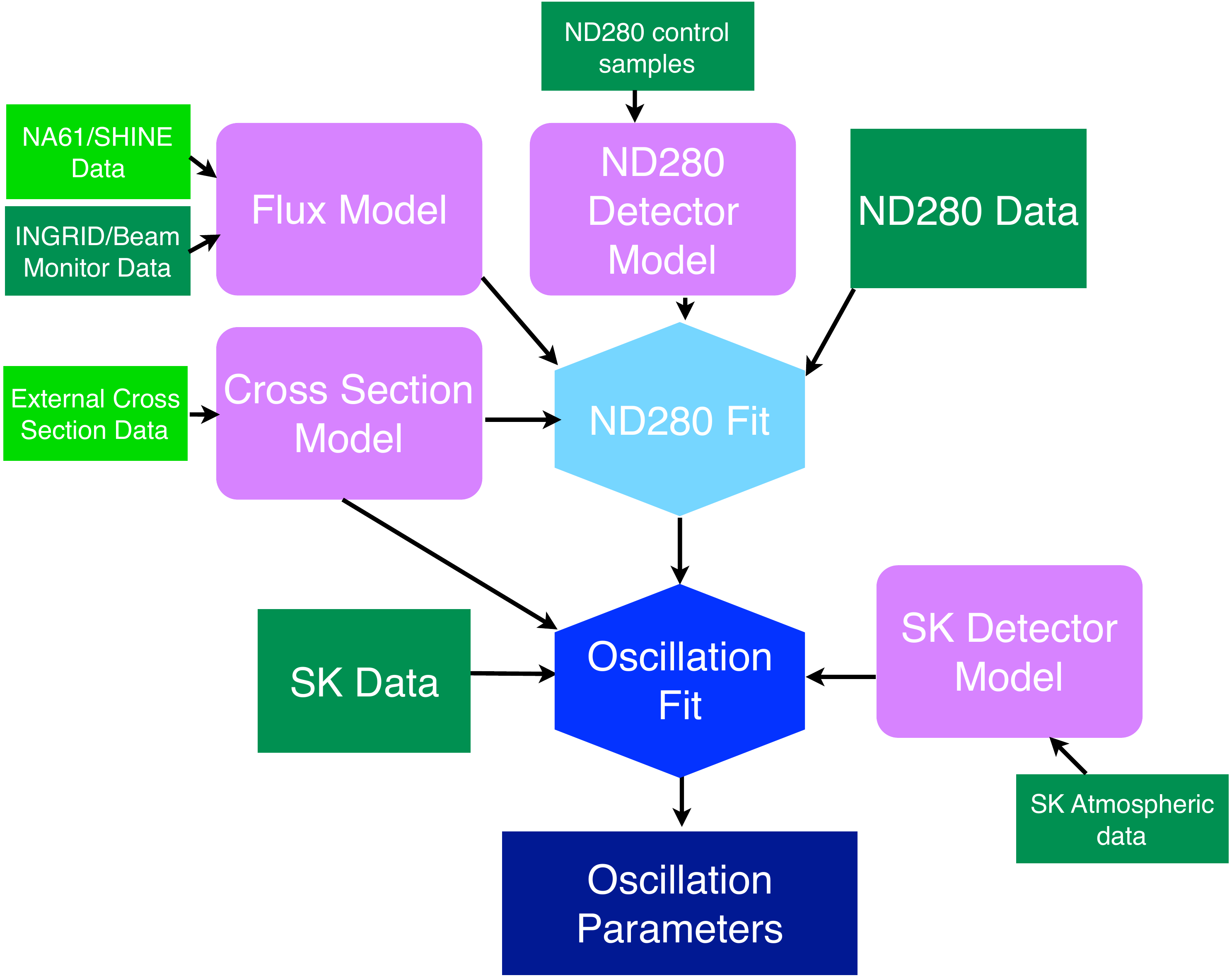} 
   \caption{The flow of oscillation analyses at T2K. Green boxes show data (lighter green for external data) which inform models (magenta boxes). A fit to ND280 data produces constraints on the models, which are fed into oscillation fits.}
   \label{fig:T2Kanalysisflow}
\end{figure}

T2K makes oscillation measurements using data samples from the near and far detectors, beamline instrumentation, and the best available measurements and physics models. Figure~\ref{fig:T2Kanalysisflow} shows the flow of T2K oscillation analyses, and what parts of the analysis are drawn from external measurements and internal measurements. 

\subsection{Flux and Cross Section Models}

\begin{figure}
    \centering
    \begin{subfigure}[b]{0.48\textwidth}
        \includegraphics[width=\textwidth]{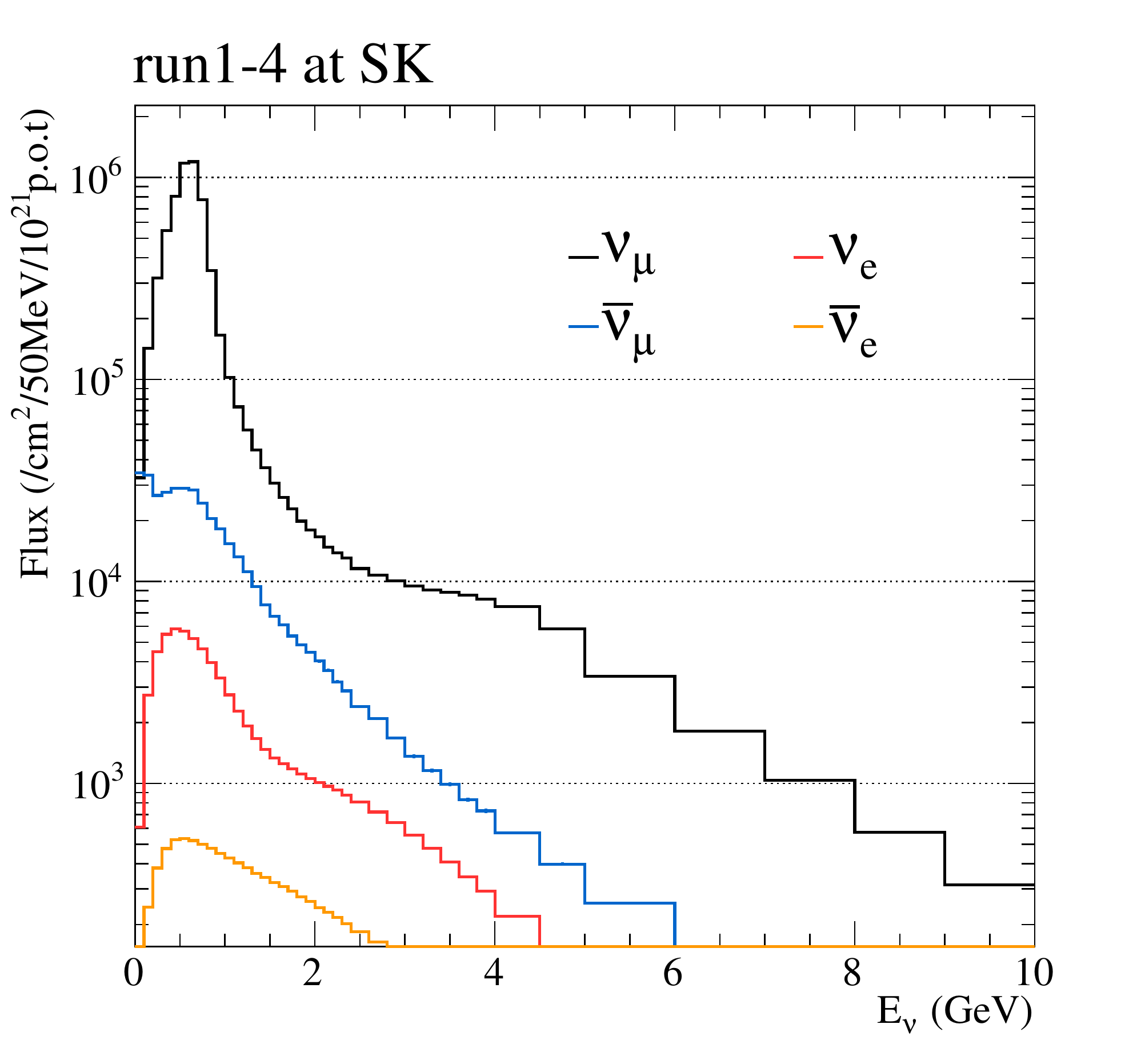}
        \caption{Neutrino-mode beam flux}
    \end{subfigure}
    \begin{subfigure}[b]{0.48\textwidth}
        \includegraphics[width=\textwidth]{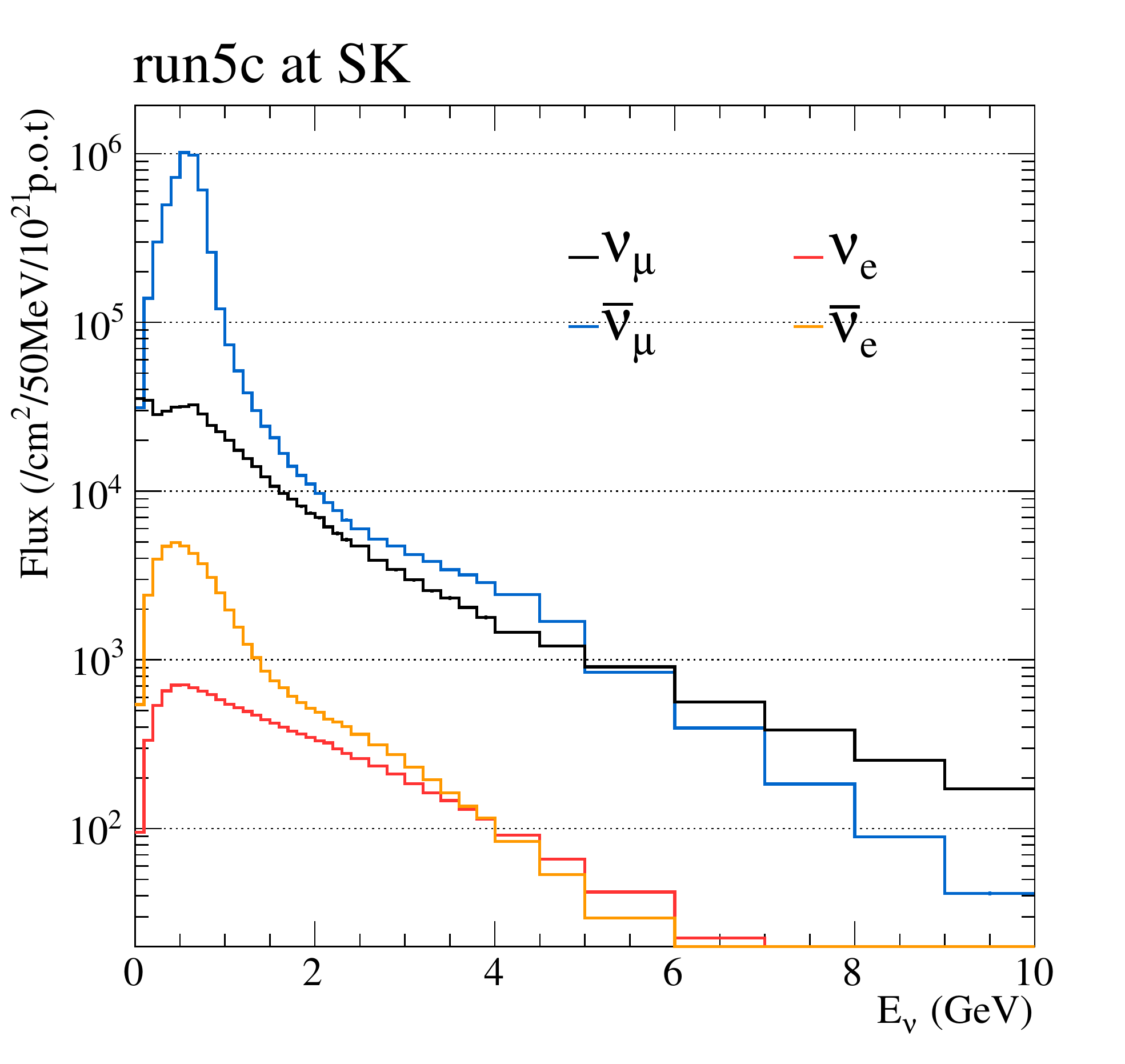}
        \caption{Antineutrino-mode beam flux} 
    \end{subfigure}
    \caption{T2K flux at the far detector.}
    \label{fig:flux}
\end{figure}

The unoscillated flux at the T2K detectors is predicted~\cite{Abe:2012av} with a simulation
of the secondary beamline using FLUKA~\cite{FLUKA1,FLUKA2} and GEANT3 with GCALOR~\cite{G3GCAL1,G3GCAL2}.
Figure~\ref{fig:flux} shows the fluxes for $\nu-$mode and $\bar{\nu}-$mode.
The NA61/SHINE thin-target data~\cite{Abgrall:2011ts,Abgrall:2011ae,Abgrall:2015hmv} through 2009 is used to tune the hadronic production
of pions and kaons in the target.
The analysis considers sources of error from the beamline, as constrained by INGRID and the beamline monitors,
as well as the uncertainties coming from the NA61/SHINE data. The uncertainties are propagated through the simulation
to form a total uncertainty on the flux, which is dominated by the hadron production uncertainty.
These uncertainties are binned by neutrino energy, flavor, and detector, and the correlations between them are calculated;
these correlations that allow near detector data to reduce the uncertainty on the flux for far detector analyses.

\begin{figure}[htbp] 
   \centering
   \includegraphics[width=0.5\textwidth]{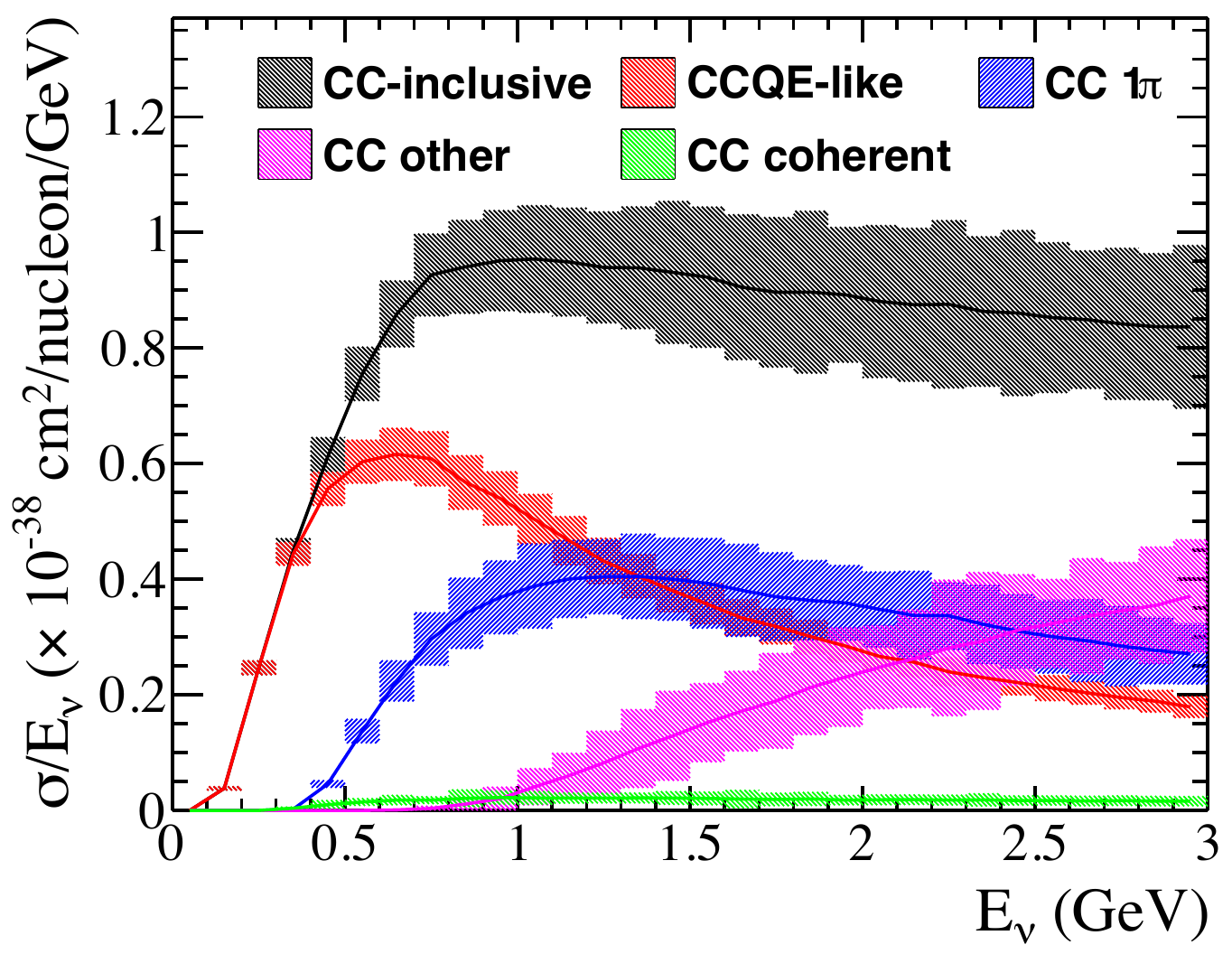} 
   \caption{Neutrino cross sections as a function of energy. The colored bands indicate to the uncertainty. The T2K beam peak is near 0.6~GeV, and so CCQE-like cross sections dominated in the peak.}
   \label{fig:nuxsec}
\end{figure}

T2K uses the NEUT neutrino interaction generator~\cite{Hayato:2009zz} to model the interaction of neutrinos on detector materials. Figure~\ref{fig:nuxsec} shows the total cross section and component cross section modes as a function of energy. The dominant cross section mode for T2K is charged current quasi-elastic (CCQE) interactions of the form $\nu_\mu + n \rightarrow \mu^- + p$ for neutrinos and $\bar{\nu}_\mu + p \rightarrow \mu^+ + n$ for antineutrinos. At slightly higher energies, CC single resonant pion and CC deep inelastic scatter events dominate. These events can still be important to the measurement as they can be reconstructed as CCQE interactions if some outgoing particles are missed. External data sets from the MINER$\nu$A, MINIBooNE, ANL and BNL experiments are used to tune the cross section model. 

\subsection{Near Detector Data}

Fourteen data samples are used from ND280 in oscillation analyses: six from $\nu$-mode running and eight from $\bar{\nu}$-mode running.
All of the data samples focus on $\nu_\mu$ charged-current interactions by selecting events with a muon in them. The samples are then further defined by the final state particles.
In the $\nu$-mode running, there are three samples in each FGD: CC0$\pi$, which has no final state pions, and is dominated by CCQE interactions; CC1$\pi^+$, which has one positive final state pion, and is dominated by resonant pion interactions; and CCOther, which contains interactions not in the other two samples, and is dominated by CCDIS interactions.
In the  $\bar{\nu}$-mode beam, there are four samples in each FGD: $\bar{\nu}$ 1 track, which has only one charged track (the muon) in the interaction, $\bar{\nu}$ N tracks, which contains all other events, and corresponding samples for interactions of neutrinos contaminating the $\bar{\nu}$-mode beam.
These last two samples provide a constraint on the ``wrong-sign" component of the $\bar{\nu}$-mode beam. In ND280, the use of both FGDs means that a sample of events on water from FGD2 is included in the analysis, which provides constraints for neutrino cross section models on oxygen. 

The ND280 samples are fit using a binned likelihood, with the samples binned according to the muon momentum and angle, including uncertainties in the flux and cross section models described above, as well as those coming from the near detector data selection and reconstruction, which are constrained using a variety of control samples. The output of this analysis is a covariance matrix which correlates flux and cross section parameters, which can be propagated to oscillation analyses. Use of the ND280 samples reduces the uncertainty due to flux and cross section uncertainties on the number of predicted events at the far detector from $\sim 10\%$ to $\sim 3\%$.

\subsection{Far Detector Data}

In the current oscillation analysis selected events are required to have vertices within the fiducial volume, defined as the 
region offset from the inner detector boundary by 200~cm, must not have particles depositing light in the veto volume, and must 
deposit more than 30~MeV of visible energy in the inner volume.
In order to reconstruct the parent neutrino energy, the event selection focuses on CCQE interactions, selecting single-ring 
events, divided into electron-like and muon-like subsamples, with either zero or not more than one decay electron, respectively.
Further cuts designed to reduce backgrounds in each subsample are described elsewhere~\cite{Abe:2015awa}.
Detector systematic uncertainties are evaluated using atmospheric data samples.

\subsection{Cross Section Results}

T2K is pursuing a complete program
of cross-section measurements of different interaction channels, on different
targets (carbon, oxygen, iron), 
for different neutrino species ($\nu_{\mu},\bar{\nu}_{\mu}$,$\nu_{e}$,$\bar{\nu}_e$) 
and at different energies  (off-axis and on-axis fluxes);
$\nu_{\mu}$ CC inclusive, CCQE and  CC0$\pi$ measurements on carbon~\cite{Abe:2013jth}\cite{Abe:2014nox}\cite{Abe:2014iza}\cite{Abe:2016tmq}
and iron~\cite{Abe:2015biq}\cite{Abe:2014nox}\cite{Abe:2015oar},
$\nu_{\mu}$ CC1$\pi$ on water~\cite{Abe:2016aoo}
and $\nu_{e}$ CC inclusive measurements on carbon~\cite{Abe:2014agb} and water~\cite{Abe:2015mxf},
CC coherent pion production on carbon~\cite{Abe:2016fic}, NC1$\gamma$ on oxygen~\cite{Abe:2014dyd},
and NC1$\pi^0$ on oxygen.
Since nuclear effects on the initial and final state have a different dependence on the target
and on the neutrino species, the comparison of these measurements will allow T2K to give an estimation
of the different nuclear effects separately. 
This effort is pursued by T2K in parallel
with a fruitful collaboration with theoreticians to improve the predictive
power and usability of the available interaction models.

\begin{figure}
    \centering
    \begin{subfigure}[b]{0.48\textwidth}
        \includegraphics[width=\textwidth]{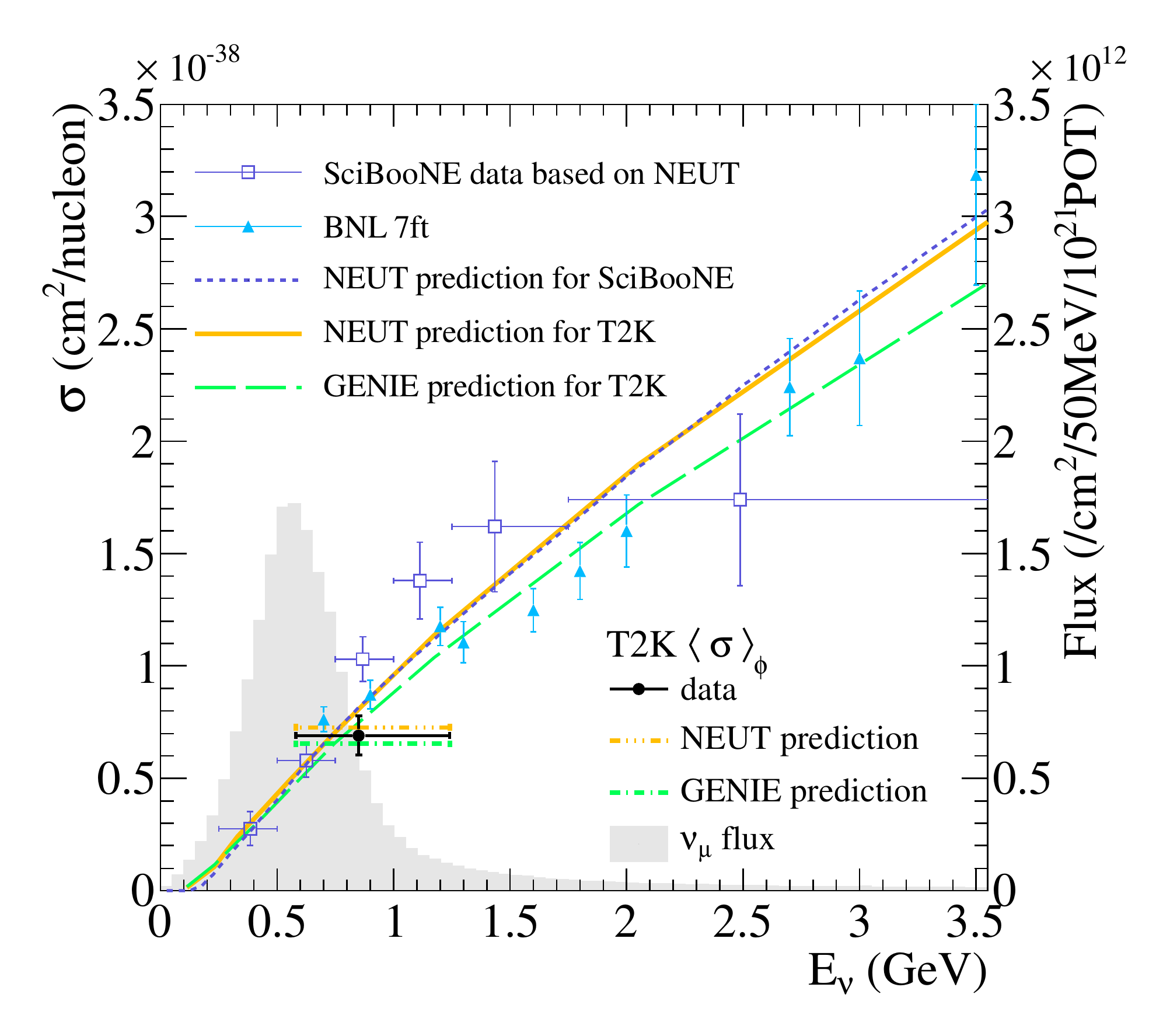}
        \caption{Total $\nu_{\mu}$ CC inclusive cross section~\cite{Abe:2013jth}. The T2K data point is placed at the mean $\nu_{\mu}$ flux. The SciBooNE $\nu_{\mu}$ data are also shown.}
        \label{fig:numu_xsec}
    \end{subfigure}
    \begin{subfigure}[b]{0.48\textwidth}
        \includegraphics[width=\textwidth]{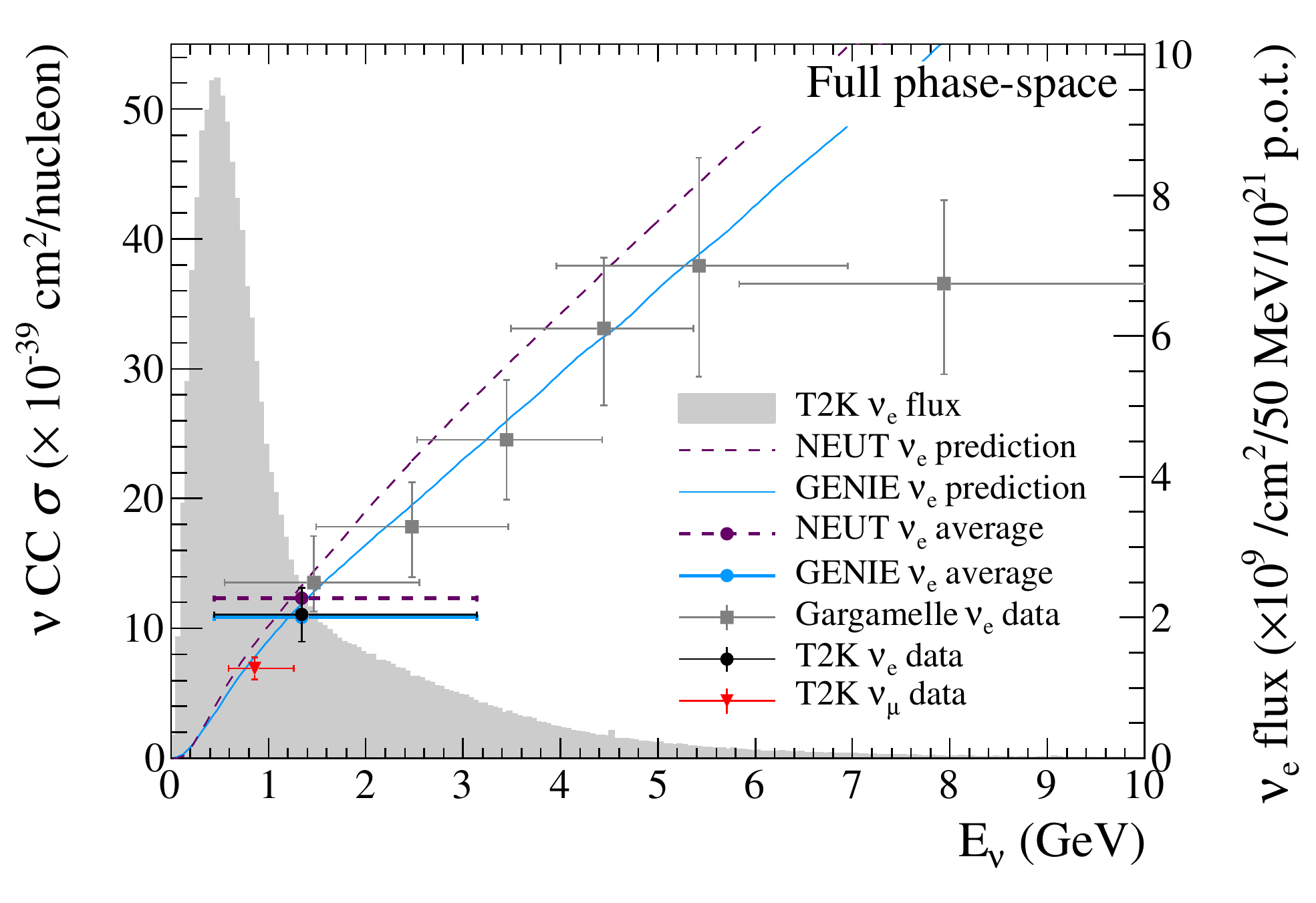}
        \caption{Total $\nu_{e}$ CC inclusive cross section when unfolding through $Q^2_{QE}$~\cite{Abe:2014agb}. The T2K data point is placed at the mean $\nu_{e}$ flux.  The Gargamelle $\nu_{e}$ and T2K $\nu_{\mu}$ data are also shown.}
        \label{fig:nue_xsec}
    \end{subfigure}
    \caption{T2K $\nu_{\mu}$ and $\nu_{e}$ CC inclusive cross section measurements on carbon.  In both plots the T2K flux prediction is shown in grey and the respective cross section predictions from the NEUT and GENIE generators are shown.  The vertical error bar represents the total uncertainty and the horizontal error bar represent 68\% of the flux each side of the mean neutrino energy.}
    \label{fig:cross_sections}
\end{figure}

T2K is focusing on producing cross-section measurements in an as model-independent way as possible and is working closely with theorists and the neutrino interaction generator groups to ensure these data can be used effectively.  This approach is demonstrated by the $\nu_{\mu}$ and $\nu_{e}$ CC inclusive measurements on carbon, the primary results of which are shown in Fig.~\ref{fig:cross_sections}.  
M. Ivanov \textit{et al.}~\cite{Ivanov:2015aya} used these data to compare to the predictions of their SuperScaling neutrino interaction model, finding that the model well reproduced the $\nu_{\mu}$ result, but underpredicted the $\nu_{e}$ cross section.  This showed that previously unconsidered interaction modes were important contributors to the cross section and has spurred further model development.

\newpage
\
\newpage
\section{Upgrades and Improvements to Maximize T2K Phase 2 Physics Sensitivity}
\label{sec:t2kupgrade}
\subsection{Projected MR Beam Power and POT Accumulation}
\label{sec:MR}

Since the start of the operation,
the J-PARC MR beam power has steadily increased.
In May 2016, 420~kW beam with 2.2$\times10^{14}$ protons-per-pulse (ppp)
every 2.48 seconds was successfully provided to the neutrino beamline.
There have been intensive discussions with the MR group regarding increasing
the beam power. 
The plan by J-PARC to achieve the design intensity of 750~kW is 
to reduce the repetition cycle to 1.3 seconds
with an upgrade to the power supplies for the MR main magnets,
RF cavities, and some injection and extraction devices by January 2019. 
Studies to increase the ppp are also in progress, with
 $2.73\times 10^{14}$ ppp equivalent beam
with acceptable beam loss already demonstrated in a test operation
with two bunches.
Based on these developments,
MR beam power prospects were updated and presented 
in the accelerator report at the PAC meeting in July 2015\cite{pac16}
and an anticipated beam power of 1.3~MW with 3.2$\times$10$^{14}$ ppp
and a repetition cycle of 1.16 seconds has been presented
at international workshops{\cite{jparcnuws,nuinfra}}.

\begin{figure}[bhtp] \centering
\includegraphics[width=0.65\textwidth]{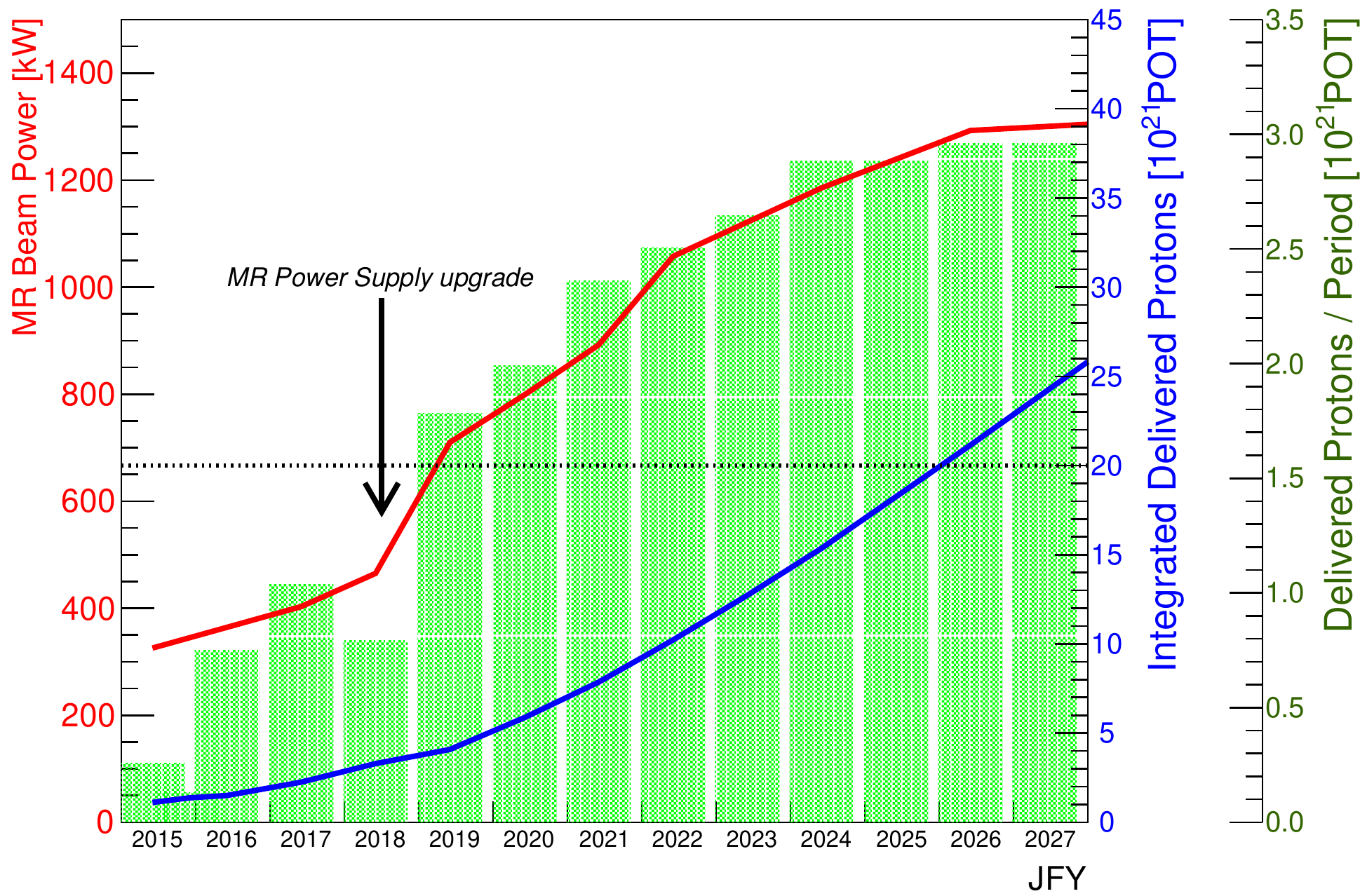}
\caption[POT]{Anticipated MR beam power and POT accumulation plan as function of calendar year.
\label{fig:POT2}}
\end{figure}

Figure \ref{fig:POT2}  shows our projected data accumulation scenario where five months of
neutrino operation each year and running time efficiency of 90\%
are assumed.
In this scenario, we expect to accumulate $\twopott$ by JFY2026
with five months operation each year and by JFY2025 with six months operation
each year.

\subsection{Beamline Upgrade for 1.3~MW Operation}

The instantaneous beam intensity acceptable in the current neutrino
beam facility is limited to $3.3\times 10^{14}$~ppp by
the thermal shock induced by the beam on the target and beam window.
The MR power upgrade plan up to 1.3~MW by increasing the repetition rate keeps
the instantaneous beam intensity within the acceptable range. However,
the heat generated by beam operation increases proportionally with the MR beam power.
The current cooling system for components such as the target and helium vessel was designed to have capacity up to 750~kW beam,
and needs to be upgraded in order to accept 1.3~MW beam.

The production target and the beam window of the Target Station (TS)
Helium vessel are cooled by helium gas.
The flow rate of the helium gas must to be increased to remove the heat from
the 1.3~MW beam, which requires reinforcement of the helium compressors.
Modification of the titanium container of the target and
titanium body of the beam window may be necessary to achieve the
higher helium flow rate. 

Several components, such as the TS helium vessel, decay volume, and beam
dump, are cooled by water. Their water circulation pumps need to be
upgraded to increase the flow rate. Replacement of all of the
heat exchangers for the cooling system with higher capacity ones is also
required.

The radioactive waste generated due to beam operation 
increases with higher beam power. The appropriate treatment of radioactive
water is a particularly important key to achieve 1.3~MW beam power. The water
disposal system for the components at the TS would be
upgraded with larger dilution tanks.

Since the strength of the aluminum alloy used in the magnetic horns
decreases dramatically above 100$^{\circ}$C, the maximum allowed temperature for
the conductors is set to 80$^{\circ}$C.
The performance of the main conductor cooling system is sufficient for 1.3 MW beam, where the maximum
 temperature is expected to be around 61$^{\circ}$C.
 The cooling of the horn striplines, currently accomplished by surface helium gas flow, would need to be improved.
 Water-cooled striplines that can accept more than 1.3~MW beam are under development.

Significant upgrades will be made for the secondary beamline components.
However, other components also need to be upgraded for the 1.3 MW beam.
In case the proton beam size is enlarged to achieve a beam intensity
of $3.3\times 10^{14}$~ppp, the aperture of the beam pipes in the primary
beamline should also be enlarged. Degradation of beam monitor elements
is an issue for high intensity beam.
Robust beam monitors such as wire-type secondary emission monitors
and beam induced fluorescent monitors are currently under development.
The beamline DAQ system would also need to be upgraded for the higher repetition
rate of 1~Hz. Safe operation of the beamline is extremely important
for such a high intensity beam. An upgrade of the beamline control system,
including reinforcement of the interlock system, should be performed.

\subsection{Improvement of the Neutrino Flux by Beamline Upgrades}
The magnetic horns were designed to be operated at 320~kA
current, but so far the operation current is limited to 250~kA because
of the limitation of the power supplies.  

Horn operation at 320~kA gives a 10\% higher neutrino flux and also
reduces contamination of the wrong-sign component of neutrinos 
({\em i.e.}, anti-neutrinos in $\nu-$mode beam or neutrinos in $\bar{\nu}-$mode beam) by 5-10\%.

The electrical system, such as power supplies, transformers, and
striplines, have been newly developed aiming for 1~Hz operation at
320~kA.  Some of these components have already been produced and
operated with satisfactory performance. At this moment, three
magnetic horns are driven by two power supplies and two transformers.
Operation with 320~kA can be realized with three power supplies and three transformers.
Therefore, it is necessary to install an additional power supply and peripherals to make
full use of the capability of the horns.

\subsection{Timeline of the Beamline Upgrade}
\begin{figure}[htbp] 
   \centering
   \includegraphics[width=\textwidth]{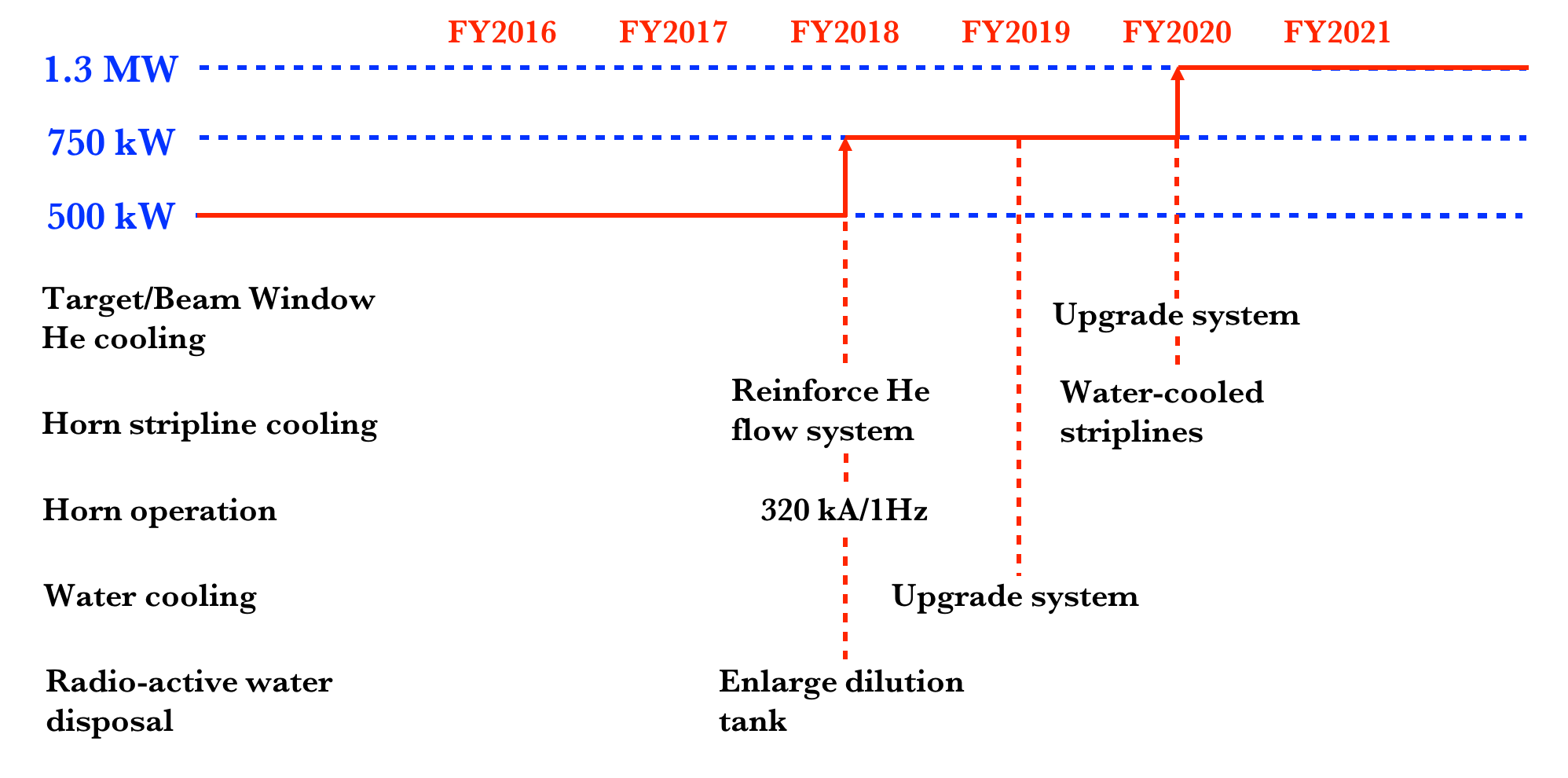} 
   \caption{Time table for beamline upgrade.}
   \label{fig:beamupgrade}
\end{figure}

  We request J-PARC to upgrade the neutrino beamline simultaneously 
with the MR upgrade as shown in Figure \ref{fig:beamupgrade} so that it can accept the maximum MR beam power that the
accelerator complex can provide.
To realize this, the installation of an additional horn power supply and peripherals should be
prepared by 2019.
The upgrades of the water cooling system for the secondary beamline components
need to be completed by 2020.
The installation of the new horns with water-cooled striplines, and
the upgrades of the helium circulation systems for the target
and the beam window need to be completed by 2021.
The upgrade of the water disposal system requires a long construction period without beam operation
and it is desirable to do this during the MR long shut down
in 2018 to minimize the beam-off period.

\subsection{Improved Super-K Sample Selection}

The current T2K selection for oscillated $\nu_e$ events in SK is shown in
Table~\ref{tab:nueselection}. Following basic requirements of containment and fiducialization (``fully contained fiducial volume'' or FCFV), $\nu_e$ charged current quasi-elastic (CCQE) scattering events, where no pions are expected (``CC$0\pi$''), are selected by identifying events with a single $e$-like Cherenkov ring.  Considering the $\nu_e$ CC interactions inclusively as the targeted sample (rather than the subset of CC$0\pi$ interactions), the main sources of inefficiency in this
selection are requiring a single ring (13.3\%), zero Michel electrons
(10.9\%), $E_\nu<1250$~MeV (4.1\%), and that the event is not
consistent with a $\pi^0$ hypothesis (8.0\%). In future analyses, many
of the signal $\nu_e$ events can be recovered by expanding the signal definition beyond the CC$0\pi$ channel to include pion production channels, and additional signal
events can be added by extending the current fiducial volume
definition. Some of these developments will be enabled by fully utilizing a new reconstruction algorithm with better vertex and kinematic resolution, and enhanced multi-ring reconstruction capabilities. So far, the use of this algorithm has been limited to improving the rejection of $\pi^0$ backgrounds in the SK $\nu_e$ selection.

\begin{table}[h!]
\begin{center}
\caption{
Event reduction for the $\nu_e$ CC selection at the far detector.  The numbers of 
expected MC events divided into four categories are shown after each 
selection criterion is applied.  The MC expectation is based upon three-neutrino 
oscillations for $\sin^{2}2\theta_{23}=1.0$, $\Delta m^2_{32}=2.4\times10^{-3}$eV$^2/$c$^4$,
$\sin^{2}2\theta_{13}=0.1$, $\delta_{CP}=0$ and normal mass hierarchy
(parameters chosen without reference to the T2K data).
}
\begin{tabular}{ll}
(1) & There is only one reconstructed Cherenkov ring \\
(2) & The ring is $e$-like \\
(3) & The visible energy, $E_{\mathrm{vis}}$, is greater than 100~MeV \\
(4) & There is no reconstructed Michel electron \\
(5) & The reconstructed energy, $E_{\nu}^{\mathrm{rec}}$, is less than 1.25~GeV \\
(6) & The event is not consistent with a $\pi^0$ hypothesis \\
\end{tabular}
\vskip 6mm
\begin{tabular}{lccccc}
\hline
\hline
&                 & \ \ $\nu_\mu+\bar{\nu}_\mu$ \ \ & \ \ $\nu_e+\bar{\nu}_e$ \ \ & \ \ $\nu+\bar{\nu}$ \ \ & \ \ $\nu_\mu\rightarrow\nu_e$ \ \ \\
& \ \ MC total\ \ & CC & CC & NC & CC \\
\hline
interactions in FV                 & 656.83 & 325.67 & 15.97 & 288.11 & 27.07 \\
FCFV              & 372.35 & 247.75 & 15.36 & 83.02  & 26.22 \\
(1) single ring                    & 198.44 & 142.44 & 9.82  & 23.46  & 22.72 \\
(2) electron-like                  & 54.17  & 5.63   & 9.74   & 16.35 & 22.45 \\
(3) $E_{\rm vis}>100{\rm MeV}$      & 49.36  & 3.66   & 9.68   & 13.99 & 22.04 \\
(4) no Michel election            & 40.03  & 0.69   & 7.87   & 11.84 & 19.63 \\
(5) $E_{\nu}^{\rm rec}<1250{\rm MeV}$& 31.76  & 0.21   & 3.73   & 8.99 & 18.82 \\
(6) not $\pi^{0}$-like             & 21.59  & 0.07   & 3.24   & 0.96  & 17.32  \\
\hline
\hline
\end{tabular}
\label{tab:nueselection}
\end{center}
\end{table}

The simplest extension to the existing $\nu_e$ selection is to select
events with exactly 1 Michel electron. The oscillated signal events in
this sample are mostly CC$\pi^+$ events where the pion was below the
Cherenkov threshold, but still produced a Michel electron from the $\pi^+\to\mu^+\to e^+$ decay chain. An internal
analysis of this sample is nearly complete, and adds 12.6\% more
$\nu_e$ events after all selection cuts.

Another significant gain in efficiency will be possible by including
multi-ring event samples. Recent developments in multi-ring event
reconstruction will allow for the identification of CC$\pi^+$ events
where both the electron and pion are above Cherenkov threshold, and
3-ring CC$\pi^0$ events. Of the oscillated CC-$\nu_e$ events removed
by the current $\nu_e$ selection, 29\% are CC0$\pi$ events, 51\%
are CC$\pi^+$ events (3/4 of which have a $\pi^+$ above Cherenkov
threshold), and 13\% are CC$\pi^0$ events. Together, the CC$\pi^+$ and
CC$\pi^0$ events could increase the total $\nu_e$ sample by as much as
35\% (including the 12.6\% from CC$\pi^+$ events with a pion below the
Cherenkov threshold). The actual gains will be somewhat smaller due to
the selection cuts applied to these samples, and the purity of these
samples is currently under study.

The $\pi^0$ cut in the current analysis was optimized for the $\nu_e$
appearance search, which required high purity to mitigate the impact
of the high systematic errors on the $\pi^0$ background rate. However,
for future CP violation analyses in which $\nu_e$ event samples will
be compared with $\bar{\nu}_e$ samples, the presence of additional
$\pi^0$ background has a smaller negative impact on the sensitivity to
CP violation than the benefit of increasing signal statistics, since
the $\pi^0$ background will be common to both samples. Making this
adjustment will recover some of the signal $\nu_e$ events removed by
the existing $\pi^0$ cut, although the precise size of this gain will
depend upon a full CP violation sensitivity optimization that is
currently under study.

\begin{figure}[h!]
\centering
\includegraphics[width=0.45\textwidth]{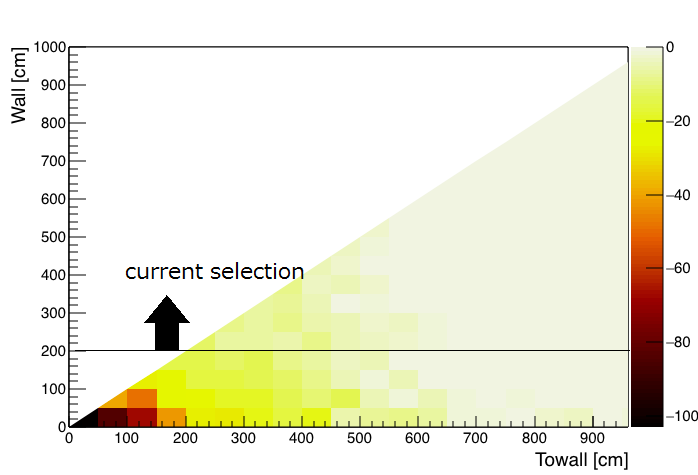}
\includegraphics[width=0.45\textwidth]{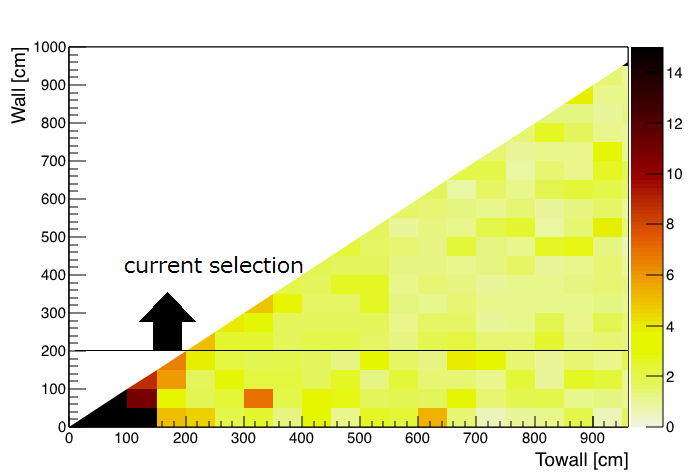}
\caption{The reconstructed momentum bias (left; in percent) and mean angle between the true and reconstructed track direction
  (right; in degrees) for single-ring electron events are shown as a function of
  {\em wall}  and
  {\em towall} (see text for definitions). The current selection requires {\em wall} $>200$ cm.}
\label{fig:walltowall}
\end{figure}

Finally, there is ongoing effort to expand the fiducial volume
definition currently used in the SK detector. The SK inner detector
volume is 36.2~m tall with a radius of 16.9~m. The current fiducial volume requirements remove the outer 2~m of this
volume, which accounts for 31\% of the total inner detector volume; 
significant gains can be made even with a small adjustment to this
requirement. The performance of the reconstruction depends on both the
distance between the reconstructed event vertex and the nearest wall
(simply called ``{\em wall}'') and the distance from the event vertex to the
wall along the reconstructed direction of the particle (called
``{\em towall}''). Figure~\ref{fig:walltowall} shows the degradation in the
reconstructed momentum bias and average direction between the true and reconstructed angle. When {\em towall} is large, the reconstruction performance is good, even at
smaller values of {\em wall}. These signal efficiency studies will be
combined with an analysis of backgrounds produced near the inner
detector wall, as well as entering backgrounds produced just outside
the inner detector wall, to optimize the final fiducial volume
definition. Preliminary studies suggest that an effective fiducial
volume gain of 10-15\% may be possible.

The combined impact of all of the aforementioned improvements can
potentially increase the efficiency of the T2K CC$\nu_e$ event sample
by as much as 40\%.

\subsection{Improvement of Systematics for T2K phase 2}
\label{sec:anaimp}

As will be described in Sec.~\ref{sec:physics},
the current systematic errors, if they are not improved,
will significantly reduce the sensitivity
to CP violation with the T2K-II statistics.
Any improvement on the systematics would enhance physics potential.
Here, we first describe the current systematic errors and then 
describe projected improvements.

Based on their source the systematic errors are categorized into  
neutrino flux, neutrino interaction model, and detector model uncertainties.
The uncertainties in the neutrino flux and 
interaction model are first constrained by external measurements
and then further constrained by a fit to data from the ND280 near detector.

The uncertainties on the total predicted number of events
in the Super-K samples encapsulate the first order impact of
systematic errors on the oscillation parameter measurements,
and the current sizes are summarized
in Table~\ref{tab:syst_sum}.
The CP phase $\delta_{CP}$ is measured
through the difference in the oscillation probabilities for $\nu_\mu\to\nu_{e}$
and $\bar{\nu}_\mu\to\bar{\nu}_{e}$.
Hence, we also show the uncertainty on the ratio of expected $\nu_e/\bar{\nu}_e$ candidates
at Super-K with neutrino ($\nu$) and antineutrino ($\bar{\nu}$) beam mode.

The uncertainty from oscillation parameters not measured by T2K-II 
is negligible for $\nu_\mu/\bar{\nu}_\mu$ events
at SK in the $\nu_\mu/\bar{\nu}_\mu$ disappearance measurements.
The 4\% uncertainties on the $\nu_e/\bar{\nu}_e$ samples arise from the precision of the $\theta_{13}$ measurement
by reactor experiments($\sin^{2}(2\theta_{13})=0.085\pm0.005$)~\cite{pdg}.
However, this uncertainty is correlated between $\nu$ and $\bar{\nu}$ beam 
mode samples and its impact on the observation of a CP asymmetry in T2K data
is small.

The numbers shown in Table~\ref{tab:syst_sum} are obtained
by varying the parameters which model each source of systematic uncertainty.
However, some of the systematics arising from the neutrino interaction model
are not yet parameterized.
Examples include the uncertainty of the nuclear Fermi gas model and $W$ dependence of the 2p-2h interactions arising from multinucleon effects (see Section \ref{sec:nuint}).
For these systematics, we evaluate the change in the results
when different models are applied and confirm whether this change 
is small compared to the total error.
This is indeed the case for the current level of statistical uncertainty,
but not a permanent solution.
We have been improving our model in the neutrino interaction generator to correctly constrain these models
and estimate the associated uncertainties as discussed in Sec.\ref{sec:nuint} and \ref{sec_xsec}.
Improved flux prediction and near detector measurements improve the oscillation analysis sensitivity
directly, but also would be useful to improve our neutrino interaction model.


\begin{table}[t]
 \caption{Errors on the number of predicted events
 in the Super-K samples from individual 
systematic error sources in neutrino ($\nu$ mode) and
 antineutrino beam mode ($\bar{\nu}$ mode).
Also shown is the error on the ratio 1R$e$ events in $\nu$ mode/$\bar{\nu}$ mode.
Uncertainties arising from multinucleon effects and the $1p1h$ model (described in Section \ref{sec:nuint}) are not included
and are handled separately as described in the text.
The uncertainties represent for preliminary T2K neutrino oscillation results in 2016.}
 \label{tab:syst_sum}
 \begin{center}
   \begin{tabular}{l|c|c|c|c|c}
     \hline \hline
                  & \multicolumn{5}{|c}{$\delta_{N_{SK}}/N_{SK}$ (\%)}  \\ \hline 
                  &  \multicolumn{2}{|c}{ 1-Ring $\mu$} & \multicolumn{3}{|c}{ 1-Ring $e$} \\ \hline
     Error Type   & $\nu$ mode & $\bar{\nu}$ mode & $\nu$ mode & $\bar{\nu}$ mode & $\nu$/$\bar{\nu}$ \\ \hline \hline
     SK Detector  &  3.9   & 3.3    & 2.5    &  3.1   & 1.6     \\ \hline
     SK Final State \& Secondary Interactions  & 1.5   & 2.1  &  2.5  & 2.5  & 3.5     \\ \hline
     ND280 Constrained Flux \& Cross-section   & 2.8  & 3.3 & 3.0 & 3.3 & 2.2     \\ \hline 
     $\sigma_{\nu_{e}}/\sigma_{\nu_{\mu}}$, $\sigma_{\bar{\nu}_{e}}/\sigma_{\bar{\nu}_{\mu}}$ &  0.0  & 0.0   & 2.6 & 1.5 & 3.1     \\ \hline
     NC 1$\gamma$ Cross-section &  0.0  & 0.0 & 1.5 & 3.0 & 1.5     \\ \hline
     NC Other Cross-section     &  0.8  & 0.8 & 0.2 & 0.3 & 0.2     \\ \hline \hline
     Total Systematic Error     &  5.1  & 5.2 & 5.5 & 6.8 & 5.9    \\ \hline \hline
     External Constraint on $\theta_{12}$, $\theta_{13}$, $\Delta m^{2}_{21}$  & 0.0 & 0.0 & 4.1 & 4.0 & 0.8     \\
     \hline \hline
   \end{tabular}
 \end{center}
\end{table}

\subsubsection{Neutrino Flux}
\label{sec:flux}
The neutrino flux prediction uncertainty is currently dominated by uncertainties
on the hadron interaction
modeling in the target and surrounding materials in the neutrino beamline
and by the proton beam orbit measurement~\cite{Abe:2012av}.
The errors on the flux can be represented as an absolute flux uncertainty,
which is relevant for neutrino cross section measurements,
and an extrapolation uncertainty, which is relevant for oscillation measurements.
The current absolute and extrapolation uncertainties
at the peak energy ($\sim 600$ MeV) are $\sim 9\%$ and 0.2
\%, respectively.  The detailed uncertainties are listed in Table~\ref{tab:flux_errors}.
While the extrapolation error is already quite small,
the oscillation analysis may benefit from
further reduction of flux systematic errors
since the interaction model can be more strongly constrained
with smaller flux uncertainties.  

The main reduction in the absolute flux uncertainty will come from the use
of NA61/SHINE measurements of the hadron production from a replica of the T2K
target.  NA61/SHINE has already measured the pion production with initial
replica target data sets~\cite{Abgrall:2012pp,Abgrall:2016jif}, and has achieved
$\sim4\%$ precision on the measurement of $\pi^{\pm}$ spectra exiting the target.  
The use of these measurements in the T2K flux calculation will eliminate the largest
source of systematic error, the uncertainty on the interaction rates of hadrons
interacting inside the T2K target.  

Another large source of uncertainty in the flux prediction is the uncertainty on 
the beam direction due to the uncertainties on the alignment of beamline components and 
the position of the proton beam on the upstream end of the T2K target.  This uncertainty
will be reduced by implementing a fit of the flux model to INGRID beam direction data to
better constrain the simulated beam direction.

We have estimated absolute and extrapolation errors on the flux model for the improvements listed above and
the results are shown in  Table~\ref{tab:flux_errors}.
The projected uncertainty on the absolute flux prediction is $\sim 6\%$ near the peak energy.

\begin{table}[t]
 \caption{The current and projected flux uncertainties for $0.4<E_{\nu}<1.2$ GeV for each neutrino species and horn operation mode.}
 \label{tab:flux_errors}
 \begin{center}
   \begin{tabular}{l|c|c|c|c}
     \hline \hline
& \multicolumn{2}{|c}{Current Uncertainty} & \multicolumn{2}{|c}{Projected Uncertainty }  \\ 
& \multicolumn{2}{|c}{(\%)} & \multicolumn{2}{|c}{(\%) }  \\ \hline 
     Neutrino species   & Absolute & Extrapolation & Absolute & Extrapolation  \\ \hline \hline
     $\nu-$mode, $\nu_{\mu}$       &  9.1    & 0.17   & 5.6    &  0.12   \\ \hline
     $\nu-$mode, $\bar{\nu}_{\mu}$ &  7.6    & 0.62   & 6.6    &  0.38   \\ \hline
     $\nu-$mode, $\nu_{e}$         &  8.8    & 0.37   & 5.2    &  0.27   \\ \hline
     $\nu-$mode, $\bar{\nu}_{e}$   &  7.2    & 0.50   & 5.0    &  0.41   \\ \hline
     $\bar{\nu}-$mode, $\nu_{\mu}$       &  7.3    & 0.61   & 6.3    &  0.31   \\ \hline
     $\bar{\nu}-$mode, $\bar{\nu}_{\mu}$ &  9.1    & 0.28   & 5.5    &  0.27   \\ \hline
     $\bar{\nu}-$mode, $\nu_{e}$         &  6.7    & 0.73   & 4.8    &  0.33   \\ \hline
     $\bar{\nu}-$mode, $\bar{\nu}_{e}$   &  8.7    & 0.43   & 5.3    &  0.16   \\ \hline
     \hline 
   \end{tabular}
 \end{center}
\end{table}


\subsubsection{Near Detector Measurement}

In the current analysis,
detector-related systematic uncertainties of $\sim2\%$ have been achieved
in $\nu_\mu/\bar{\nu}_\mu$ charged-current samples selected in ND280.
The main sources of uncertainty are the TPC particle identification and
track-finding efficiency, backgrounds from neutrino interactions outside of the fiducial volume,
the FGD mass, and pion reinteraction modelling.
Among these, the first three, related to
reconstruction efficiencies and backgrounds, are expected to be reduced
with ongoing analysis and software development efforts.
By far the largest uncertainty, however, arises from pion secondary interaction uncertainties,
which may be reduced by external measurements or by studying pion interactions within ND280 itself.
We expect to reduce this uncertainty and achieve $\sim1\%$ overall systematic error in the ND280 samples.
This reduction of the pion secondary interaction uncertainty requires two different improvements.
First are improved constraints from including previously unavailable external data on pion interactions.
Second, the models in the simulation must be improved since some of the cross section models are
clearly in disagreement with the data.
We are confident that we can reduce by a factor of 2 the pion secondary interaction uncertainty because
we are already implementing improvements to both the external data with latest DUET measurements \cite{duet:2015}
and the simulation models by replacing the GEANT4 models with the much improved NEUT cascade models \cite{ElderTalk:2015}.

\begin{figure} \centering
\includegraphics[width=0.65\textwidth]{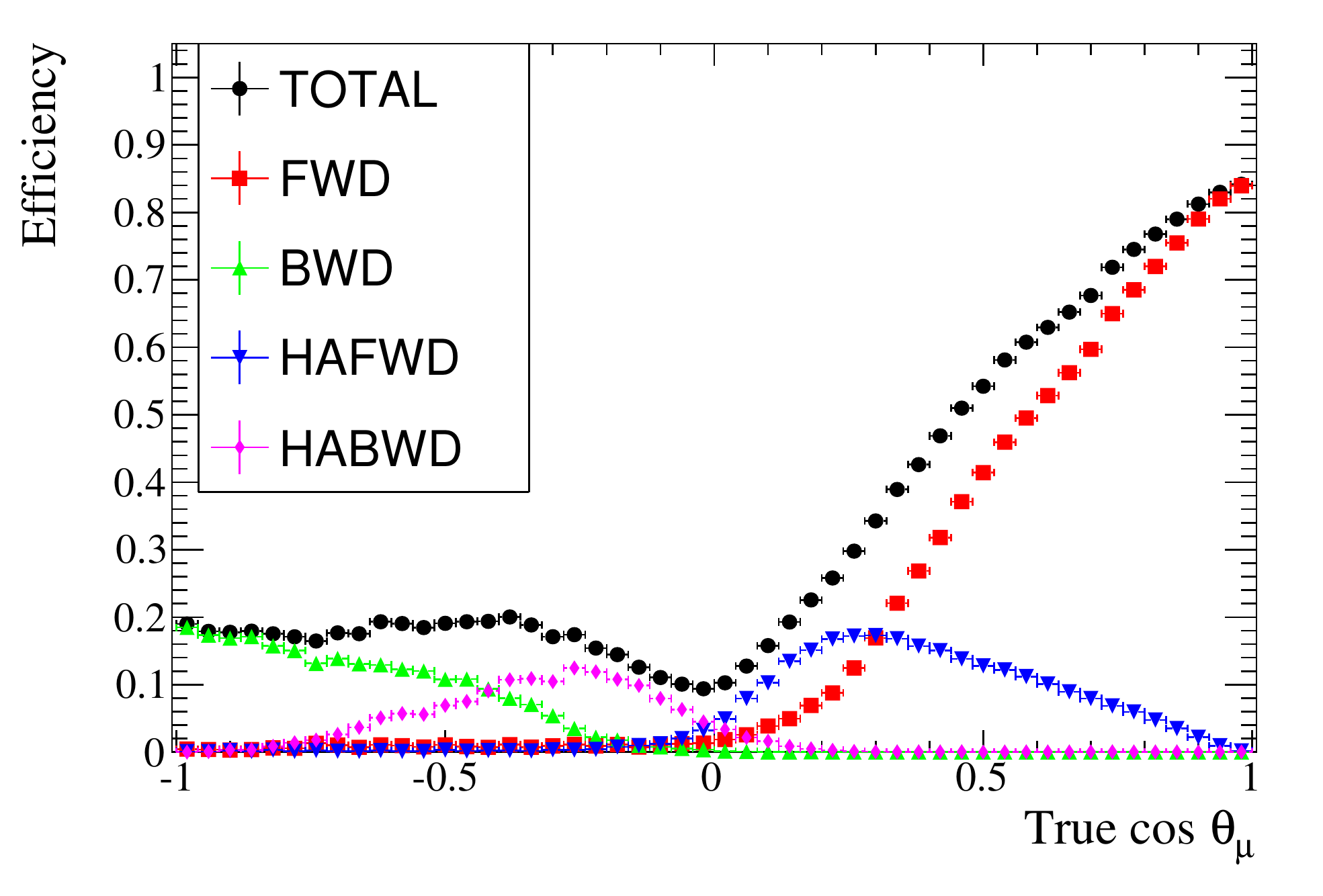}
\caption[]{Reconstruction efficiency of the muon track as function of angle against beam axis.
  Depending on the combination of sub-detectors, forward-going (FWD), backward-going (BWD)
  high-angle forward (HAFWD) and high-angle backward (HABWD) tracks are reconstructed.
  \label{fig:mueff}}
\end{figure}

Additional near detector samples will be added for the flux and cross section constraint for the oscillation analysis. 
For example, the $\nu_e$ event rate has been measured with a precision of 8\%\cite{Abe:2014usb}.
Since the systematic error is 5\%, it will be improved with more statistics and would provide an important cross-check on the flux and cross section of $\nu_e$'s.
In the $\bar{\nu}$-mode, charged current interaction events accompanied by pion tracks, which
are now treated as the `CC-Ntrack' sample, will be separated into the `CC-$1\pi^-$' and `CC-other' samples.
The angular phase space coverage for the muon track has been extended in recent years.
The reconstruction efficiencies for the muon track with various combination of
sub-detectors are shown in Fig.~\ref{fig:mueff}.
The current oscillation analysis only uses the forward-going muon samples,
but the backward-going tracks will be used in the next update.
These samples help to place tighter constraints on neutrino interaction uncertainties in the oscillation analysis.

\subsubsection{Neutrino Interaction}
\label{sec:nuint}

The uncertainties of neutrino interactions and secondary interactions could be the largest errors in various analyses in T2K-II. 
There are several attempts to model neutrino-nucleus interactions by combining the neutrino-nucleon
interaction with various corrections in the nuclear medium.
However, the existing data on the neutrino-nucleon interaction is statistically limited especially in the T2K energy region
and it is not a simple task to evaluate systematic uncertainty using them. 
Therefore, it is crucial to extensively use the neutrino-nucleus scattering data, both from the T2K ND280 and 
the other recent experiments, to minimize the uncertainty.

\begin{figure} \centering 
\includegraphics[width=7cm]{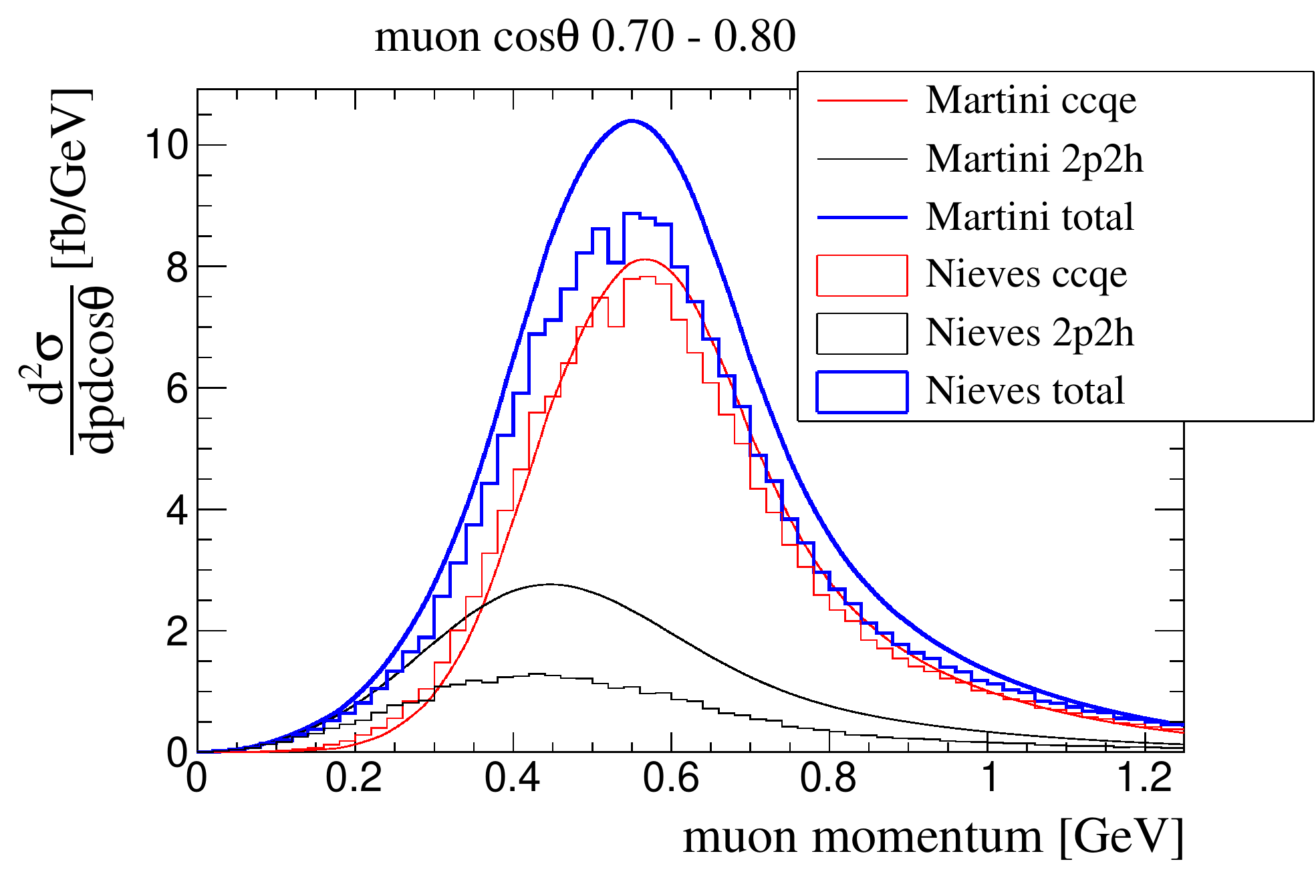}
\includegraphics[width=7cm]{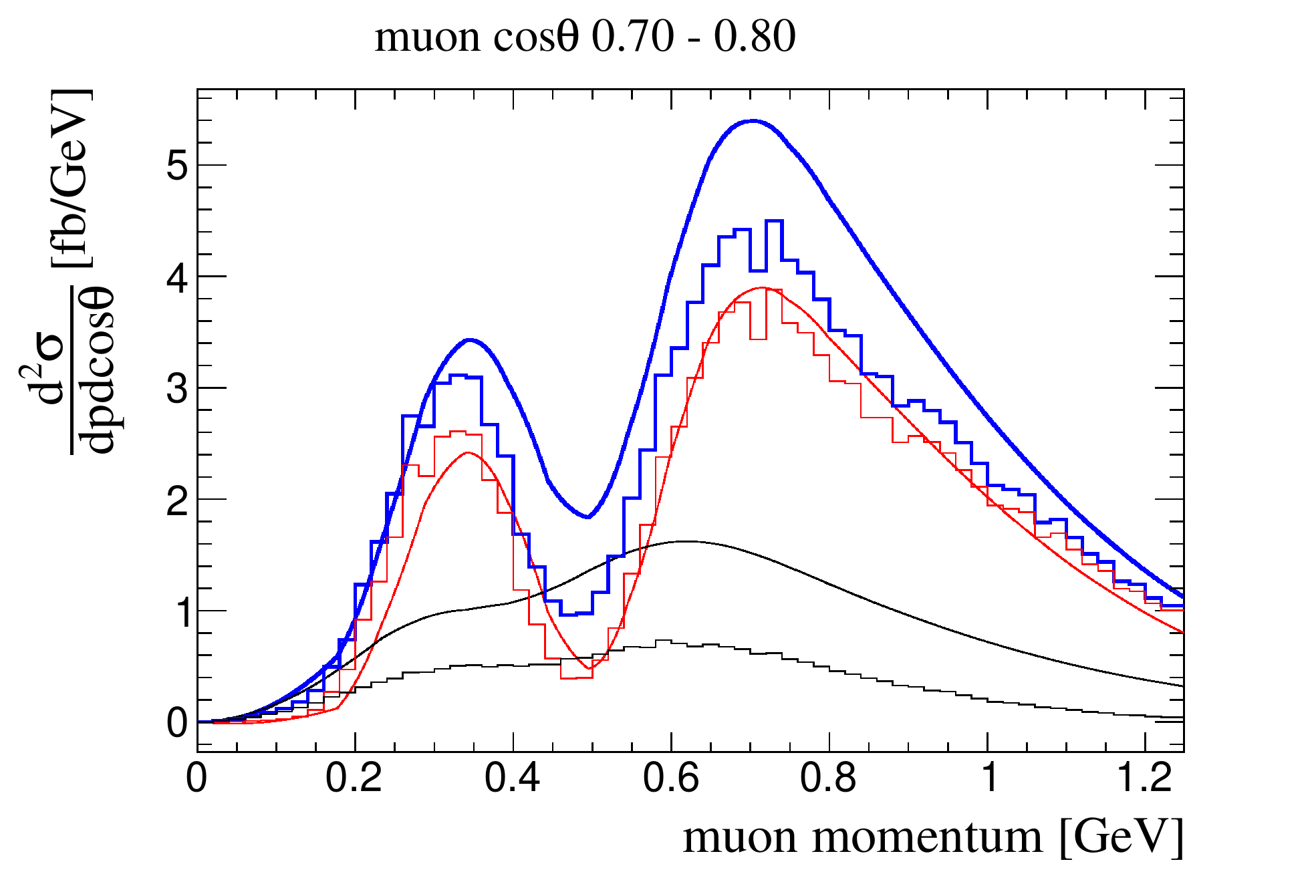}
\caption{Distribution of CCQE and 2p2h contributions as a function of muon momentum in the angular range $\cos\theta=[0.7, 0.8]$
at ND280 (left) and Super-K (right) as predicted in the models of Martini {\em et al.}\cite{Martini:2009} (continuous line) 
and Nieves {\em et al.}\cite{Nieves:2011pp, Nieves:2004wx} (histogram).}
\label{NIWG_2p2h}
\end{figure}

The systematic uncertainties due to the modelling of neutrino-nucleus interactions are dominated by
various nuclear effects: short-range and long-range multi-nucleon correlations (also known as 2p2h and RPA) 
and hadron final state interactions. The problem is not simple because experimental
disentanglement of various neutrino primary interactions and secondary interactions is difficult in most of the cases.
Also, kinematical acceptances of the recent experiments are rather limited and it is difficult to reduce uncertainties
from those less experimentally explored kinematic regimes.
In the T2K oscillation analysis, the uncertainties in modeling these effects are constrained by the near detector, 
but such constraints are limited by the differences in the neutrino energy spectrum and 
the differences in acceptances between the near detector and Super-K. 
For instance, Figure~\ref{NIWG_2p2h} shows the distribution of expected CCQE+2p2h events  for $\cos\theta_\mu$ from 0.7 to 0.8
at the near and far detector for the models of Martini et al.\cite{Martini:2009} and Nieves et al.\cite{Nieves:2011pp, Nieves:2004wx}: 
the 2p2h component manifests itself at ND280 mainly as an overall increase in the
cross-section normalization, while at Super-K 2p2h events tend to bias the neutrino oscillated energy spectrum, 
filling the oscillation dip. There is moreover a large difference (around a factor of two)
between the prediction of the two considered models which is an indication of the scale of the large uncertainties on the 2p2h modelling.
Differences between the models are now under study by theorists but we also have to find the way to resolve the situation experimentally,
as explained in Sec.\ref{sec_xsec}.

\begin{figure} \centering 
\includegraphics[width=7cm]{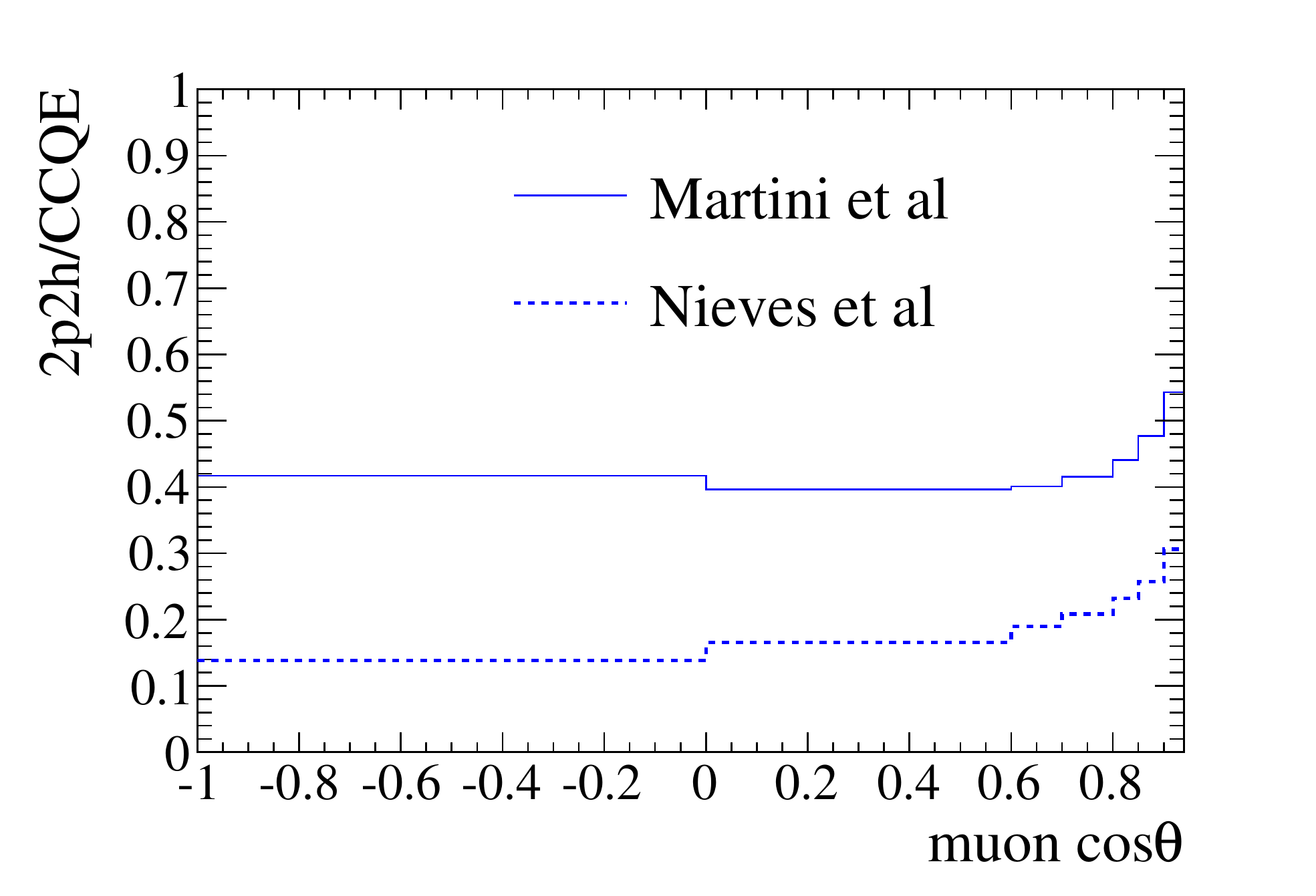}
\includegraphics[width=7cm]{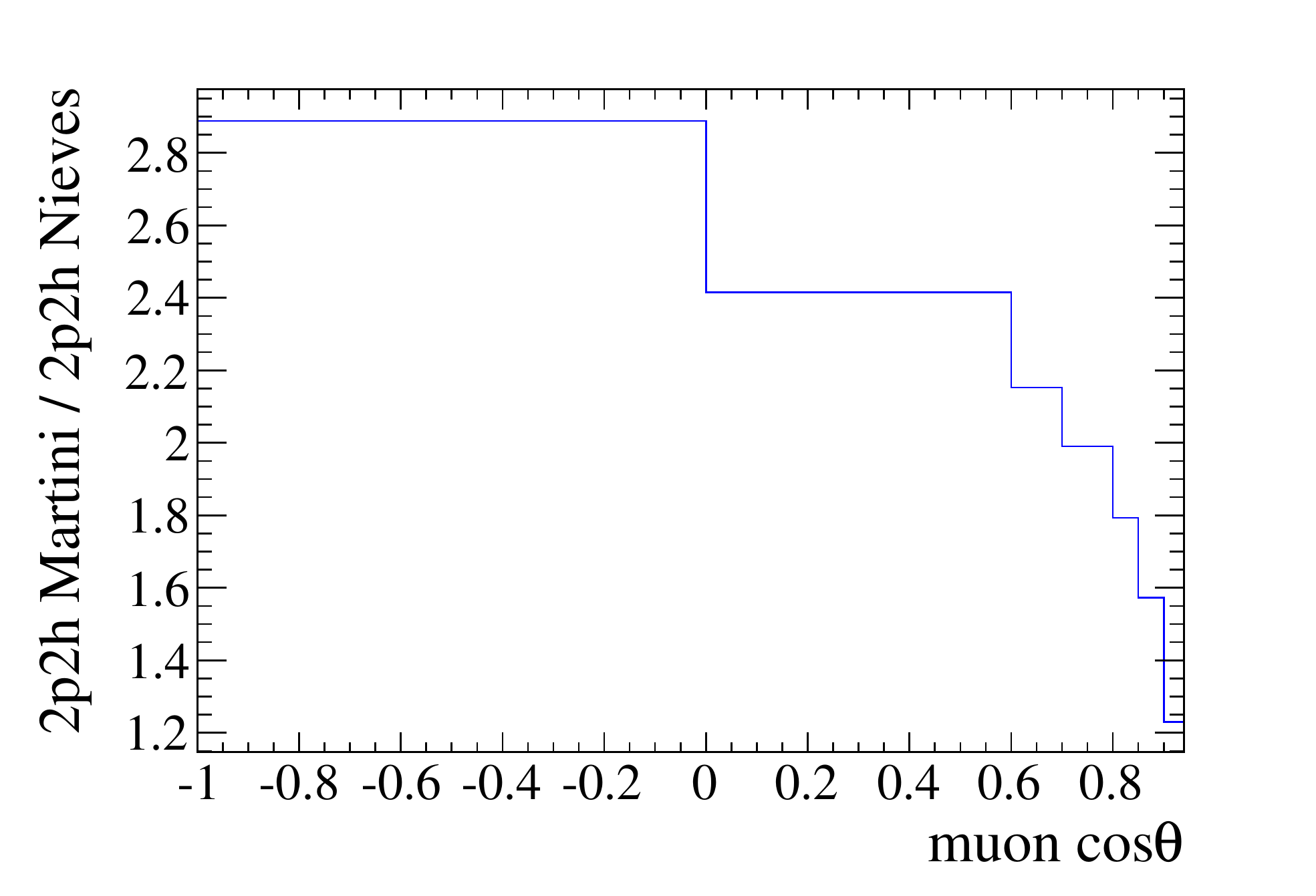}
\caption{Left: ratio between 2p2h and CCQE events at ND280 as a function of muon angle, 
as predicted by the models of Martini et al.\cite{Martini:2009} (continuous line)  and Nieves et al.\cite{Nieves:2011pp, Nieves:2004wx} (shaded line). 
Right: ratio between 2p2h contribution between the two models as a function of muon angle at ND280.
Cuts are applied to stay in the region of validity of the models (muon momentum below 1.2 GeV and muon cos$\theta$
below 0.94).}
\label{NIWG_2p2h_2}
\end{figure}

T2K has maintained a significant  neutrino-nucleus interaction modelling effort in order to properly parametrize
and optimally constrain the related uncertainties, in tight collaboration with Monte Carlo experts and model builders. 
In order to minimize such uncertainty T2K has engaged in a continuous effort to reduce the flux uncertainties, enlarge the ND280 acceptance and introduce additional samples.
Events with neutrino interactions on water in ND280 have been recently included and such water target sample
will be further extended in the future. 
The phase space coverage of the ND280 measurements will be extended 
to backward going tracks, particularly helpful to better constrain the uncertainties in the high $Q^2$ region.
In Fig.\ref{NIWG_2p2h_2} the relative contribution of CCQE and 2p2h events is shown for the models mentioned above; 
the 2p2h models have different angular distributions.
It is particularly relevant to improve the angular acceptance of ND280 in the backward region in order
to cover the full acceptance of Super-Kamiokande, since differences in acceptance may be a source of possible biases 
in the neutrino interaction modelling in the extrapolation from the near to the far detector.

Thanks to the increased statistics and improved flux uncertainty, T2K-II will be more sensitive
to finer details of the nuclear models. In particular, T2K-II will improve the precision on the constraints
in the backward region which are today limited by statistical uncertainties.
The increased statistics will also permit  more exclusive and differential measurements constraining
also the kinematics of outgoing hadrons and the correlations between the hadron and the muon kinematics.

Another important systematic uncertainty for next-generation long-baseline experiments is due to the
difference between electron and muon neutrino cross-section. 
In the fundamental neutrino-quark interaction, the difference between the electron and muon mass
has a small impact on the kinematics of the outgoing lepton: it changes the limits for the allowed value of 
$Q^2$ at fixed neutrino energy and, most importantly, affects the radiative corrections to the interaction process.
All these effects are in principle calculable but uncertainties rise from the convolution of such effects with nucleon
form factors and with nuclear effects which are not well known. 
From an approximate calculation~\cite{Day:2012gb}, the difference between $\nu_e$ and $\nu_\mu$ cross-sections 
due to radiative corrections should be smaller than 10\% (12\%) for neutrino (antineutrino) at T2K's energy. 
Such effects are not yet included in Monte Carlo generators and more complete calculations including box diagrams
which are expected to cancel, at least partially, this effect as a consequence of the Kinoshita-Lee-Nauenberg theorem are needed.
These calculations have not yet been performed for exclusive elastic or quasielastic scattering.
In neutrino-nucleon cross-section calculations, second class currents are typically assumed to be negligible but the data still allow for a relatively large contribution which would cause a difference between $\nu_e$ and $\nu_\mu$ cross-sections 
of the order of a few percent at T2K's energy and with opposite sign for neutrino and antineutrino~\cite{Day:2012gb}.
A further source of uncertainty comes from the interplay in the cross-section modelling of lepton kinematics 
factors and nuclear response functions, as explained in~\cite{Martini:2016eec}: the ratio
of $\nu_e$ and $\nu_\mu$, as well as $\bar{\nu}_e$ and $\bar{\nu}_\mu$ cross-sections, 
are different for CCQE and 2p2h processes, therefore the uncertainty
on the relative amount of these contributions causes an uncertainty on the  $\nu_e$ and $\nu_\mu$ cross-section difference.
Such uncertainties on the ratios $\sigma_{\nu_e}/\sigma_{\nu_\mu}$ and $\sigma_{\bar{\nu}_e}/\sigma_{\bar{\nu}_\mu}$ are 
a primary source of systematics on the measurement of CP violation. T2K is working to include $\nu_e$ and $\bar{\nu}_e$ samples
in ND280 analysis but this approach is limited by low statistics. 
The increased sample of $\nu_e/\bar{\nu}_e$ events in T2K-II will improve the constraints on the model, which in turn will allow better constraints on the cross section ratios.

In order to disentangle  different nuclear effects and improve the constraints on uncertainties due to interaction modelling, 
T2K not only relies on internal datasets but also  exploits externally published measurements. 
Figure~\ref{NIWG_Minerva}  shows the comparison of
the MINER$\nu$A measurement of low-energy recoil data to different interaction models. 
This approach will be similarly pursued with ND280 data.
The comparison of such measurements at the different MINER$\nu$A and T2K energies is expected to shed light 
on nuclear effects, and will benefit from the narrow-band T2K neutrino flux.
\begin{figure} \centering 
\includegraphics[width=15cm]{figs/minervaLowEneRecoil.png}
\caption{Comparison of MINER$\nu$A low recoil CC-inclusive data with the NEUT MC generator as a function of available hadronic energy and three-momentum transfer. These data can be used to improve the modelling of CCQE interactions and the of 2p2h contribution.}
\label{NIWG_Minerva}
\end{figure}

\subsubsection{Super-K Systematics Improvement}

Systematic errors arising from uncertainties in the response of the Super-K detector
are derived by comparing the atmospheric neutrino data and cosmic ray muon control samples
to the MC simulations. The targeted samples in the T2K analysis, namely single-ring electron- and muon-like events, are identified in the atmospheric neutrino sample in Super-K along with sideband regions defined by events which fail one or more of the selection criteria.
The values of the selection cuts in the simulation are varied to fit to the data.
Simultaneously, other systematic parameters related to neutrino interaction and flux modeling are varied as nuisance parameters. The resulting offset in the cut values and their uncertainty are translated into a systematic error in the selection efficiency. In the cosmic ray muon samples, the range of the muon is estimated by using the decay electron to mark the muon end point, and the observed energy/range is compared with Monte Carlo simulation to obtain a systematic error. The mean energy of the decay electron spectrum is similarly used to determine the energy scale uncertainty at lower energies.

While the atmospheric neutrino data provide an all-encompassing constraint on detector systematics, these errors 
nonetheless occupy a sizable portion of the total T2K error budget : 4.0\% relative uncertainty on the number of electron-like candidate events and 2.7\% for muon-like candidate events~\cite{Abe:2015awa}.
In order to reduce these uncertainties upgrades to the current error evaluation are essential
to the high statistics measurements planned at T2K-II. 

Improvements to the atmospheric neutrino fit to include cross sections constraints from  ND280 can in principle allow for tighter 
systematic constraints on detector systematics.
Additionally, while the current fit essentially fits the total event rate in each of the signal and sideband samples, more precise 
constraints are expected from fitting the shape of the likelihood distributions underlying the event selections.
Both of these developments are currently in progress with the expectation to reduce detector uncertainties for the T2K sample in the next year or so. 

While the atmospheric neutrino fitting scheme has adequately estimated SK detector errors so far, its use in the future may be limited by our understanding of the atmospheric neutrino flux and cross-sections. An alternative method for estimating detector errors using fundamental detector performance parameters is therefore under development.
This model would parametrize the detector response in terms of quantities such as the water transparency, the reflectivity of the detector surfaces, and the charge response of the photosensor which can be constrained by low level calibration data. 
Using precision calibrations, the ultimate size of the detector systematic can be carefully controlled and propagated directly to the 
T2K analysis without an atmospheric neutrino intermediary, in principle. 
Considering the manpower necessary to introduce and make this program successful, its implementation is longer-term but could be realized 
in the next few years. 

It should be noted that this type of low level error parametrization will be beneficial immediately after the SK-Gd upgrade to Super-K, planned to occur a few years from now,
 when the atmospheric neutrino statistics available will be insufficient to use the current 
error estimation method.
Dissolving gadolinium sulfate into the detector water to improve its ability to detect neutrons is the focus of the SK-Gd project, 
and will represent a fundamental change in the detector environment. 
Neutron sensitivity will enable better separation of neutrino from antineutrino interactions, since the latter produces more neutrons 
on average, and can therefore be utilized to improve T2K's sensitivity to CP violation.
At the same time the introduction of the gadolinium compound will change and potentially degrade the optical properties of the 
detector water. 
While current estimates indicate the there will be minimal loss of oscillation sensitivity due to changes in transparency with SK-Gd,
the impact on detector systematic uncertainties cannot be evaluated \textit{a priori}.
For this reason realizing a low level detector error parametrization that can be controlled by 
calibration data will be essential to understanding and constraining systematic errors during the T2K-II era.

\subsection{Near Detector Plan}
\subsubsection{Longevity of the Current Near Detectors}
All scintillator based detectors such as the INGRID, ECal,  SMRD, P$\O$D, and FGD have experienced gain decreases on the order of a few per cent per year. This rate is small enough that we do not expect  significant degradation in the physics performance of these detectors over the next decade. The readout channel failure rate is a few per year in a total channel count of over ten thousand. At this rate enough spare electronics exist to maintain the readout needs of these detectors for several years. Both the P$\O$D and FGD use water targets. For the FGD no water issues are expected and for the P$\O$D, using the present water bag design, we expect $\sim$1 bag failure/leak per year. Enough spare bags exist for at least a decade. In addition, a new bag design is being tested with an anticipated lower failure rate. For the TPC no serious issues are expected, though the gas system will need maintenance. The TPC electronics failure rates are also very low. Assuming no catastrophic failures of high voltage, no longevity issues are expected. Finally, the magnet system expects no long term issues except for possible future maintenance concerns. The primary concern across all ND280 detectors is the anticipated loss of current expertise as some collaborators move on to other projects.

\subsubsection{Possibility of the ND280 Upgrade}
The reduction of systematic uncertainties is desirable to enhance the physics sensitivity of T2K-II.
Measurements using near detectors provide an essential ingredient to control the systematic uncertainties due to the neutrino flux and cross sections. 
A study of the possibility of a major upgrade of ND280 to enhance the capabilities of the near detectors is under way.
This study aims at significantly improving the acceptance of the near detector
for high angle and backward tracks, in order to better match the acceptance of ND280 to that of the far detector. Moreover, an increased efficiency to low momentum protons will allow to better discriminate between different models for the neutrino cross section. 

\begin{figure}[htbp]
\centering
\includegraphics[width=0.8\textwidth]{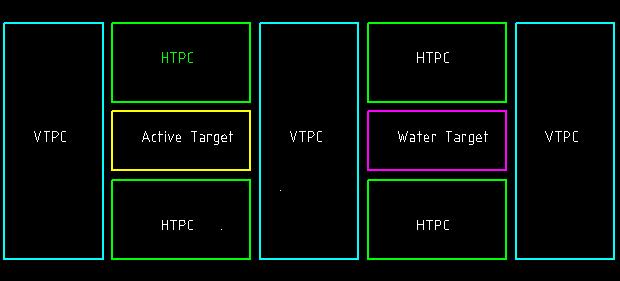}
\caption{Schematic configuration of ND280 upgrade under study.
  The VTPC(HTPC) are time projection chambers placed above/below (upstream/downstream)
  target detectors.
}
\label{Fig:ndupgrade}
\end{figure}

The reference design currently under consideration is based on the idea of improving the performance of the current tracker, which has been working very successfully, by reconfiguring and adding TPCs around two improved active targets.
The active targets will have larger angular acceptance than the current FGD, utilizing a 3D structure such as a grid structure with thin scintillators~\cite{Koga:2015iqa} or scintillator bars with a 3-axis structure.
A part of the target will contain water in order to constrain the neutrino cross section on water.
TPCs will be placed above and below the active targets in addition to upstream/downstream, to improve tracking and particle identification capabilities with a larger angular acceptance.
The configuration currently under investigation is shown in Fig.~\ref{Fig:ndupgrade}.

As a second option, we also investigate the concept based on a high pressure gas TPC, which will be able to achieve a low momentum threshold with $4\pi$ acceptance.

A Geant4-based MC simulation is under development to evaluate the performance of the proposed configuration and to determine basic parameters such as the target mass, size and type of TPC, and the segmentation of the active target.
The effect of additional capability, such as larger angular acceptance and lower momentum threshold, on the oscillation analysis will be estimated using the framework currently used to incorporate the  ND280 data into the oscillation analysis.
The study is expected to be completed in the fall of 2016 and will contain a
quantitative evaluation of the enhancement of the physics reach for T2K-II. 
Based on this report, the collaboration will discuss and decide on the next steps for this upgrade.

\subsubsection{Possibility of the Intermediate Detectors}
Discussion of the possibility of an intermediate detector at $\sim$1~km to enhance the T2K-II physics reach has started in the T2K collaboration.

The NuPRISM detector~\cite{nuprism1,nuprism2} is one promising candidate,
which has been developed by the independent NuPRISM collaboration, and uses neutrino interaction measurements on water
over a range of off-axis angles to address critical systematic 
uncertainties in the neutrino interaction model related to neutrino 
energy reconstruction, and the interaction rates of electron 
(anti)neutrinos.
Its aim is to measure kinematics of muons(electrons)
from the $\nu_\mu (\nu_e)$-water interaction as a function of neutrino energy
and make a prediction of observables at the far detector that minimizes the dependence on interaction models. 
In the course of trying to achieve the T2K-II systematic error goal,
  we may find that the existing neutrino interaction model
  is not sufficient to describe all the underlying physics processes relevant to the T2K measurement.
  Then, there could remain biases in the prediction at the far detector
  that degrades the final sensitivity.
  The NuPRISM concept would allow T2K to make
  a prediction at the far detector largely free from these biases,
  which would improve T2K systematics
  and hence the T2K-II physics reach.

Studies of an intermediate water Cherenkov detector have also
taken place within the Hyper-Kamiokande collaboration, including an alternative design called TITUS~\cite{titus}
that uses both the location, full containment of the event, neutron identification and large statistics of electron and
muon neutrinos and antineutrinos to reduce uncertainties on the predicted spectrum.
T2K is informed that the NuPRISM and TITUS groups are now merging into a single intermediate detector group
which will include additional physics enhancements from TITUS,
like the Gd neutron capture studies, to the program described above.

\subsection{Summary of Upgrades and Improvements}
Effective statistics per POT for CP violation studies will be
improved by up to 50\% by analysis improvements and beamline upgrades. 
The number of events expected at the far detector
for an exposure of \(20\times10^{21}\) POT with 50\% statistical
  improvement is given in Table \ref{tab:detevts2} for  \(\delta_{CP} = 0\) or \(-\pi / 2\). 

\begin{table}[htpb]
\begin{center}
\caption[Number of Expected Events]{Number of events expected to be observed
        at the far detector for
\(10\times10^{21}\)~POT \(\nu\)- + \(10\times10^{21}\)~POT \(\bar{\nu}\)-mode
with 50\% improvement in the effective statistics.
Assumed relevant oscillation parameters are:
\(\sin^22\theta_{13}=0.085\), \(\sin^2\theta_{23}=0.5\), 
\(\Delta m^2_{32}=2.5\times10^{-3}\) eV\(^2\), and normal mass hierarchy (NH).
\label{tab:detevts2}}
\begin{tabular}{c | c | c | c | c | c | c | c   } \hline
& & & Signal & Signal & Beam CC & Beam CC & \\
& True \(\delta_{CP}\) & Total & \(\nu_{\mu} \rightarrow \nu_e\) & \(\bar{\nu}_{\mu} \rightarrow
\bar{\nu}_e\) & \(\nu_e + \bar{\nu}_e \) & \(\nu_{\mu} + \bar{\nu}_{\mu} \) & NC\\ \hline\hline
\(\nu\)-mode & 0  & 467.6 & 356.3 &  4.0 & 73.3 & 1.8 & 32.3 \\ \cline{2-8}
$\nu_e$ sample & \(-\pi/2\) & 558.7 & 448.6 &  2.8 & 73.3 & 1.8 & 32.3 \\ \hline \hline
\(\bar{\nu}\)-mode & 0          & 133.9 & 16.7 &  73.6 & 29.2 & 0.4 & 14.1 \\ \cline{2-8}
$\bar{\nu}_e$ sample & \(-\pi/2\) & 115.8 & 19.8 &  52.3 & 29.2 & 0.4 & 14.1 \\ \hline 
\end{tabular} 
\vskip 0.4cm
\begin{tabular}{  c | c | c | c | c | c | c   } \hline
& & Beam CC & Beam CC & Beam CC & \(\nu_{\mu} \rightarrow \nu_e +\) & \\
& Total & \(\nu_\mu\) & \(\bar{\nu}_\mu\) & \(\nu_e + \bar{\nu}_e \) & \(\bar{\nu}_{\mu} \rightarrow \bar{\nu}_e\) & NC\\ \hline\hline
\(\nu\)-mode $\nu_\mu$ sample & 2735.0 & 2393.0 &  158.2 & 1.6 & 7.2 & 175.0 \\ \hline \hline
\(\bar{\nu}\)-mode $\bar{\nu}_\mu$ sample & 1283.5 & 507.8 &  707.9 & 0.6 & 1.0 & 66.2 \\ \hline
\end{tabular}
\end{center}
\end{table}


The current systematic error on the far detector prediction is from
5.5 to 6.8\%. Considering the present situation and projected
improvements, we consider that 4\% systematic error is a reachable
and reasonable target for T2K-II.
In case some uncertainties prevent us from achieving this goal, we are preparing to pursue necessary actions.
For example, though we have been improving our model of the neutrino-nucleus interaction
with our near detector data and data from other experiments
by working closely with theorists, in case these uncertainties are not resolved,
we are investigating whether near detector upgrades are absolutely needed to
resolve uncertainties from  neutrino-nucleus interaction modelling.
The pion interaction uncertainty, either from secondary or final-state
interactions, is one of the major sources contributing to all
error categories: flux prediction, near detector measurements and far detector
measurements.
A new dedicated experiment to measure the pion interaction can be an option
to reduce these uncertainties.

In what follows, this improvement in systematic error is modeled
by scaling the covariance matrix that reflects the current systematic error
to obtain an uncertainty in the far detector prediction that is 2/3 of its current size.
Whether a near detector upgrade is needed to achieve this goals
will be investigated on the one year time scale.

\newpage
\section{T2K-II Expected Physics Outcomes}
\label{sec:physics}
\subsection{Search for CP-violation in the Lepton Sector}
In this section, we describe the sensitivity to CP violation induced by a CP-odd phase in the three-flavor mixing matrix.
We assume that the full T2K-II exposure of 
\(20\times10^{21}\) POT is divided equally in \(\nu\)-mode and \(\bar{\nu}\)-mode.
A study of different ratios of $\nu$- and $\bar{\nu}$-mode running is shown later in this section.
This ratio will eventually be optimized
over the course of the experiment.
Sensitivities were initially calculated with the current T2K (2016 oscillation analysis) event rates and systematics, and the effect of the statistical enhancements from beamline and analysis improvements and systematic error reduction were implemented by a simple scaling of the event rates and covariance matrices. In what follows, unless otherwise noted, a 50\% increase in the effective statistics from horn and far detector selection improvements is assumed, and the
relevant oscillation parameters are:
\(\sin^22\theta_{13}=0.085\), \(\sin^2\theta_{23}=0.5\), 
\(\Delta m^2_{32}=2.509\times10^{-3}\) eV\(^2\), and normal mass hierarchy (MH).
Cases for $\sin^2\theta_{23}$ at the edge of the current 90\% C.L. regions
($\sin^2\theta_{23}= 0.43,\; 0.60$) are also studied.

\begin{figure} [H]
\centering 
\begin{subfigure}[H]{7.2cm}
\includegraphics[width=7.2cm]{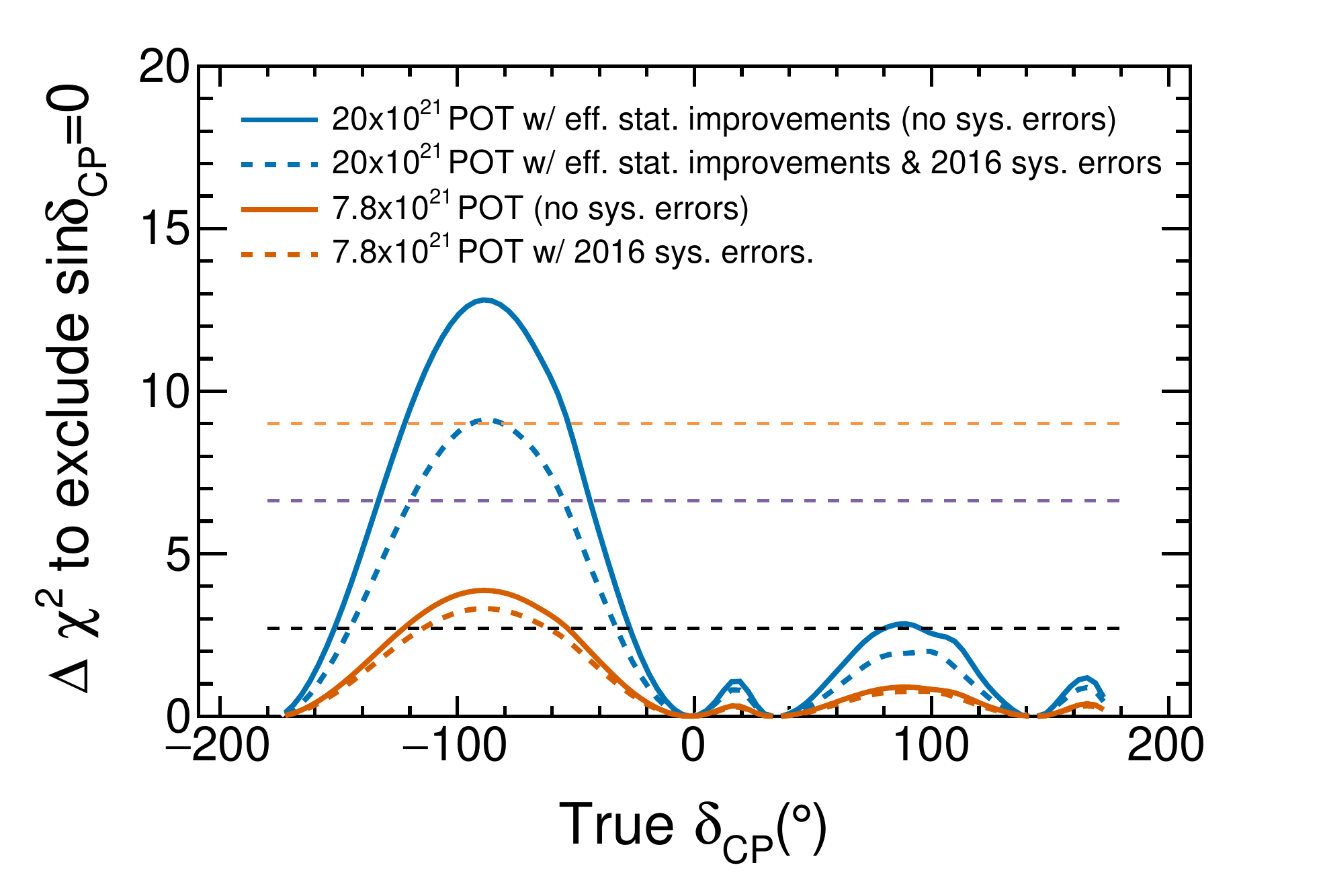}
\caption{Assuming the MH is unknown.} \label{fig:CPVvsdCP_unknownMH2016comp}
\end{subfigure} \quad 
\begin{subfigure}[H]{7.2cm}
\includegraphics[width=7.2cm]{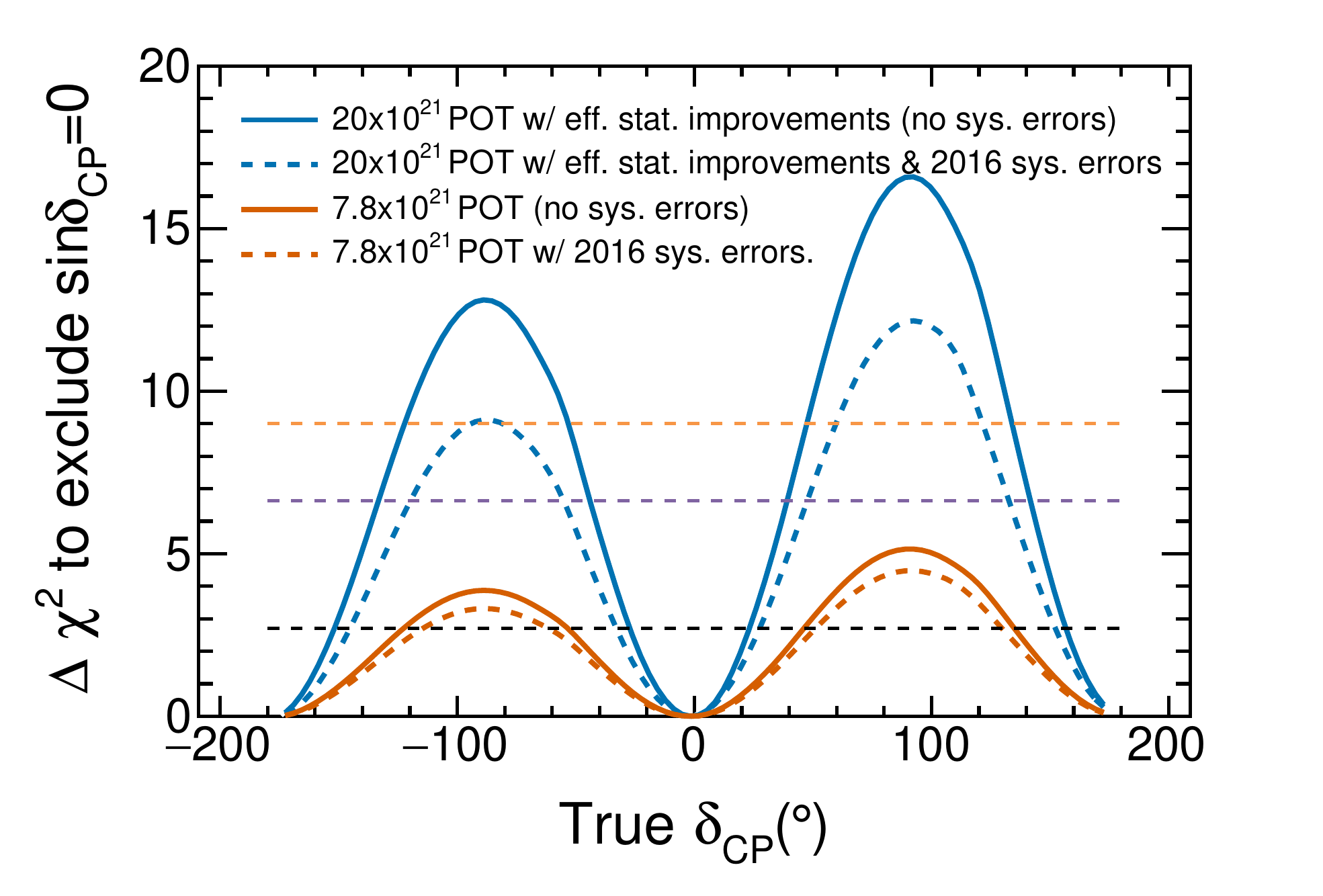}
\caption{Assuming the MH is known -- measured by an outside experiment.} \label{fig:CPVvsdCP_knownMH2016comp}
\end{subfigure} \quad
\caption[CPV vs dCP]{Sensitivity to CP violation as a function of true $\delta_{CP}$   with 2016 T2K systematic errors. The normal mass hierarchy and $\sin^2\theta_{23}=0.5$ are assumed. The left plot assumes unknown mass hierarchy and the right assumes
  known mass hierarchy.
\label{fig:CPVvsdCP2016comp}} \end{figure}

\begin{figure} [H]
\centering 
\begin{subfigure}[H]{7.2cm}
\includegraphics[width=7.2cm]{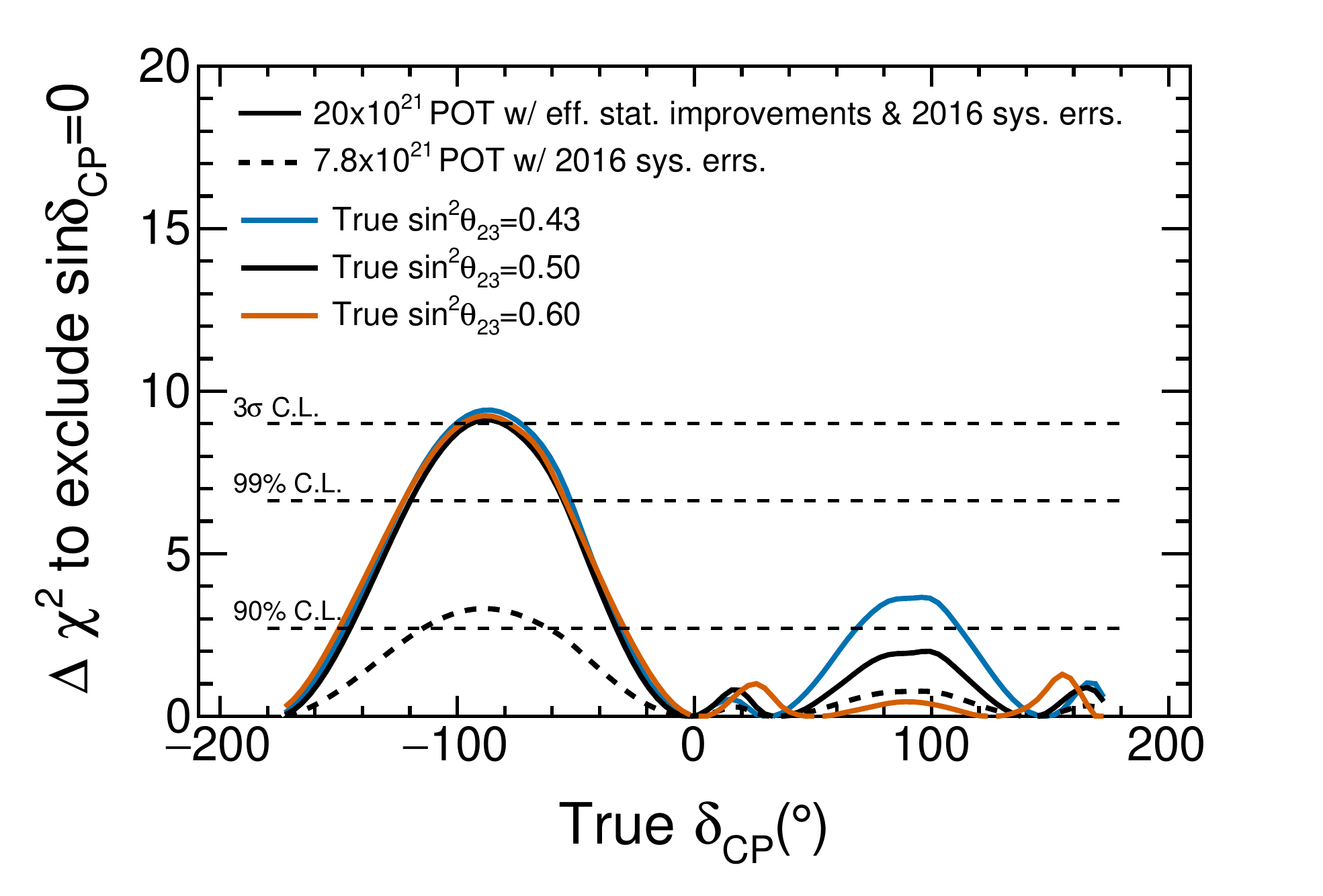}
\caption{Assuming the MH is unknown.} \label{fig:CPVvsdCP_unknownMH2016}
\end{subfigure} \quad 
\begin{subfigure}[H]{7.2cm}
\includegraphics[width=7.2cm]{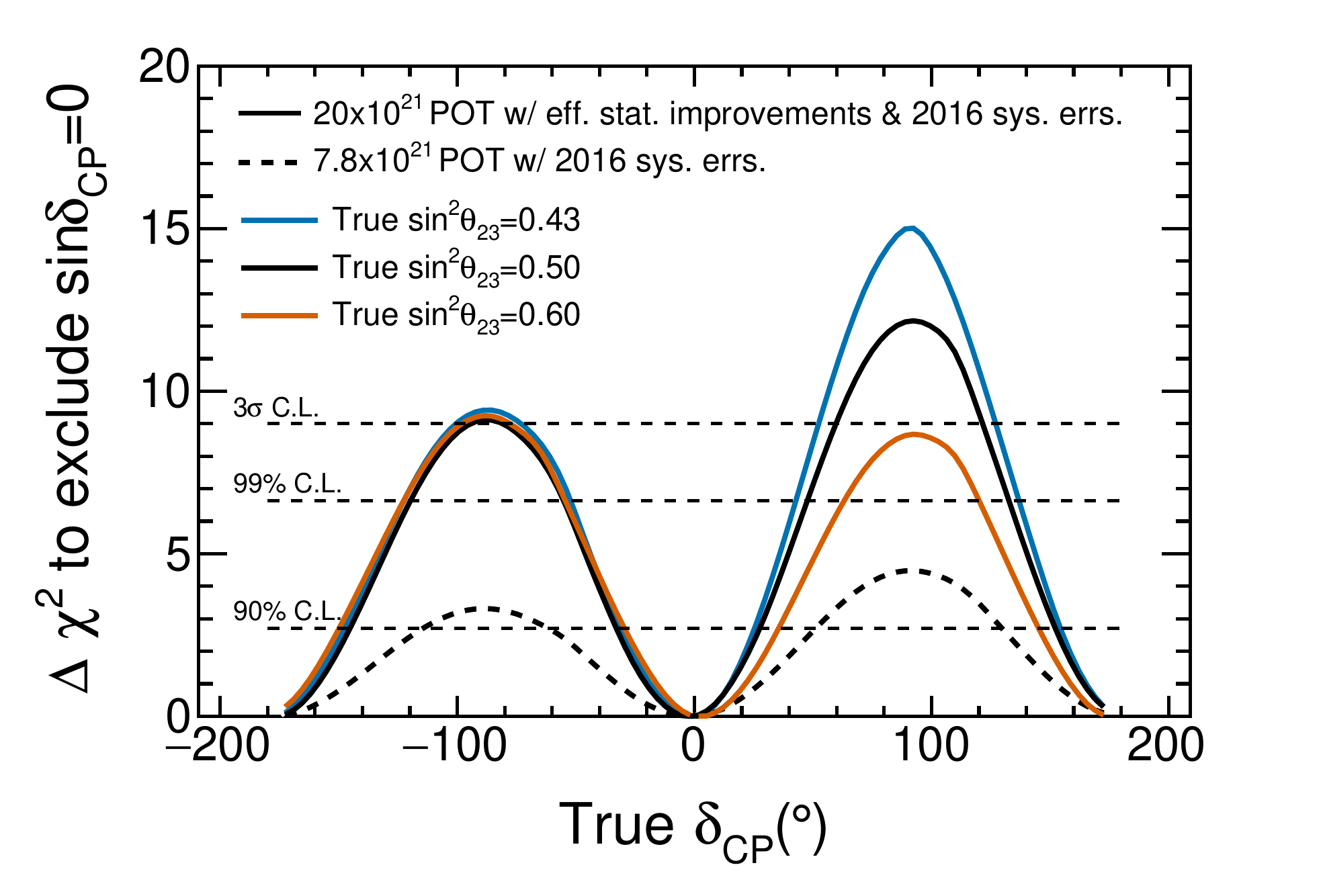}
\caption{Assuming the MH is known -- measured by an outside experiment.} \label{fig:CPVvsdCP_knownMH2016}
\end{subfigure} \quad
\caption[CPV vs dCP]{Sensitivity to CP violation as a function of true
\(\delta_{CP}\) with three values of $\sin^2\theta_{23}$ (0.43, 0.50, 0.60) and normal hierarchy for the full T2K-II exposure of $\twopott$ and 2016 T2K systematic errors.
The left plot assumes that the mass hierarchy is unknown and the right assumes it is known.
\label{fig:CPVvsdCP2016}} \end{figure}

\begin{figure} [H]
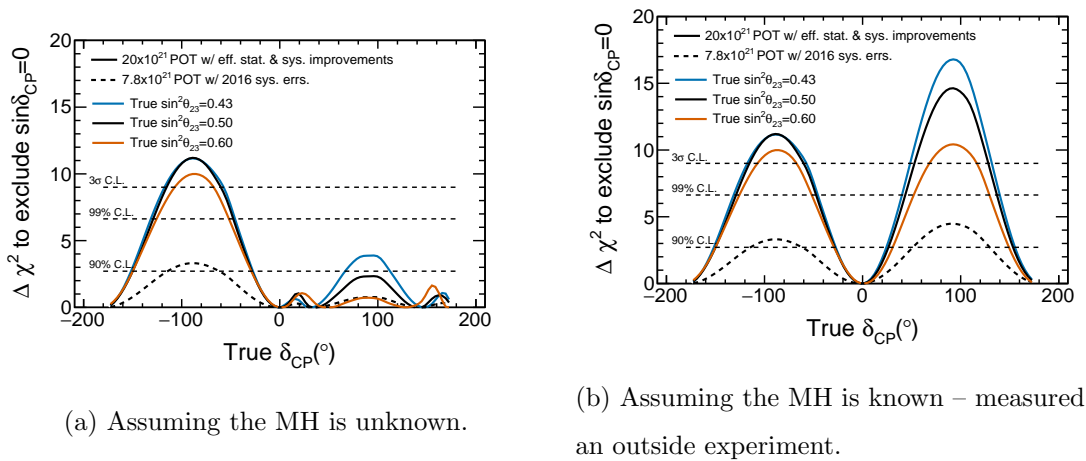

\centering 
\begin{subfigure}[H]{7.2cm}
\includegraphics[width=7.2cm]{figs/t2kpre_dcp_point1_100k4check_100ksensi_wreactorthrow_optv2s13off_truedcp_unknownMH_fakesyst_lohidcpExclusive.pdf}\caption{Assuming the MH is unknown.} \label{fig:CPVvsdCP_unknownMH}
\end{subfigure} \quad 
\begin{subfigure}[H]{7.2cm}
\includegraphics[width=7.2cm]{figs/t2kpre_dcp_point1_100k4check_100ksensi_wreactorthrow_optv2s13off_truedcp_fakesyst_lohidcpExclusive.pdf}\caption{Assuming the MH is known -- measured by an outside experiment.} \label{fig:CPVvsdCP_knownMH}
\end{subfigure} \quad
\caption[CPV vs dCP]{Sensitivity to CP violation as a function of true
\(\delta_{CP}\) with three values of $\sin^2\theta_{23}$ (0.43, 0.50, 0.60) and normal hierarchy for the full T2K-II exposure of $\twopott$.
and a reduction of the systematic error to 2/3 of the 2016 T2K uncertainties.
\label{fig:CPVvsdCP}} \end{figure}

When calculating sensitivities, the values of \(\sin^2\theta_{23}\), \(\Delta m^2_{32}\), and \(\delta_{CP}\) are
assumed to be constrained by  T2K data only, while \(\sin^22\theta_{13}\) is constrained by reactor measurements to $\sin^22\theta_{13}=0.085\pm0.005$
\cite{pdg}.
There is a degeneracy in the expected $\nu_e$ event rate if the mass hierarchy is NH (IH) and $\delta_{CP}>0$ $(<0)$, 
and the sensitivity is quite different for $\delta_{CP}>0$ compared to $\delta_{CP}<0$ if the mass hierarchy is not known.
Several experiments (NOvA, JUNO, ORCA, PINGU) are expected or plan to determine the mass hierarchy before or during
the proposed period of T2K-II \cite{An:2015jdp,Patterson:2012zs,Aartsen:2014oha,Katz:2014tta}.
Hence both MH-unknown and -known cases are studied. 
Figure~\ref{fig:CPVvsdCP2016comp} shows a comparison of sensitivity to CP violation (\(\Delta\chi^2\) for resolving \(\sin\delta_{CP}\neq0\))  at $\sin^2\theta_{23}=0.50$ plotted 
as a function of true \(\delta_{CP}\) for two cases: the approved T2K statistics ($\onepott$) without an effective statistical improvement and the full T2K-II exposure ($\twopott$) with the improvement. The sensitivity without systematic errors and with 2016 T2K systematic errors is shown; a significant degradation in sensitivity is observed due to these uncertainties.
The sensitivities to CP violation at different true values of $\theta_{23}$ are compared in Fig.~\ref{fig:CPVvsdCP2016} using  2016 T2K systematic errors. Sensitivity close to 3$\sigma$ for $\delta_{CP}=-\pi/2$ is achieved in all cases. If the systematic error is reduced to 2/3 of its current magnitude, sensitivity of at least 3$\sigma$  is possible over a significant range of possible true values of $\delta_{CP}$ as shown in Fig.\ \ref{fig:CPVvsdCP}.

The fractional region for which \(\sin\delta_{CP}=0\) can be excluded at the 99\% (3\(\sigma\)) 
C.L.\ is 49\% (36\%) of possible true values of \(\delta_{CP}\) assuming the systematic errors are reduced to 2/3 of the 2016 T2K uncertainties and that the MH has been determined by an another experiment. If the 2016 T2K systematic errors are assumed, the corresponding fractions are 42$\%$ (21$\%$).
If systematic errors are set to zero, the fractional region where CP violation can be observed at 99\% (3\(\sigma\)) C.L. becomes 53\% (43\%).
The coverage fraction is slightly larger for the case of lower octant $\sin^2\theta_{23}=0.43$
and slightly lower for the case of upper octant  $\sin^2\theta_{23}=0.60$. More details of coverage at different values of $\sin^2\theta_{23}$ can be found in Table~\ref{tab:coverage}.
 
\begin{table}[th]
  \caption{Table of $\delta_{CP}$ fractional coverages (\%) with three options of systematic treatment: no systematic error (statistical only), 2016 systematics and improved systematics. The coverages are calculated at three different values of $\sin^2\theta_{23}$ (0.43, 0.5, and 0.60) and it is assumed that the MH has been determined by an outside experiment.}
  \label{tab:coverage}
  \centering
  \begin{tabular}{c|c|c|c|c|c|c}
    \hline
    \multirow{2}{*}{} & \multicolumn{2}{c|}{$\sin^2\theta_{23}=0.43$   } &\multicolumn{2}{c|}{$\sin^2\theta_{23}=0.50$   } &  \multicolumn{2}{c}{$\sin^2\theta_{23}=0.60$  } \\
    \hline{~-----}
    & 99$\%$ C. L. & 3$\sigma$ & 99$\%$ C. L. & 3$\sigma$  & 99$\%$ C. L.  & 3$\sigma$  \\
    \hline
    Stat. Only & 57.5 & 47.9 & 53.3 & 43.1 & 49.1 & 36.7\\ 
    \hline
    2016 systematics & 45.6 & 28.3 & 41.6 & 20.5 & 34.7 & 5.2\\ 
    \hline 
    Improved systematics & 51.5 & 39.7 & 48.6 & 36.1 & 41.8 & 23.9\\ 
    \hline
  \end{tabular}
  
\end{table}

\begin{figure}[H] \centering
\begin{subfigure}[H]{7.2cm}
\includegraphics[width=7.2cm]{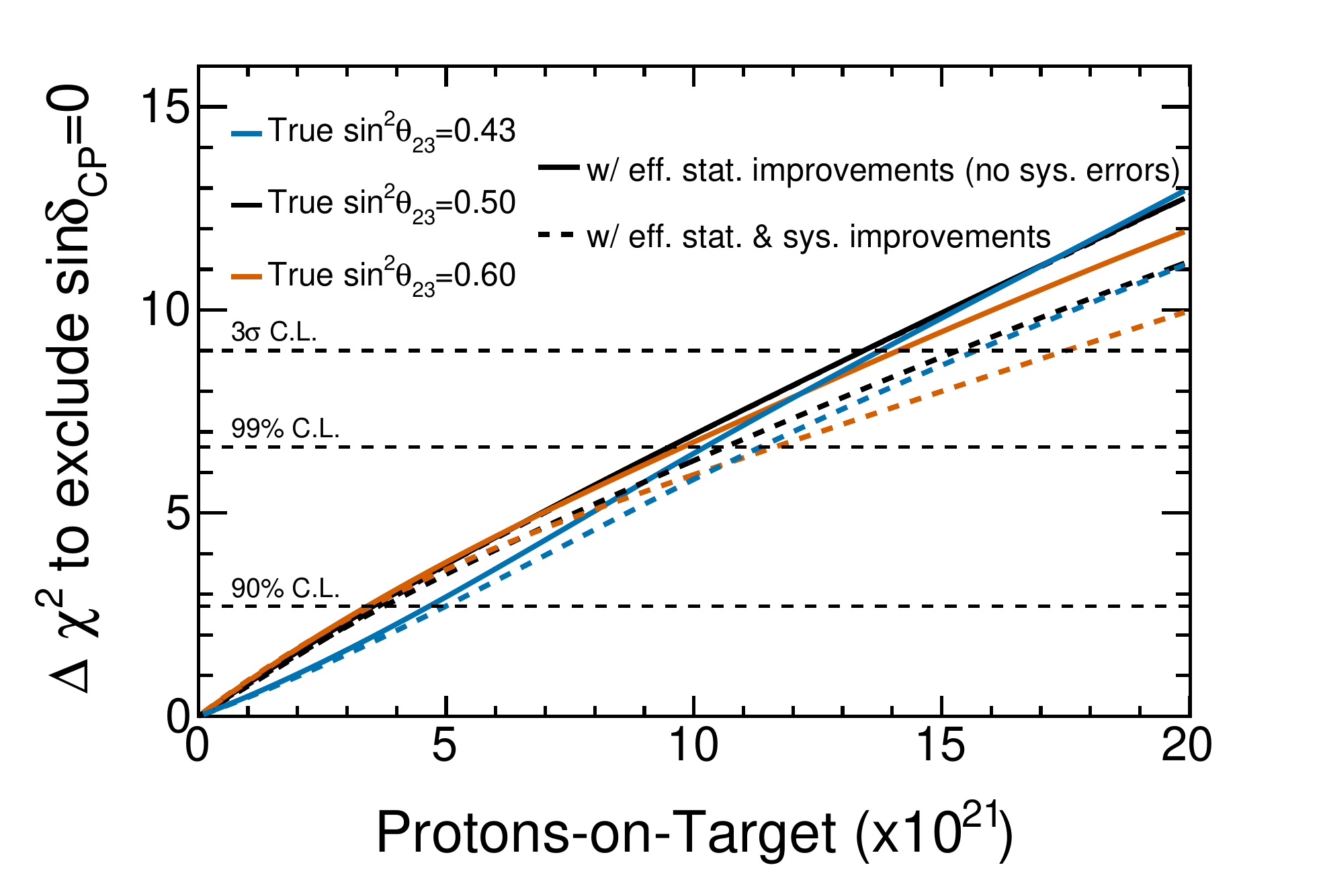}
\end{subfigure} \quad 
\begin{subfigure}[H]{7.2cm}
\includegraphics[width=7.2cm]{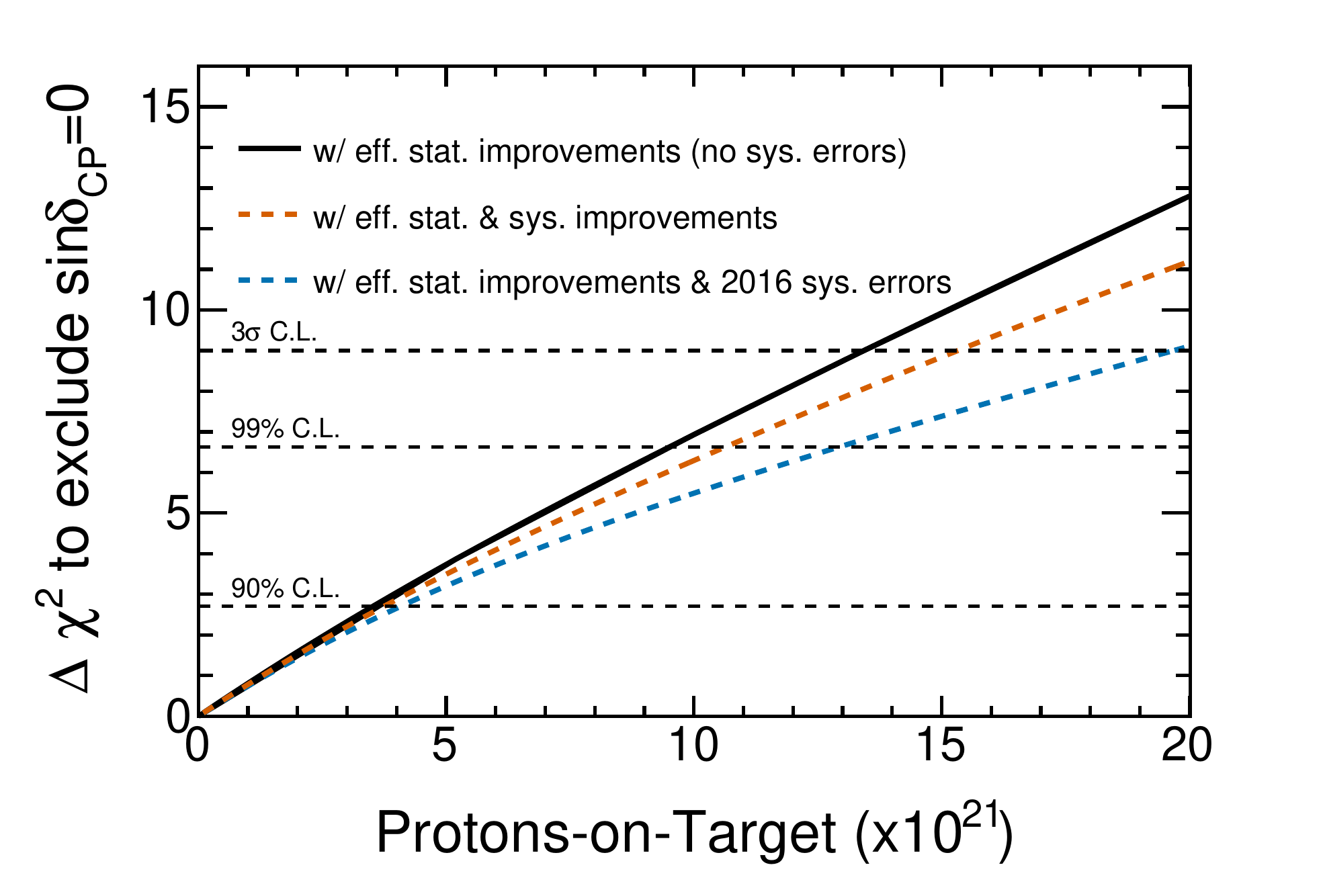}
\end{subfigure} \quad 
\caption[CPV vs POT]{Sensitivity to CP violation as a function of POT with a 50\% improvement
  in the effective statistics, assuming the true MH is the normal MH but unknown and the true value
  of \(\delta_{CP}=-\pi/2\). 
  The plot on the left compares different true values of \(\sin^2\theta_{23}\),
  while that on the right compares different assumptions for the T2K-II systematic errors
  with $\sin^2\theta_{23}=0.50$.
\label{fig:CPVvsPOT}} \end{figure}

The expected evolution of the sensitivity to CP violation as a function of POT assuming
that the T2K-II data is taken in roughly equal alternating periods of
\(\nu\)-mode and \(\bar{\nu}\)-mode (with true normal MH 
and \(\delta_{CP}=-\pi/2\)) is given in Fig.~\ref{fig:CPVvsPOT}.

\begin{figure}[H]
\centering
\includegraphics[width=0.47\textwidth]{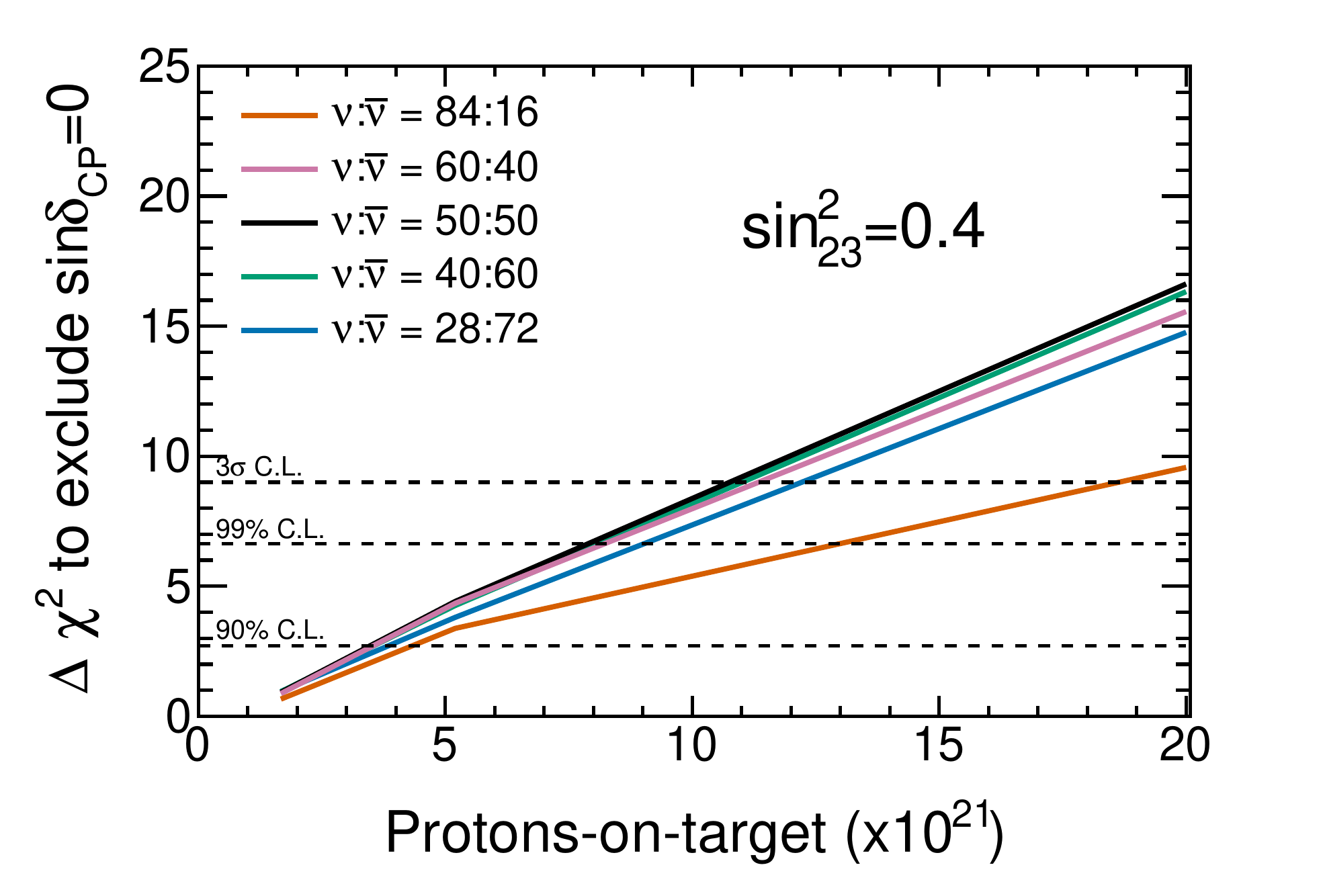} 
\includegraphics[width=0.47\textwidth]{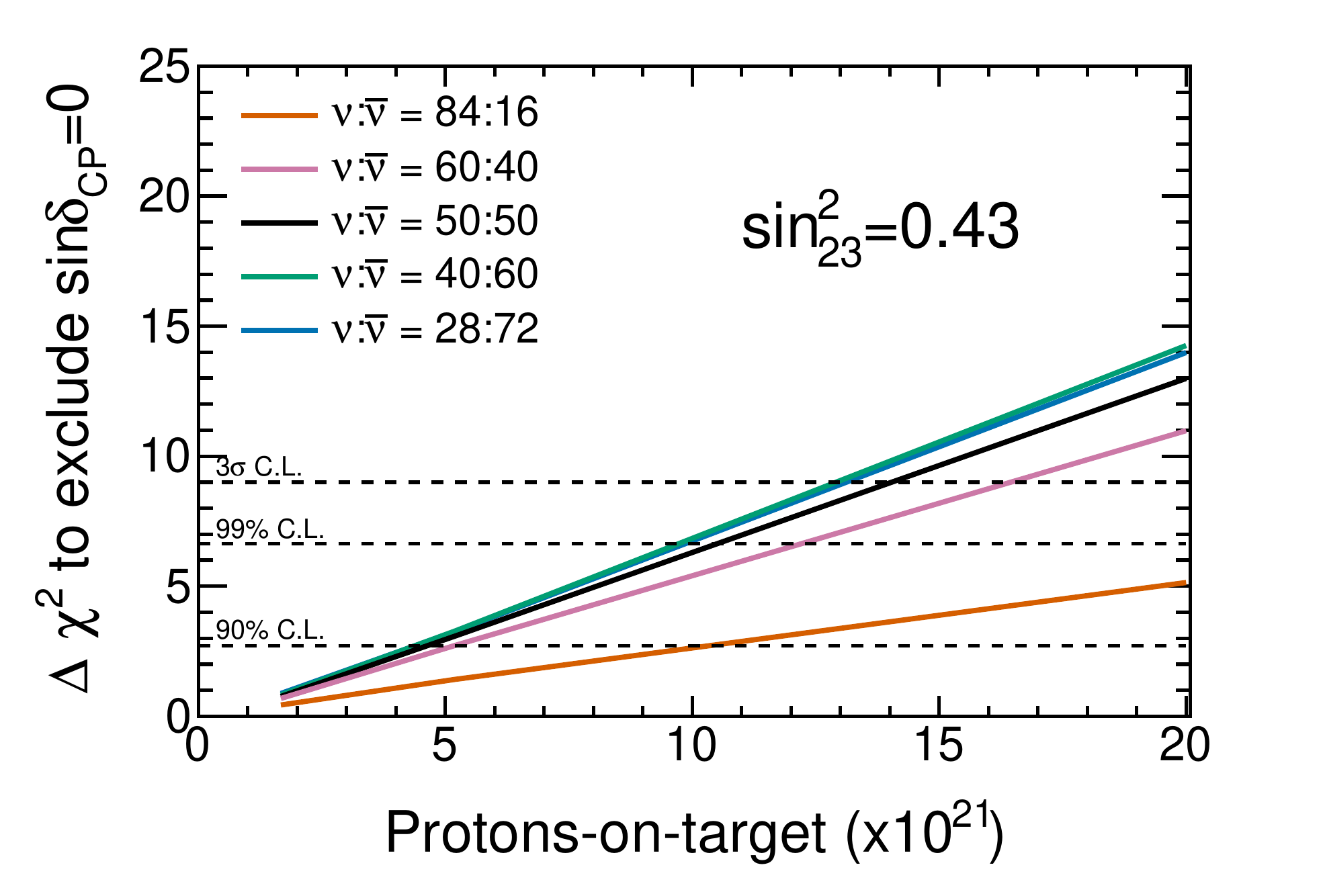} 
\includegraphics[width=0.47\textwidth]{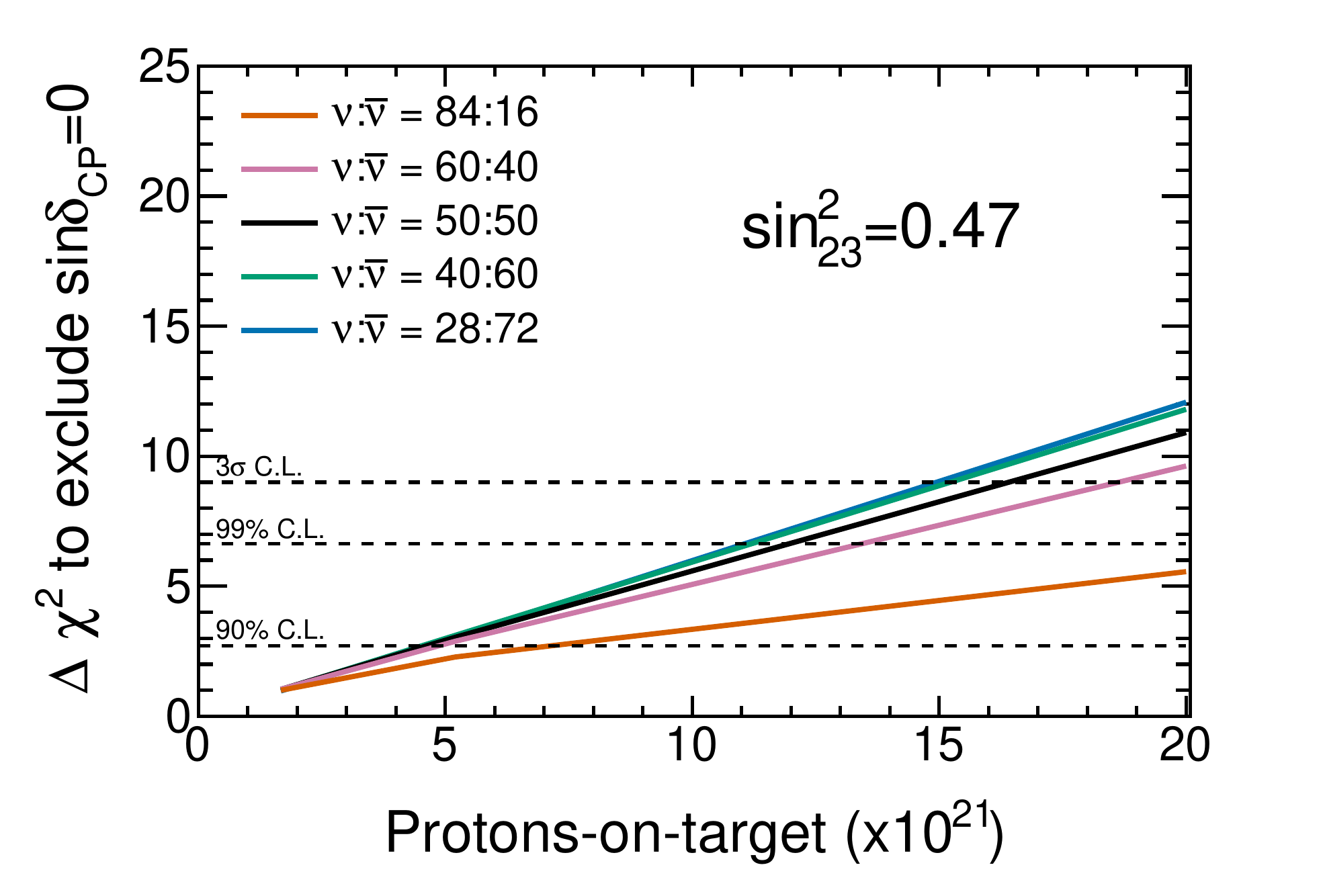} 
\includegraphics[width=0.47\textwidth]{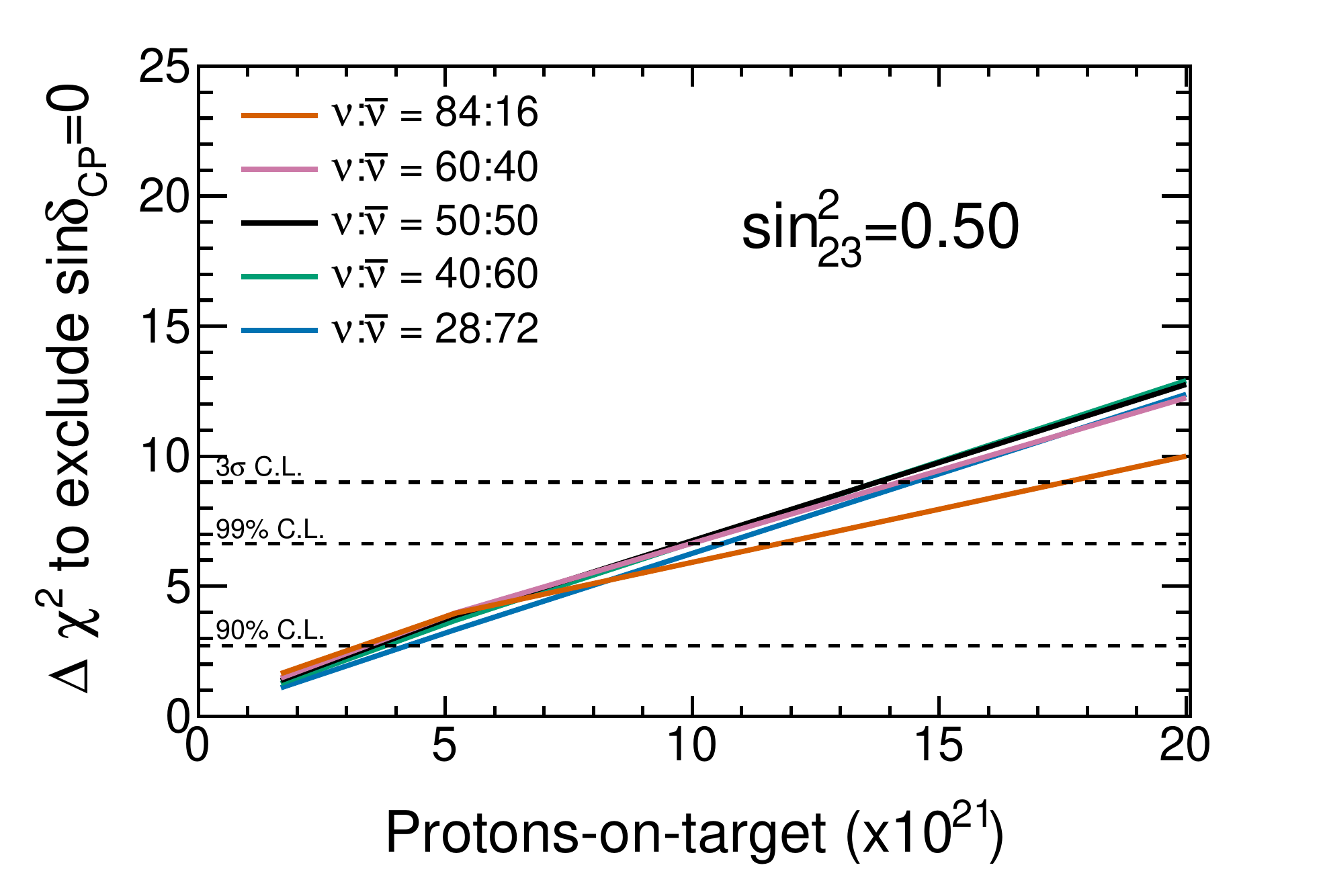} 
\includegraphics[width=0.47\textwidth]{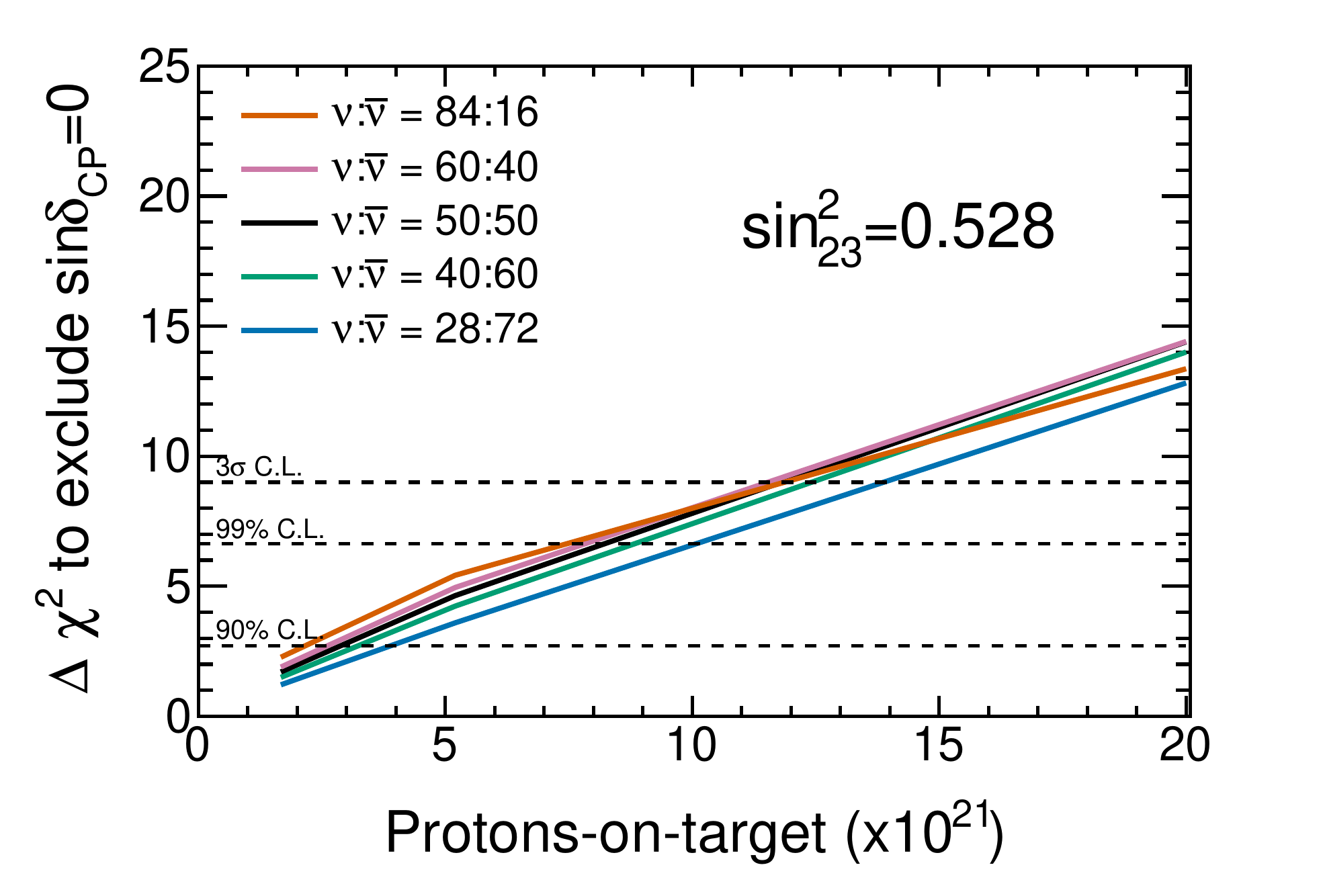} 
\includegraphics[width=0.47\textwidth]{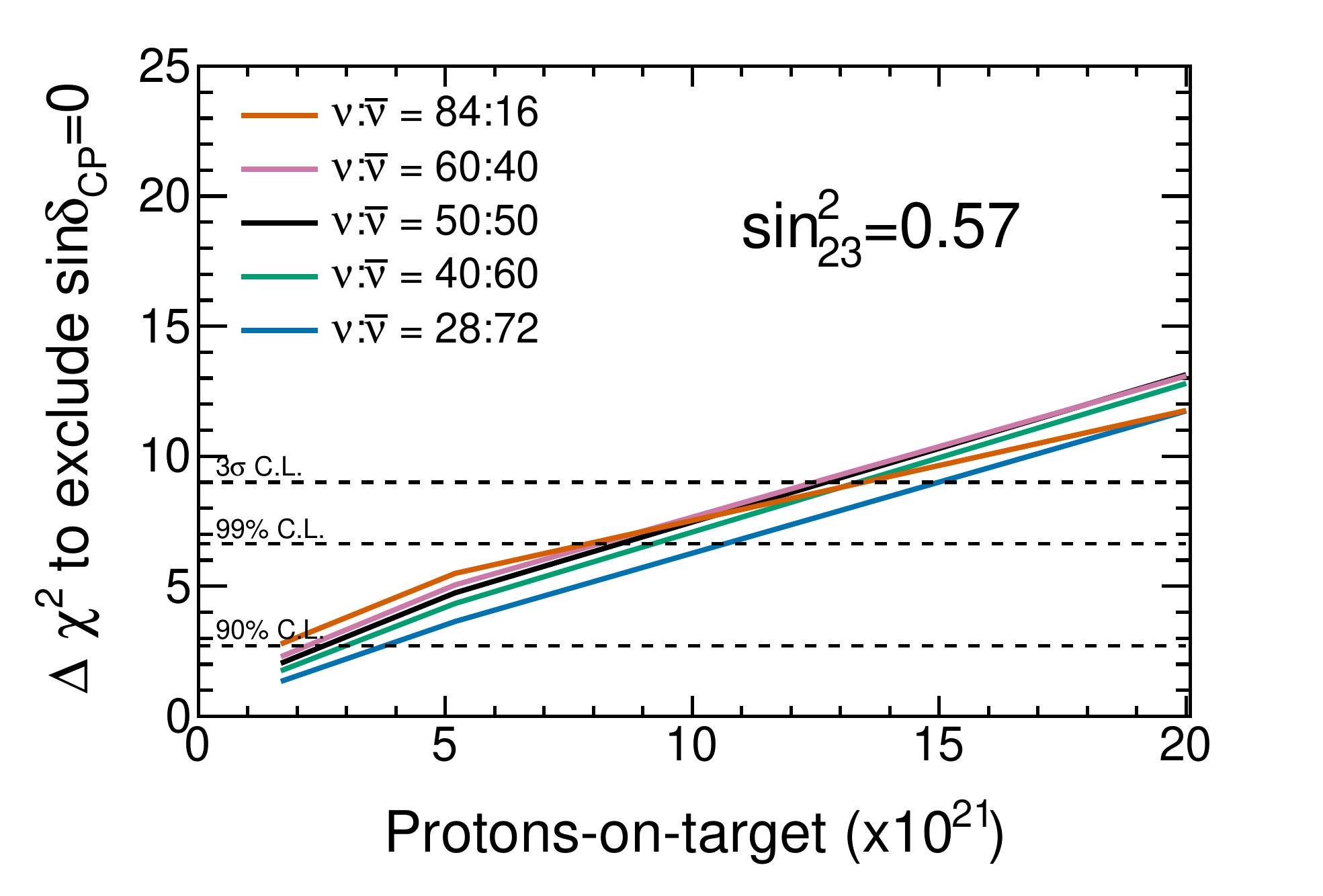} 
\includegraphics[width=0.47\textwidth]{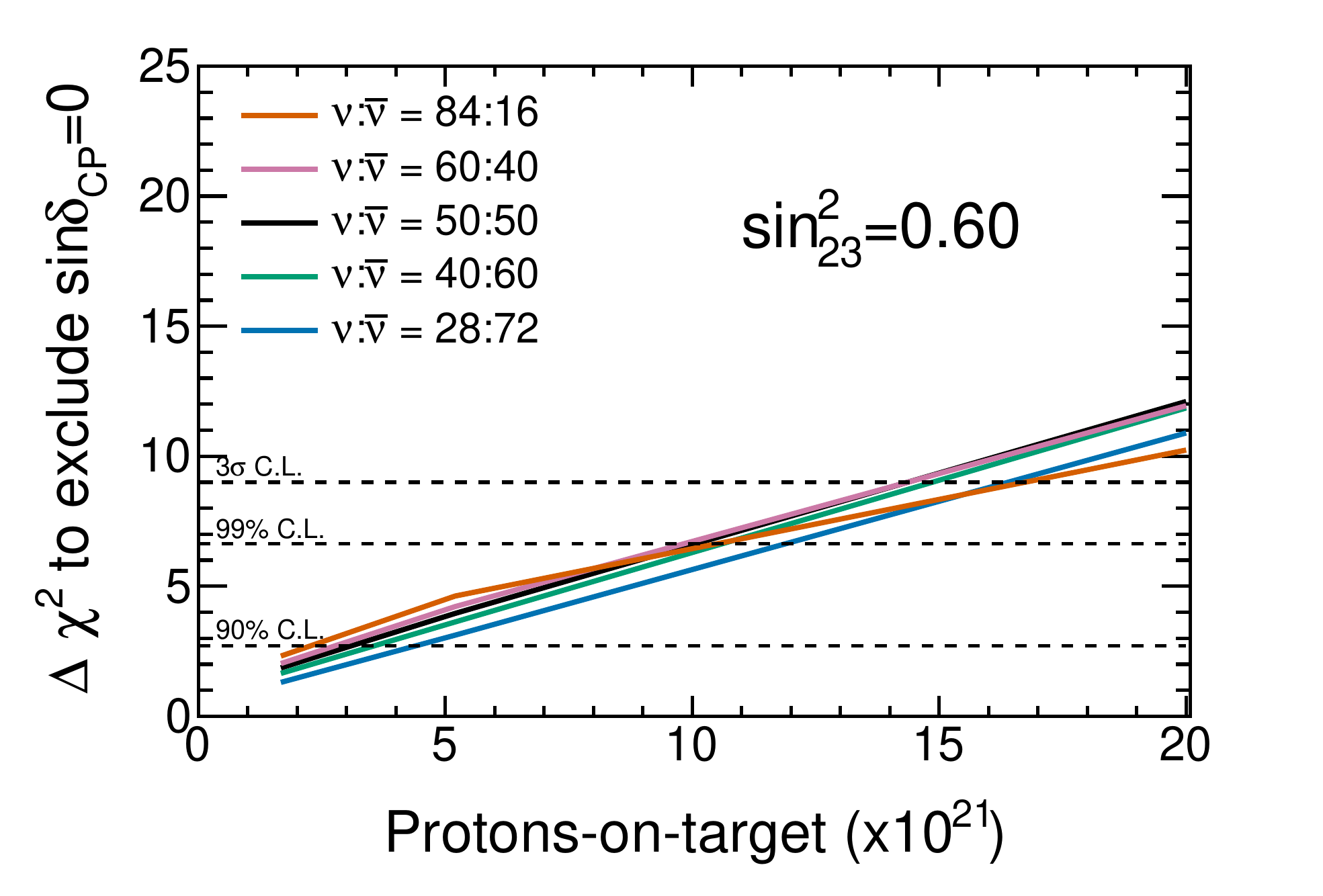} 
\caption{Sensitivity to CP violation plotted as a function of POT with various values of $\sin^2\theta_{23}$
  and various option of $\nu:\bar{\nu}$ running time ratios.
  Only statistical errors  are considered. Other conditions are same as those
  in Fig.~\ref{fig:CPVvsPOT} caption.
  The ''84:16'' or ``28:72'' are ratio when new data are taken
  only in one mode on top of existing statistics at some point during the study.} 
\label{fig:truedcpruntimerat}
\end{figure}

\begin{figure}[H]
\centering

\includegraphics[width=0.47\textwidth]{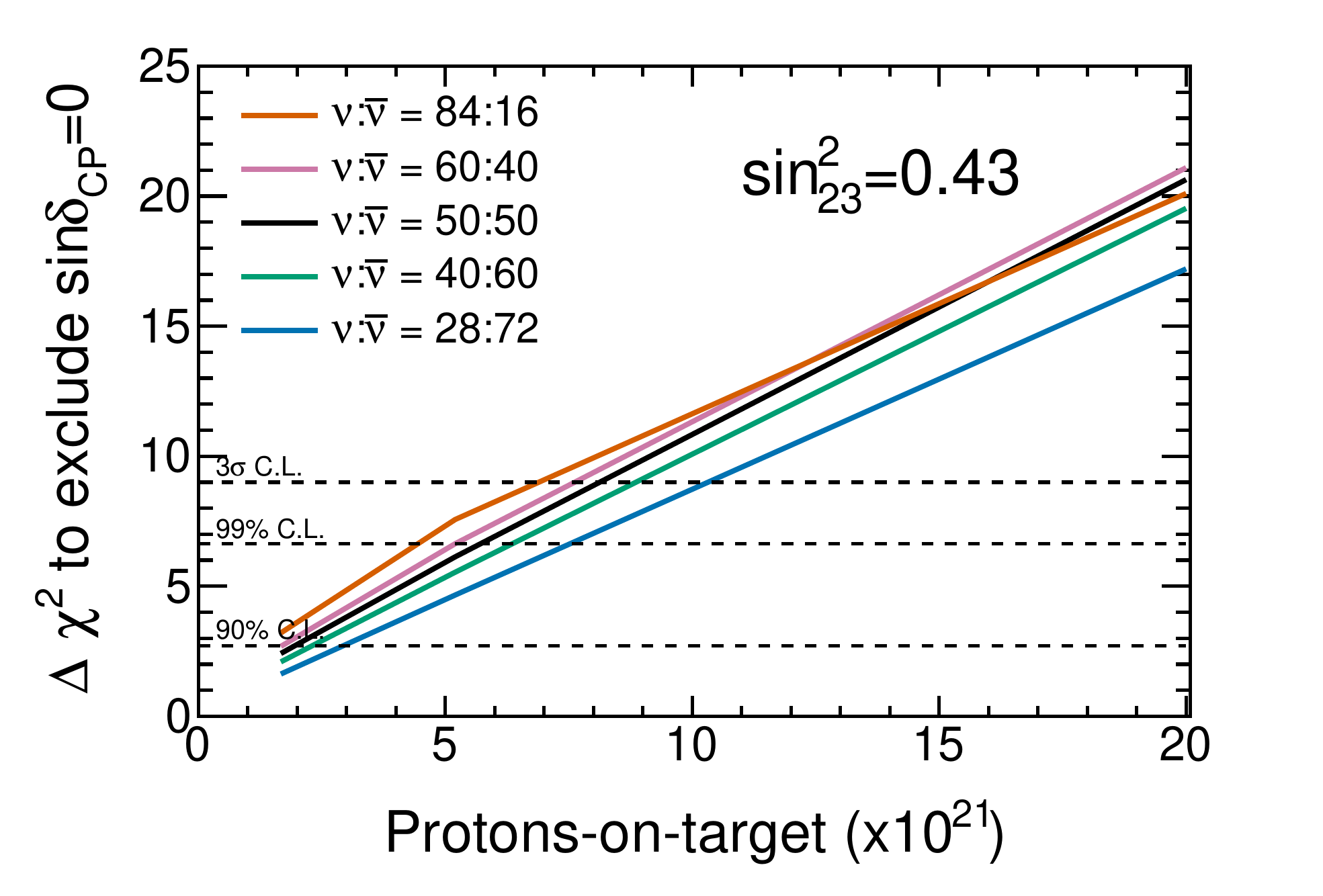} 
\includegraphics[width=0.47\textwidth]{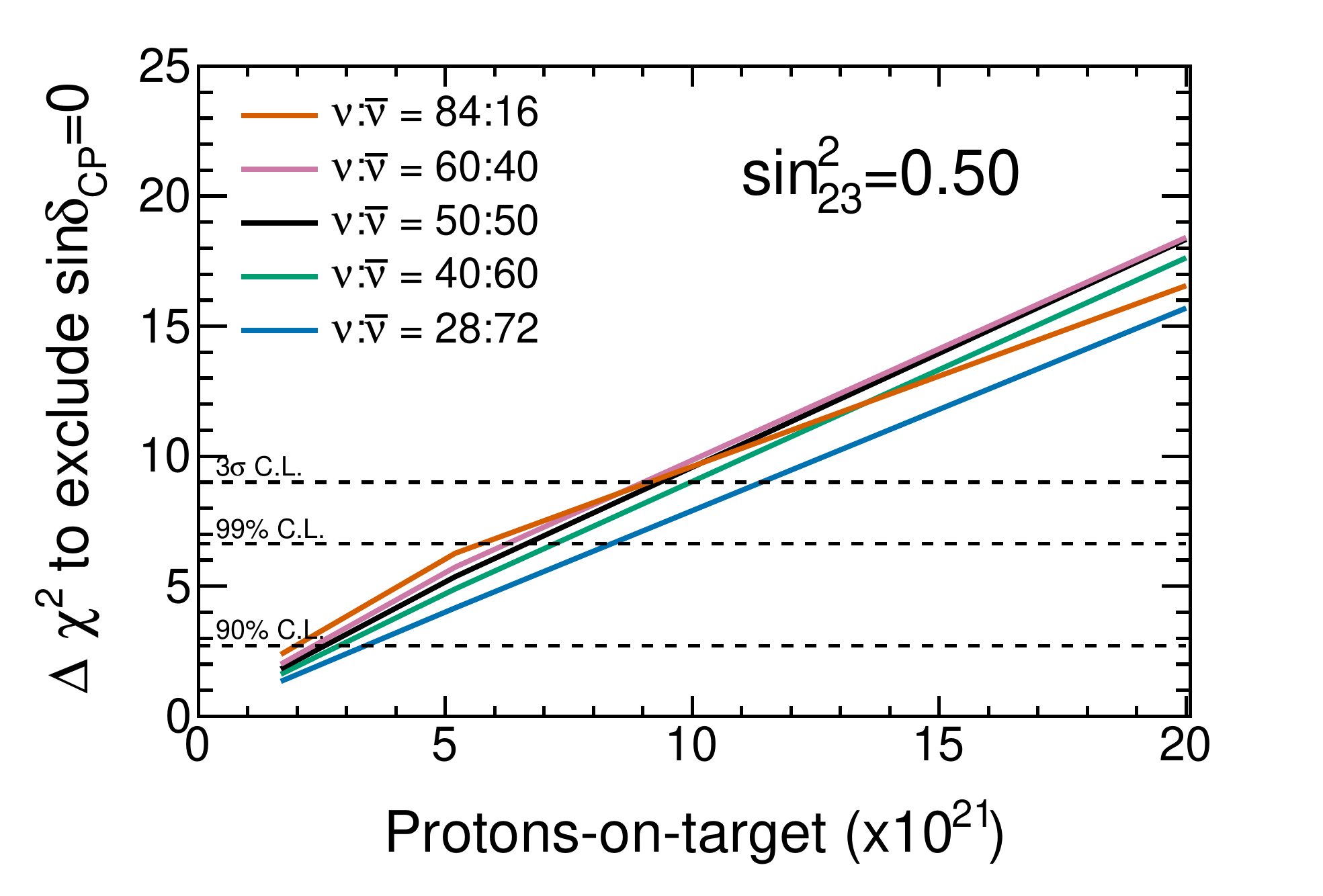} 
\includegraphics[width=0.47\textwidth]{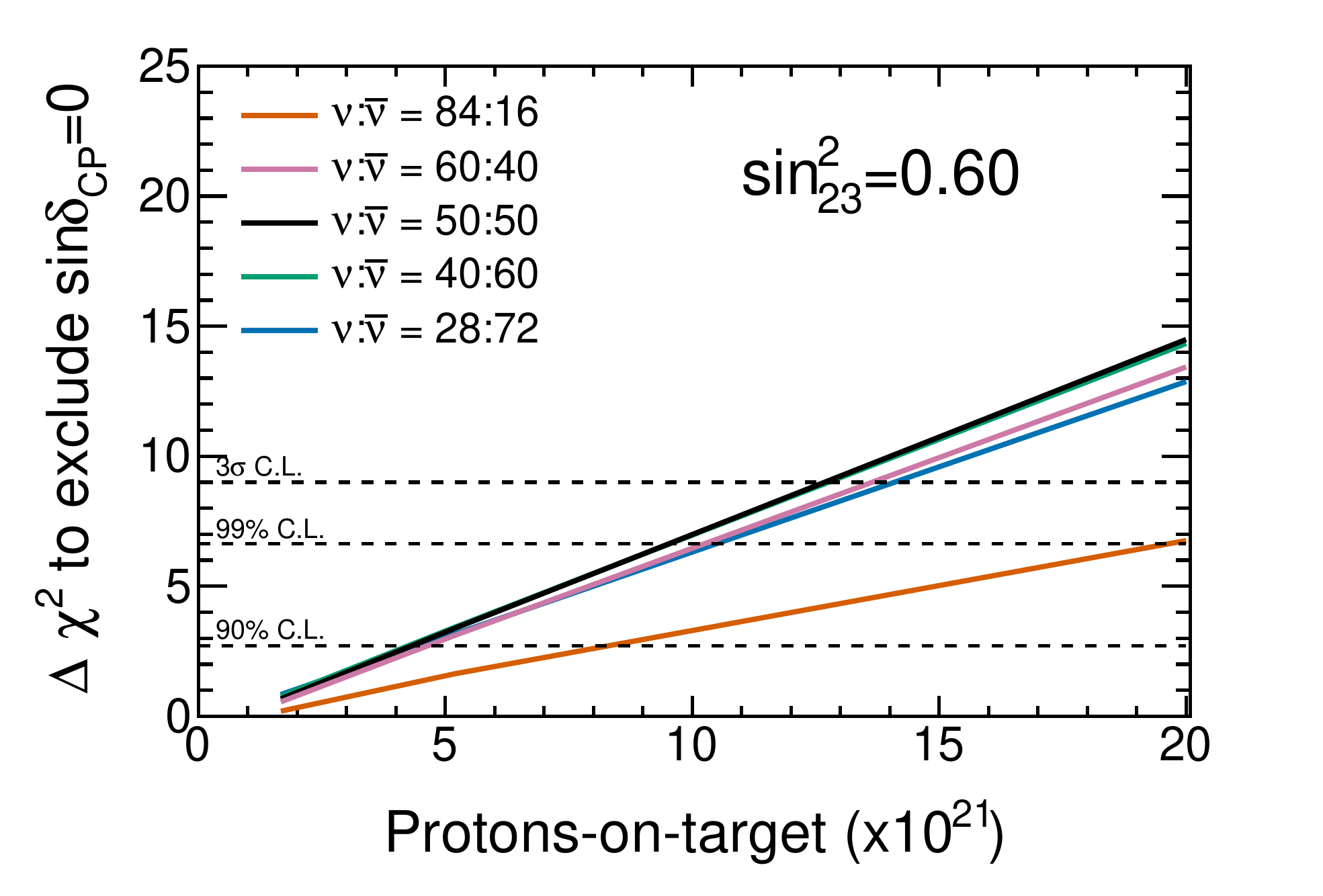} 

\caption{
  Sensitivity to CP violation as a function of POT with various values of $\sin^2\theta_{23}$ and options of $\nu:\bar{\nu}$ exposures assuming inverted hierarchy and $\delta_{CP}=+\frac{\pi}{2}$. Only statistical errors are considered.
}
\label{fig:truedcpruntimeratimh}
\end{figure}

The above study assumes that equal POT are accumulated in $\nu$-mode and the $\bar{\nu}$-mode. The balance could be optimized to enhance the significance for observing CP violation. Sensitivity to CP violation depends on resolving degeneracies such as the mass hierarchy and the $\theta_{23}$ octant. Thus, this optimization requires a detailed consideration over a large space of neutrino oscillation parameters and the outcome of future measurements.  Here, we  verify that $\nu:\bar{\nu}=50:50$ running, while not optimal in all cases, is a reasonable option that achieves sensitivities close to optimal across a range of underlying parameters.
Figure~\ref{fig:truedcpruntimerat} shows the sensitivity to CP violation plotted as a function of POT with seven true values of $\sin^2 \theta_{23}$ and five options of the $\nu:\bar{\nu}$ running time ratios (in percentage).
In this study, only  statistical uncertainty is considered and the statistical enhancement is assumed throughout.
It can be observed that the configuration where  data is taken dominantly in $\nu$-mode  gives the worst sensitivity to CP violation if the true value of $\theta_{23}$ is in the lower octant. This is because $\nu$-mode running alone has limited power to resolve the $\theta_{23}$ octant.
On the other hand, while more  $\bar{\nu}$-mode running improves the ability to resolve the $\theta_{23}$ octant, it suffers from decreased statistics. We conclude that taking data equally in $\nu$-mode and $\bar{\nu}$-mode, while not the optimal configuration for all values of $\sin^2 \theta_{23}$, consistently gives high sensitivity to  CP violation overall.
The discussion so far has concentrated on normal mass hierarchy and $\delta_{CP}=-\frac{\pi}{2}$.
Due to the symmetry of the oscillation probabilities, it is expected that the above conclusions  also hold for the case of inverted mass hierarchy and $\delta_{CP}=+\frac{\pi}{2}$ and switching the octant of $\theta_{23}$. 
Figure~\ref{fig:truedcpruntimeratimh} shows the sensitivity 
with statistical errors only for various data taking configurations and at three different values of $\sin^2\theta_{23}$ (0.43, 0.5 and 0.60)
for the case of 
inverted hierarchy and $\delta_{CP}=+\pi/2$.
It can be seen from this plot that running primarily in $\nu$-mode leads to worse sensitivity in the case that $\theta_{23}$ is in the upper octant.
 This is opposite to the case when normal mass hierarchy and $\delta_{CP}=-\frac{\pi}{2}$ are assumed. Also, taking data equally in $\nu$-mode and $\overline{\nu}$-mode gives high sensitivity over the full possible range of $\sin^2 \theta_{23}$ values. Compared to the case of normal mass hierarchy and $\delta_{CP}=-\frac{\pi}{2}$, the sensitivity to CP violation is significantly higher in the case of lower octant and maximum mixing ($\theta_{23}\sim\pi/4)$.

\subsection{Precision Measurement of $\Delta m^2_{32}$ and $\sin^2\theta_{23}$}
\begin{figure}[H] \centering
\begin{subfigure}[H]{7.2cm}
\includegraphics[width=7.2cm]{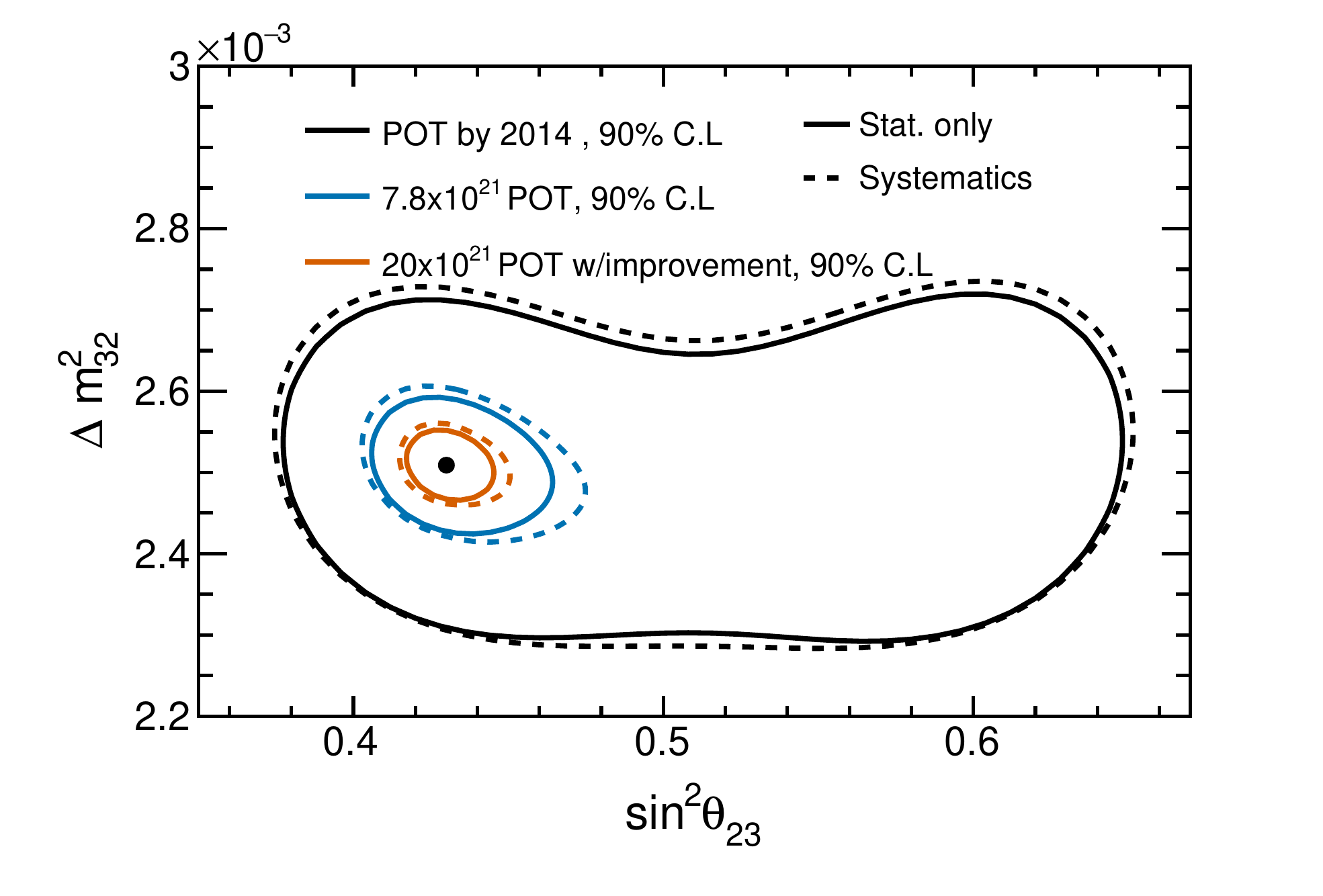}\caption{Assuming true \(\sin^2\theta_{23}=0.43\).}
\end{subfigure} \quad 
\begin{subfigure}[H]{7.2cm}
\includegraphics[width=7.2cm]{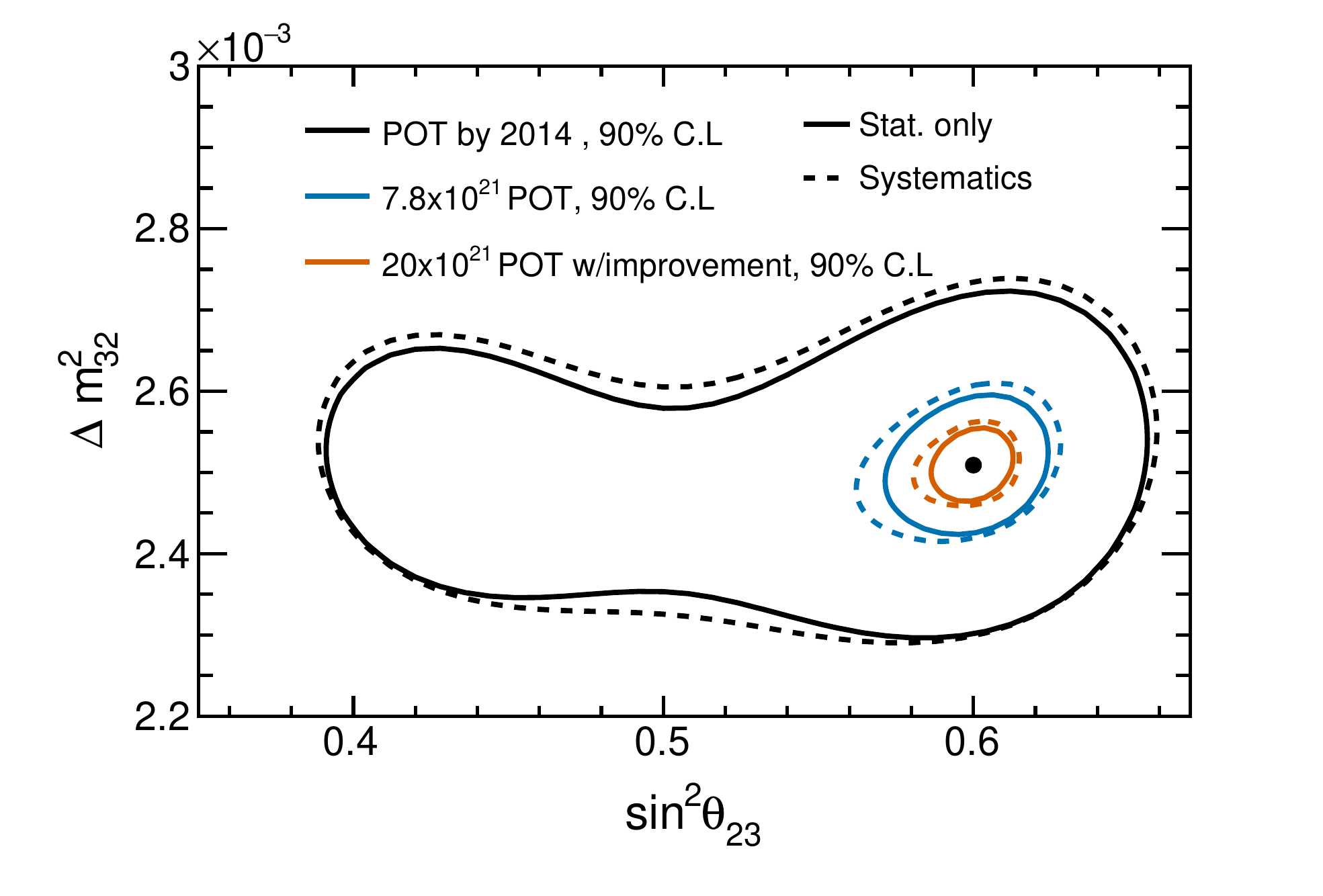}\caption{Assuming true \(\sin^2\theta_{23}=0.60\).}
\end{subfigure} \quad 
\begin{subfigure}[H]{7.2cm}
\includegraphics[width=7.2cm]{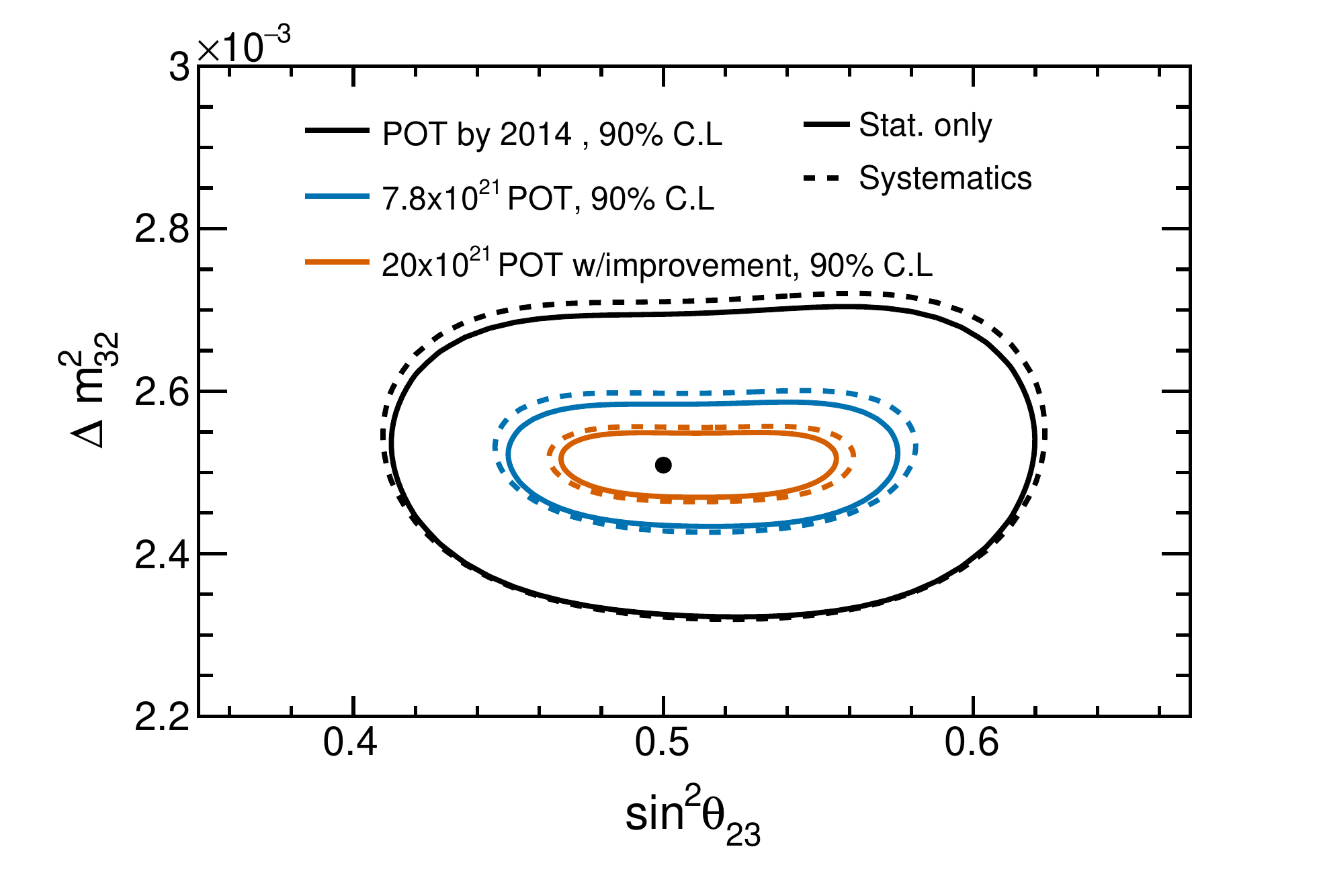}\caption{Assuming true \(\sin^2\theta_{23}=0.50\).}
\end{subfigure} \quad 
\caption[$\Delta m^2_{32}$ vs $\sin^2\theta_{23}$]{Expected 90\% C.L. sensitivity to $\Delta m^2_{32}$ and $\sin^2\theta_{23}$
with the 2016 systematic error.
The POT exposure accumulated by 2014 corresponds to \(6.9\times10^{20}\) POT \(\nu\)- + \(4.0\times10^{20}\) POT \(\bar{\nu}\)-mode.
For the T2K-II exposure of $20\times 10^{21}$ POT, a 50\% increase in effective statistics is assumed.
\label{fig:dm2vss2th23}} \end{figure}

The expected 90\% C.L. contour for $\Delta m^2_{32}$ vs $\sin^2\theta_{23}$ for the full T2K-II exposure is shown in 
Fig.\ \ref{fig:dm2vss2th23}.

\begin{figure}[H] \centering
\begin{subfigure}[H]{7.2cm}
\includegraphics[width=7.2cm]{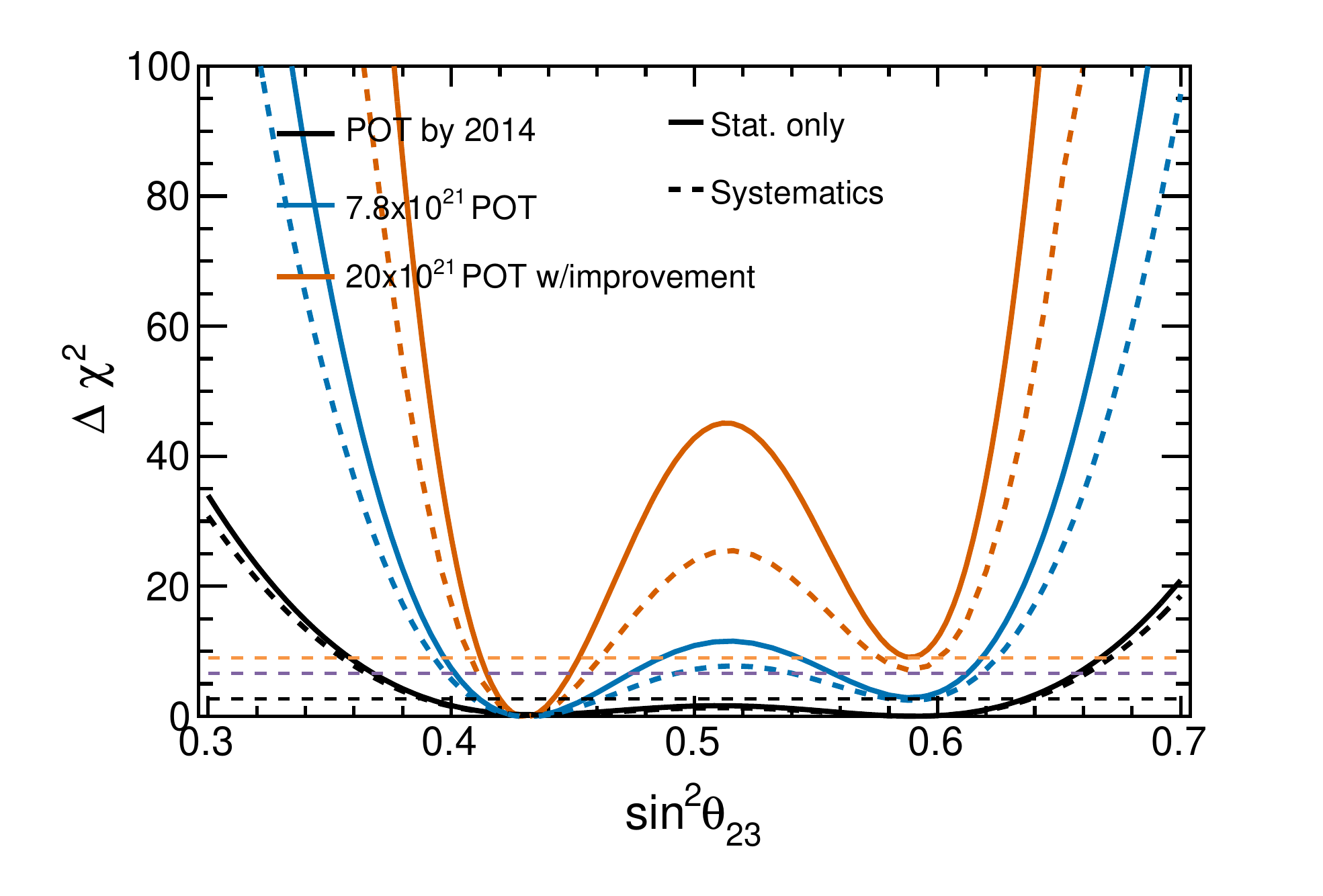}\caption{Assuming true \(\sin^2\theta_{23}=0.43\).}
\end{subfigure} \quad 
\begin{subfigure}[H]{7.2cm}
\includegraphics[width=7.2cm]{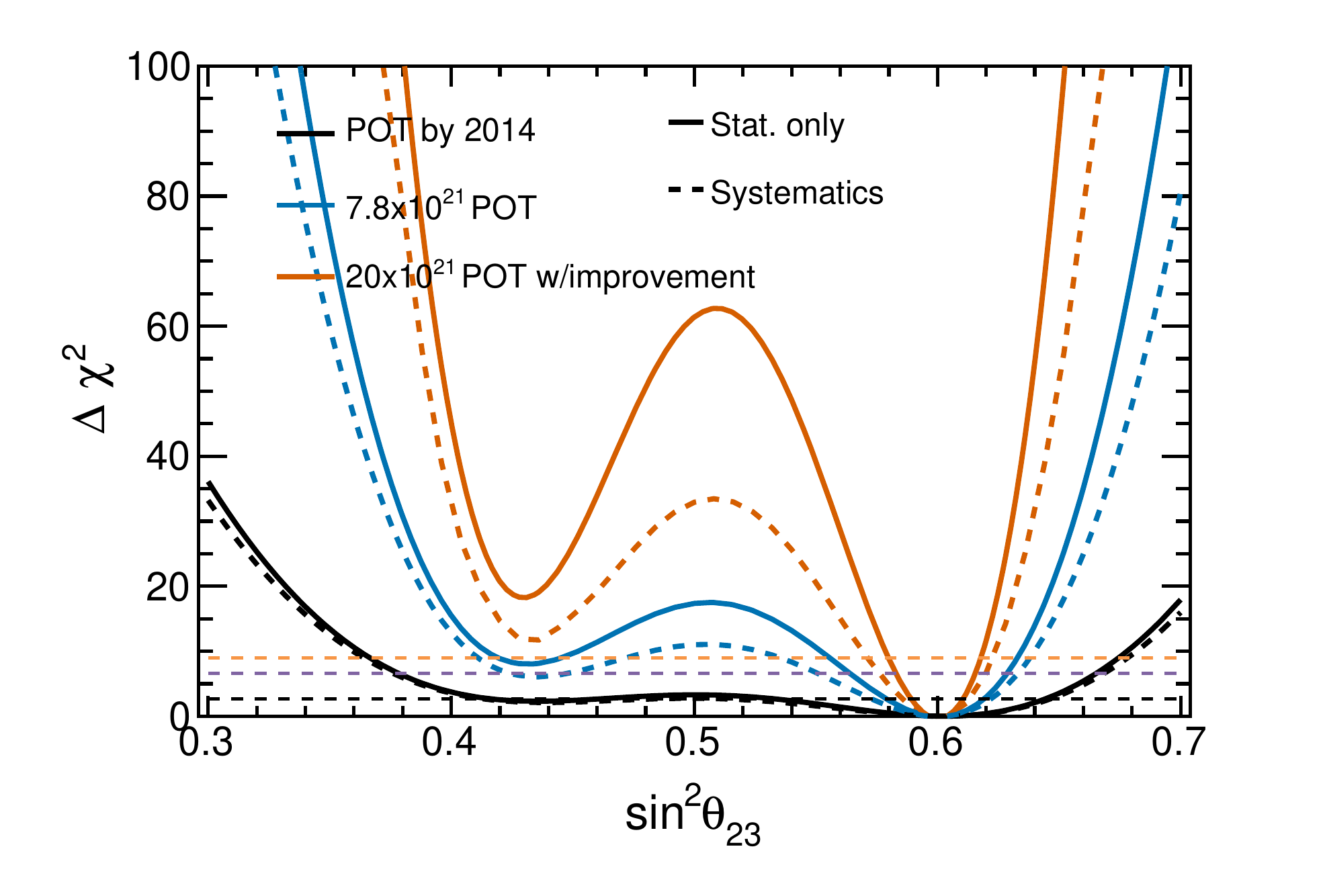}\caption{Assuming true \(\sin^2\theta_{23}=0.60\).}
\end{subfigure} \quad 
\begin{subfigure}[H]{7.2cm}
\includegraphics[width=7.2cm]{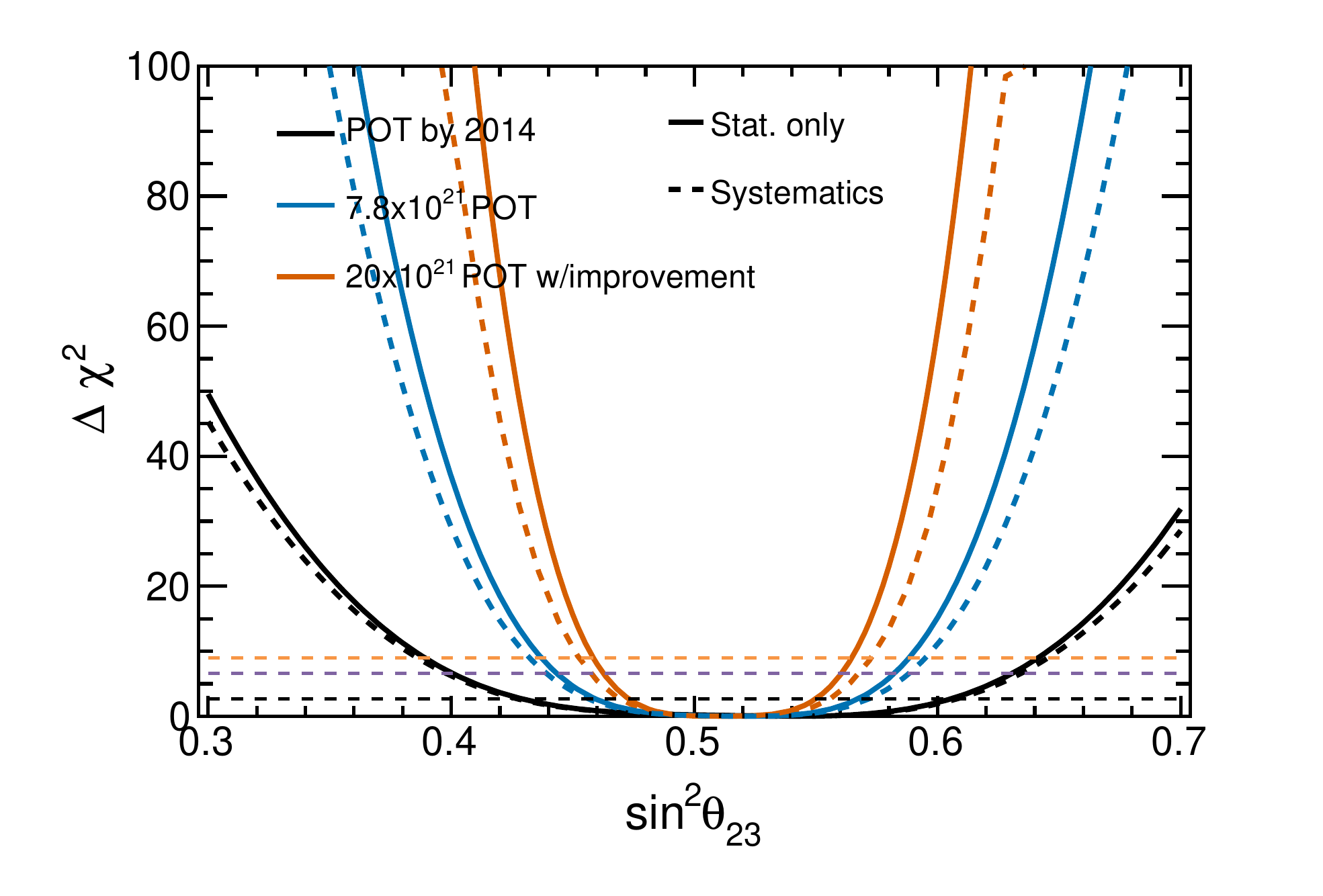}\caption{Assuming true \(\sin^2\theta_{23}=0.50\).}
\end{subfigure} \quad 
\begin{subfigure}[H]{7.2cm}
\includegraphics[width=7.2cm]{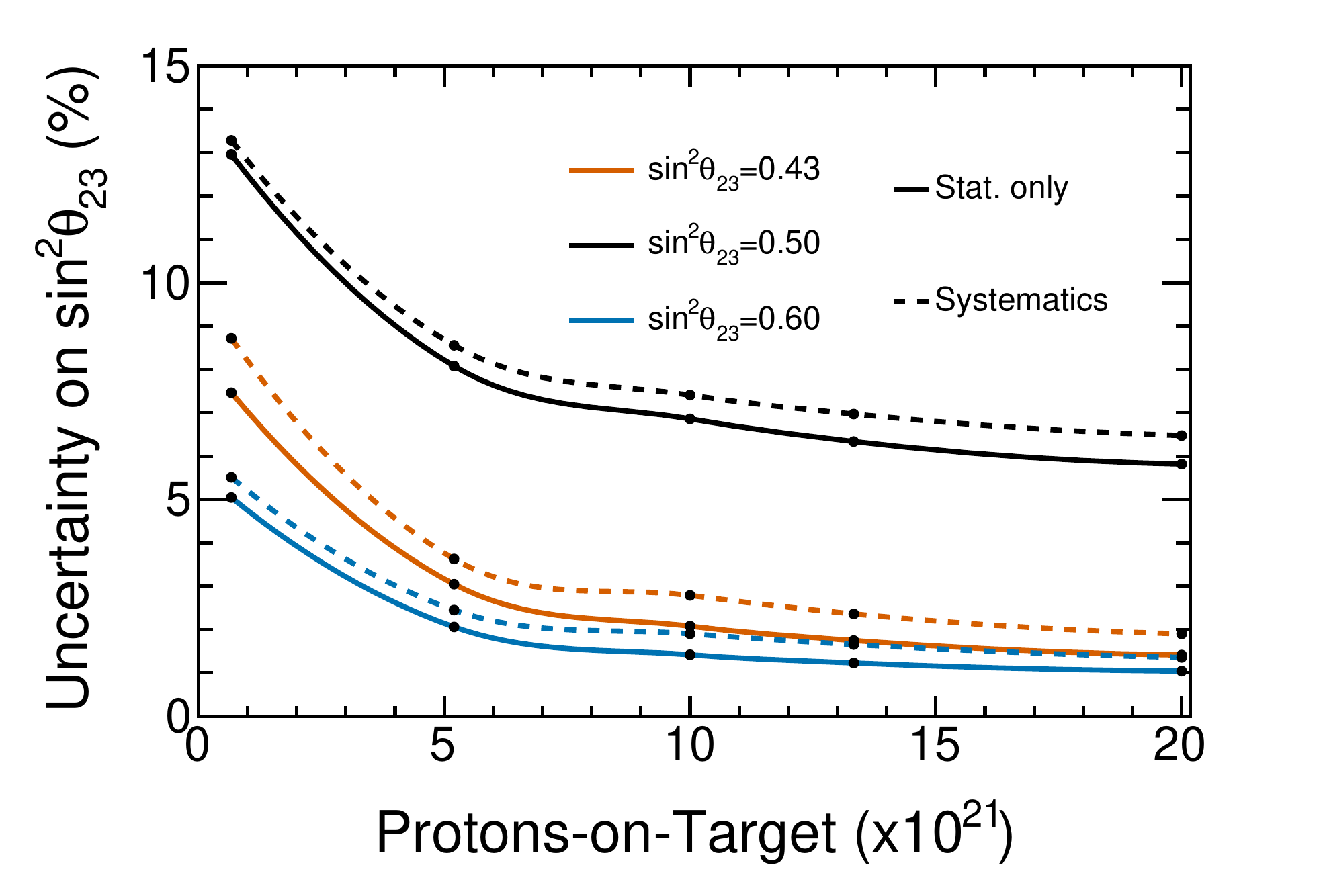}\caption{1$\sigma$ uncertainty of $\sin^2\theta_{23}$}
\end{subfigure} \quad 
\caption[$\sin^2\theta_{23}$]{$\Delta \chi^2$ vs. $\sin^2\theta_{23}$ assuming 2016 T2K systematic errors
  for a) $\sin^2\theta_{23}=0.43$, b) $\sin^2\theta_{23}=0.60$, and c) $\sin^2\theta_{23}=0.50$.
  The full T2K-II exposure of $20\times10^{21}$ POT with a 50$\%$ effective statistical improvement is compared to the approved T2K exposure and the $6.9\times 10^{20}$ POT $\nu$- and $4.0\times 10^{20}$ POT $\bar{\nu}$-mode accumulated  by 2014. (d) shows the expected uncertainty on $\sin^2\theta_{23}$  as a function of POT with different values of true $\sin^2\theta_{23}$  assuming a $50\%$ improvement in the effective statistics.
\label{fig:sin23}} \end{figure}

\noindent The plots indicate that for  $\theta_{23}$ values at the edge of the current 90\% CL regions,  T2K-II data can resolve the $\theta_{23}$ octant degeneracy.
Specifically, Fig.~\ref{fig:sin23} shows that the octant degeneracy can be resolved at more than 3$\sigma$ if  $\theta_{23}$ is in the upper octant with $\sin^2\theta_{23}$=0.60. For the lower octant case, $\sin^2\theta_{23}$=0.43, the significance of resolving the octant degeneracy is also close to 3$\sigma$. Fig.~\ref{fig:sin23} shows the uncertainty on $\sin^2\theta_{23}$ as a function of POT. If $\sin^2\theta_{23}$ is maximal, the expected $1\sigma$ precision of  $\sin^2\theta_{23}$  is 1.7$^\circ$. For the case of $\sin^2\theta_{23}=0.43$ and $0.60$ the uncertainty is 0.5$^\circ$ and 0.7$^\circ$ respectively. The uncertainty in the case of maximum mixing is much higher than the other cases since the $\nu_\mu$ survival probability at $\sin^2\theta_{23} \sim 0.50$ is nearly independent of $\theta_{23}$.

\begin{figure}[H] \centering
\begin{subfigure}[H]{7.2cm}
\includegraphics[width=7.2cm]{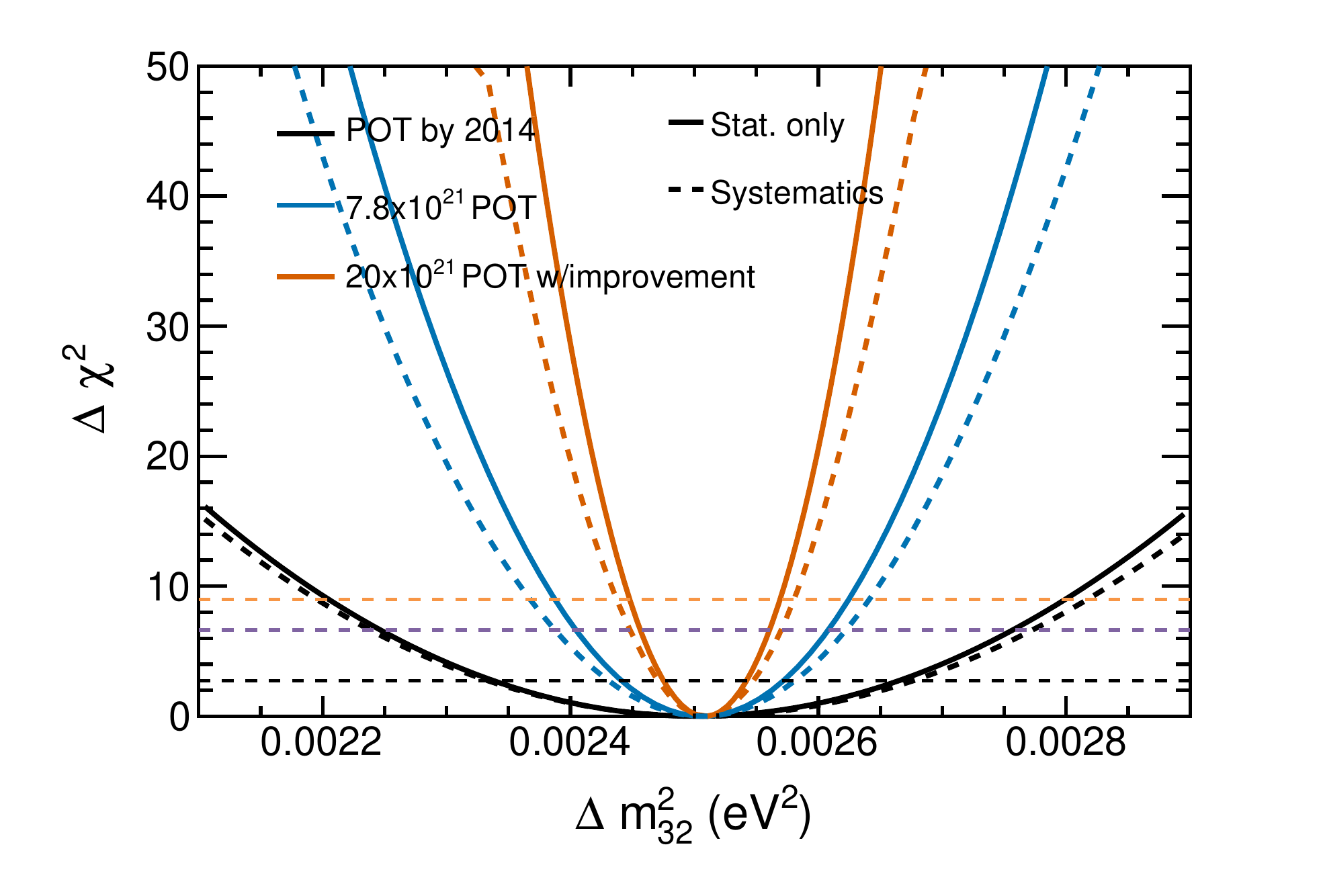}\caption{Assuming true \(\sin^2\theta_{23}=0.43\).}
\end{subfigure} \quad 
\begin{subfigure}[H]{7.2cm}
\includegraphics[width=7.2cm]{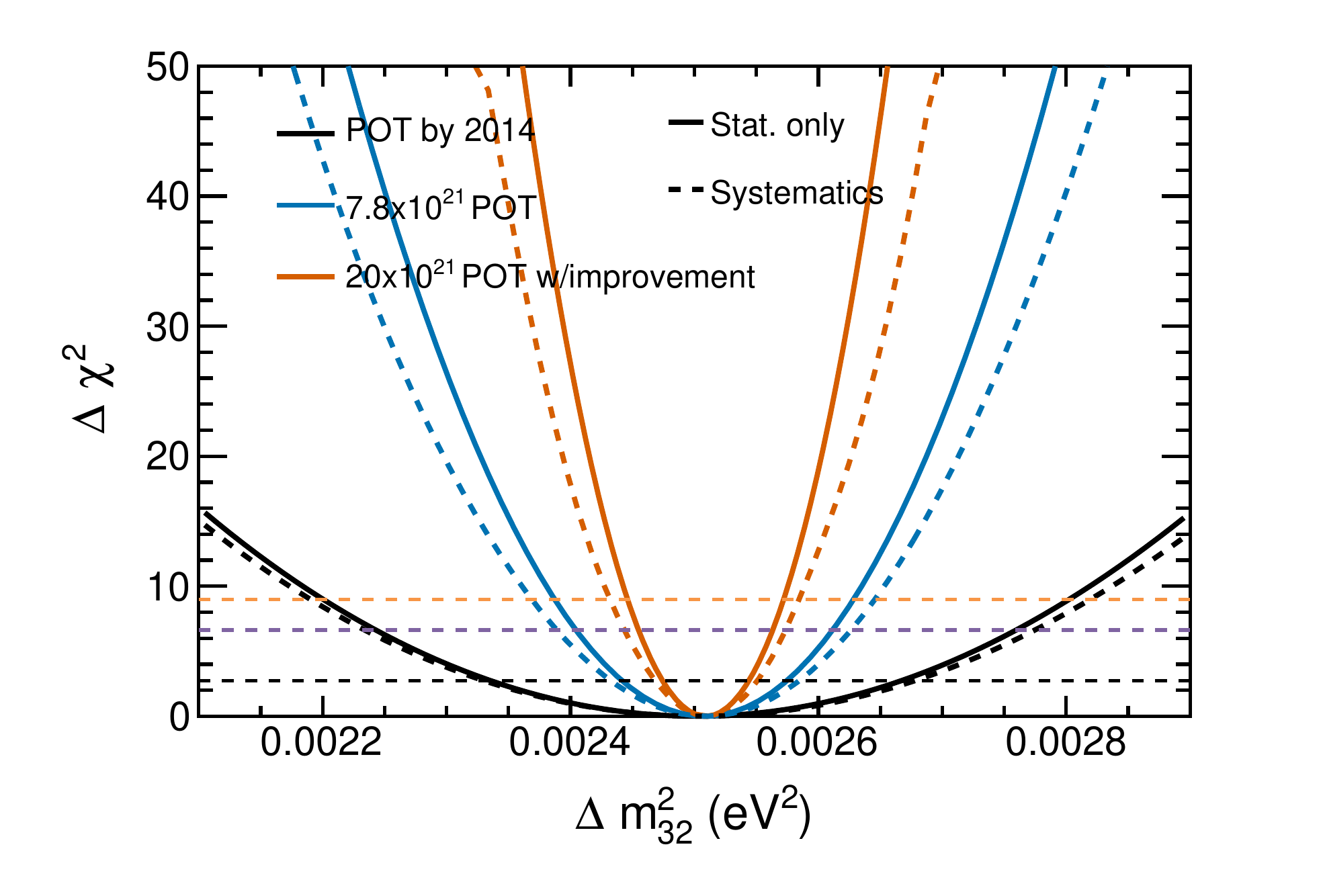}\caption{Assuming true \(\sin^2\theta_{23}=0.60\).}
\end{subfigure} \quad 
\begin{subfigure}[H]{7.2cm}
\includegraphics[width=7.2cm]{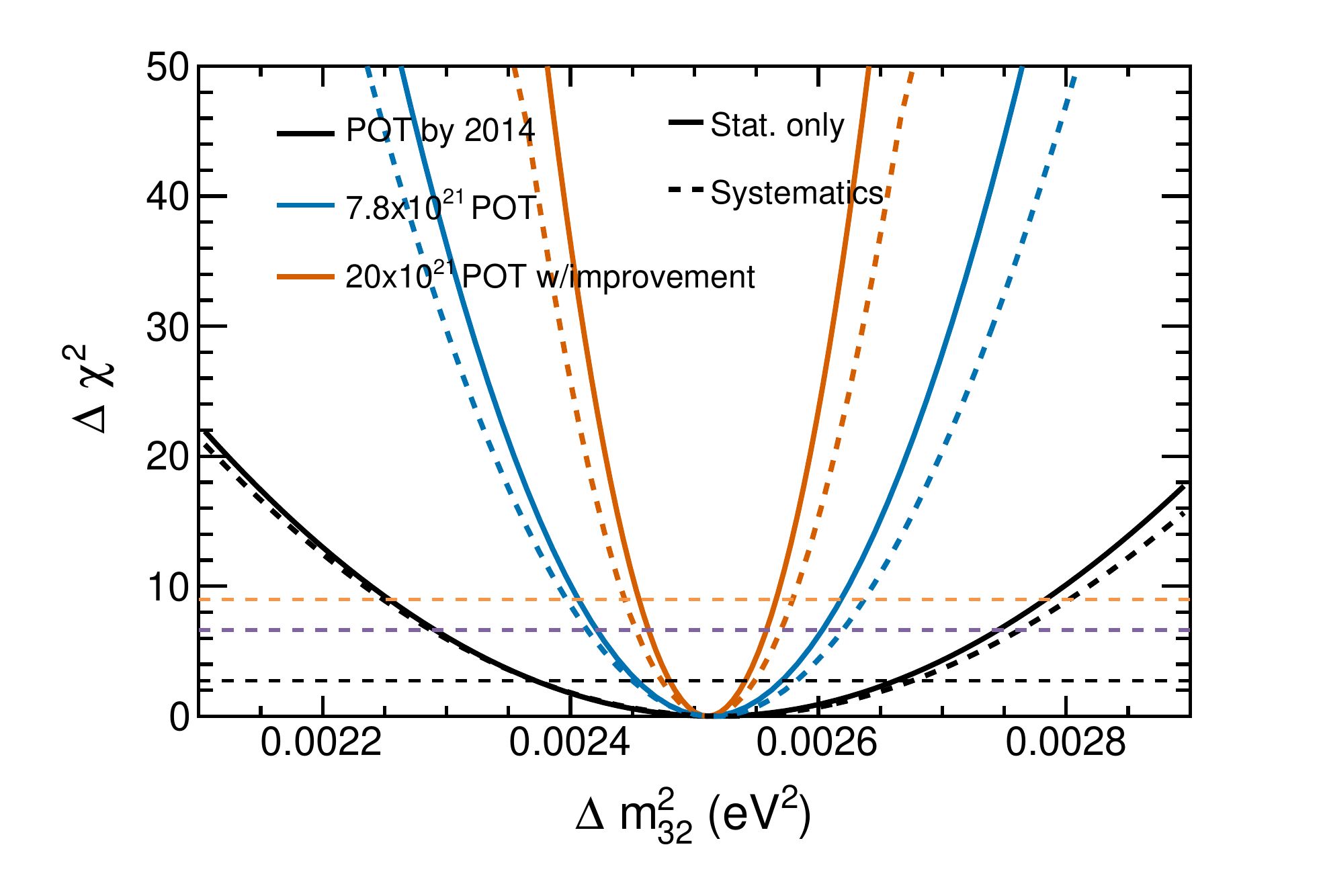}\caption{Assuming true \(\sin^2\theta_{23}=0.50\).}
\end{subfigure} \quad 
\begin{subfigure}[H]{7.2cm}
\includegraphics[width=7.2cm]{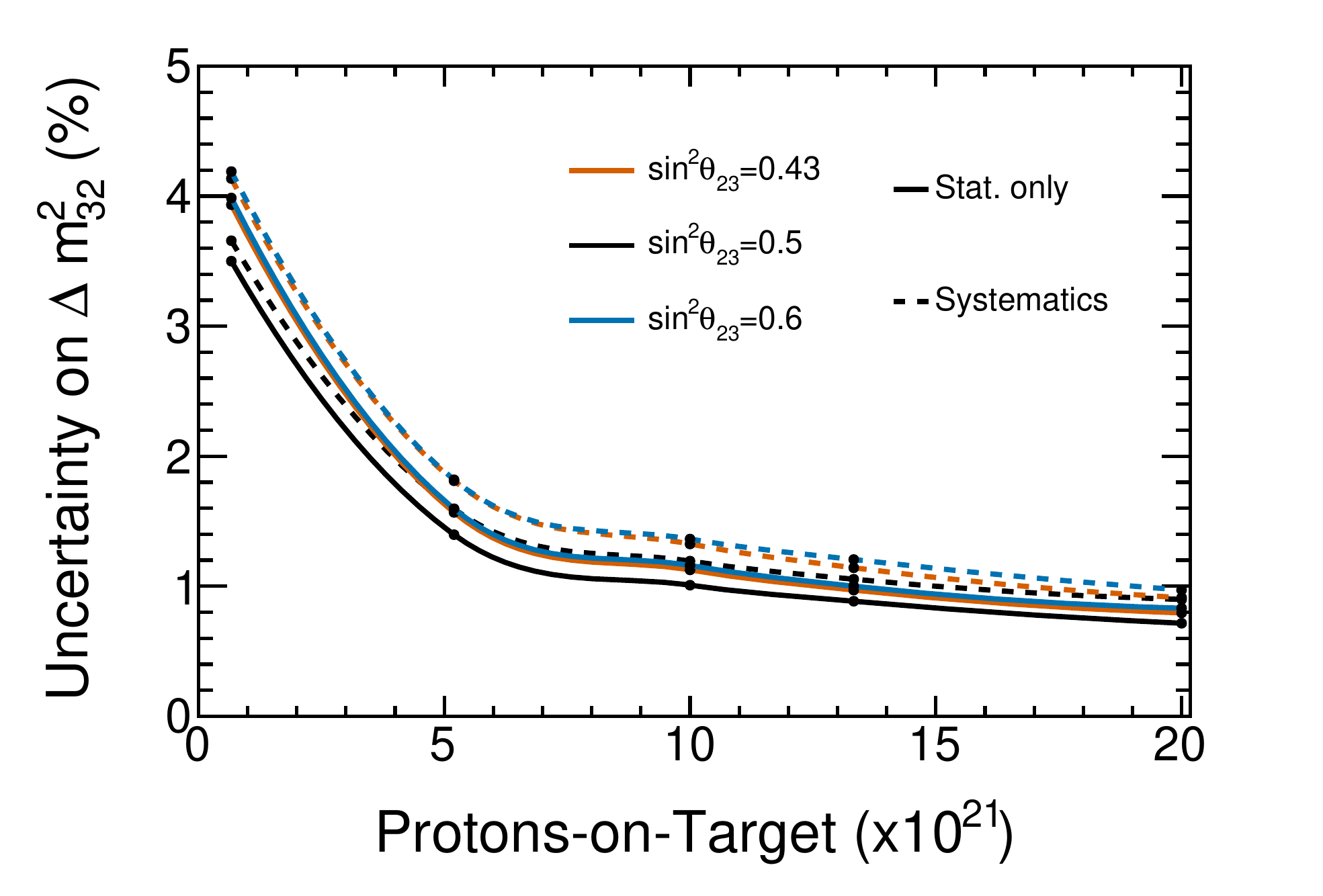}\caption{1$\sigma$ uncertainty of $\Delta m^2_{32}$}
\end{subfigure} \quad 
\caption{
$\Delta \chi^2$ vs. $\Delta m^2_{32}$ assuming 2016 T2K systematic errors
  for a) $\sin^2\theta_{23}=0.43$, b) $\sin^2\theta_{23}=0.60$, and c) $\sin^2\theta_{23}=0.50$.
  The full T2K-II exposure of $20\times10^{21}$ POT with a 50$\%$ effective statistical improvement is compared
  to the approved T2K exposure and the $6.9\times 10^{20}$ POT $\nu$- and $4.0\times 10^{20}$ POT $\bar{\nu}$-mode accumulated  by 2014.
  (d) shows the expected uncertainty on $\Delta m^2_{32}$ as a function of POT with different values of true $\sin^2\theta_{23}$
  assuming a $50\%$ improvement in the effective statistics.\label{fig:dm32}}
\end{figure}

Fig.~\ref{fig:dm32} shows $\Delta \chi^2$ plotted as a function of $\Delta m^2_{32}$ for three different values of $\sin^2\theta_{23}$ and  the uncertainty of $\Delta m^2_{32}$ as a function of POT. A precision of $\sim1\%$ 
on $\Delta m^2_{32}$ can be achieved 
in all cases.

%

\subsection{Neutrino Interaction Studies} 
\label{sec_xsec}
The additional run time of T2K-II will provide improved measurements
of neutrino and antineutrino scattering,
which probe nuclear structure through the axial vector current.
In the T2K energy region the largest contribution is due to Charged-Current Quasi-Elastic (CCQE)
interactions (50-60\%) and single pion production, mainly from $\Delta$ resonance, (about 25\%), with the rest being
due to multi-pion production and Deep Inelastic Scattering.
Actually, in modern experiments, like T2K, where the neutrinos interact with
relatively heavy nuclei (oxygen and carbon), there are important complications with respect to the
simple interpretation based on neutrino scattering modeling on free nucleons.
Indeed, neutrino-nucleus cross-section measurements are
affected by various nuclear effects, on the initial and final state,
which are difficult to disentangle experimentally and difficult to model theoretically.
There are long-standing disagreements between previous measurements in different experiments
and there is a flourishing of theoretical works trying to explain these discrepancies through 
improved modeling of nuclear effects.
The T2K datasets will help to solve these issues, which otherwise may become dominant systematics
in the future higher-statistics oscillation measurements.

\begin{figure} \centering 
\includegraphics[width=6.2cm]{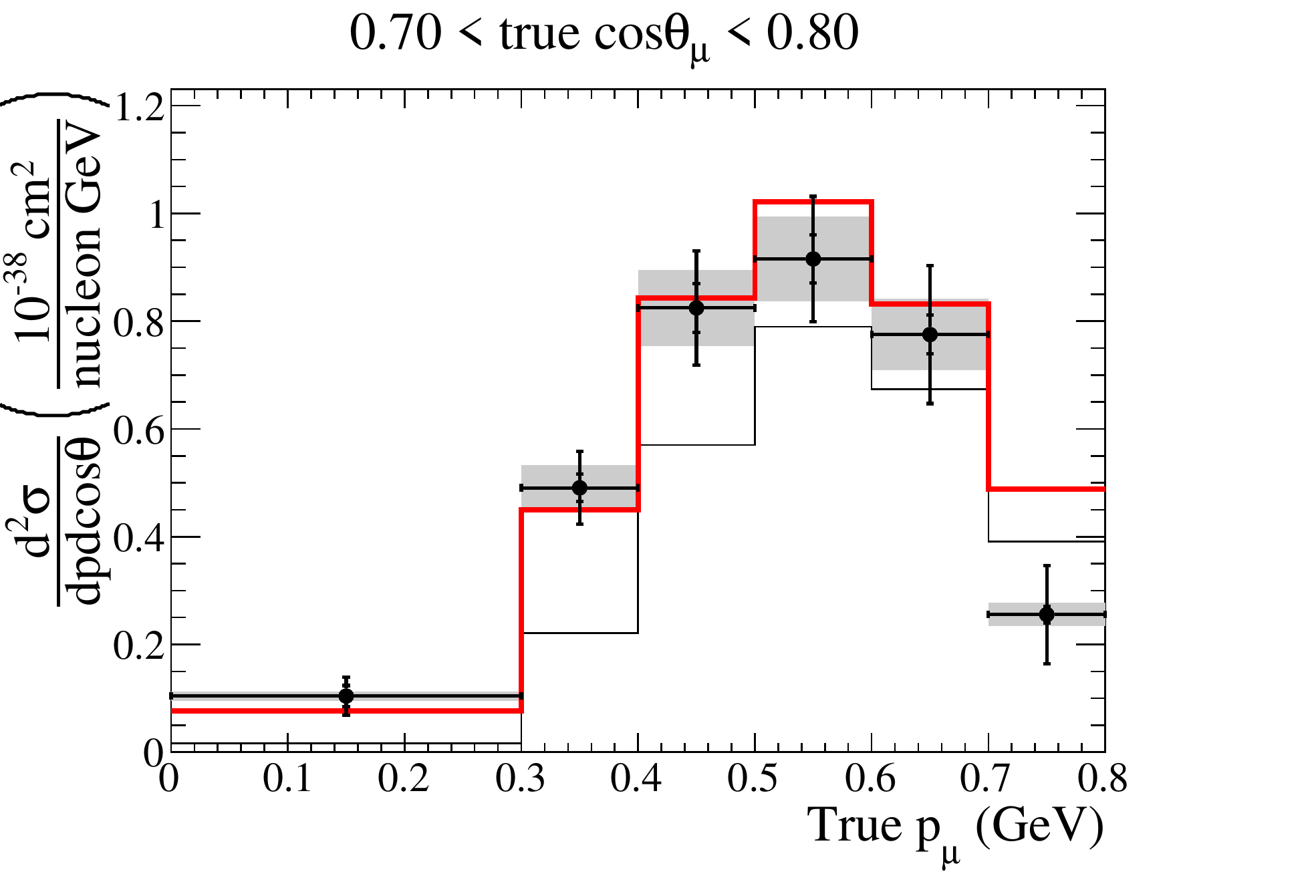}
\includegraphics[width=8.2cm]{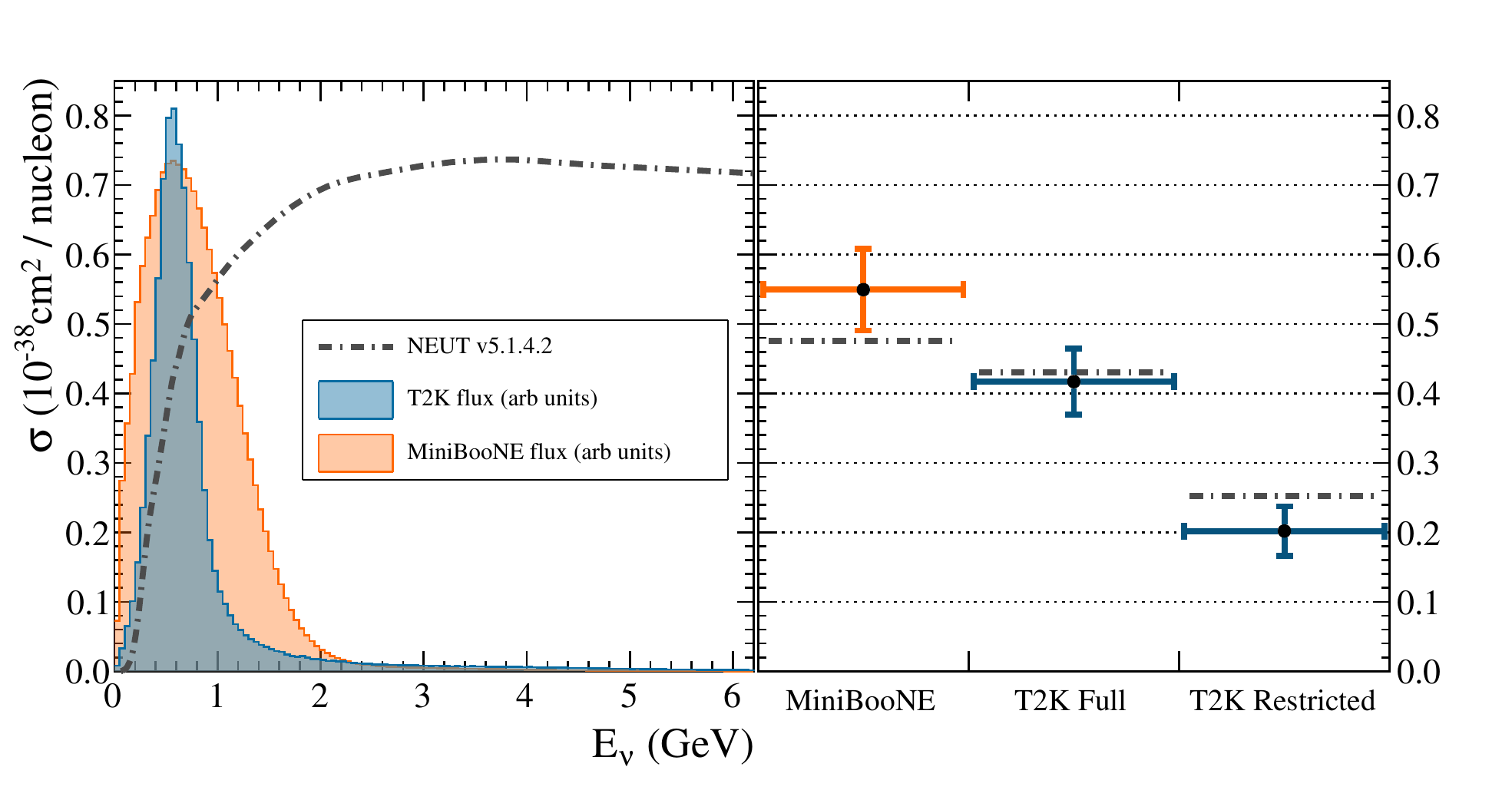}
\caption{Left: CC0$\pi$ measurement compared with predictions with (red) and without (black) 
2p2h contribution from Martini et al.\cite{Martini:2009}. 
Right: measured CC0$\pi$ flux integrated cross-section from MiniBooNE
and T2K measurement using the full and restricted phase space,
compared to NEUT predictions.}
\label{CC0piXsecMeasurement}
\end{figure}

For instance, the recent T2K measurement of charged-current events
without a pion in the final state~\cite{Abe:2016tmq} (CC0$\pi$), 
shown in Fig.\ref{CC0piXsecMeasurement}, suggests
the presence of a quasi-elastic-like component due to multi-nucleon
correlations (also known as 2p2h). The measurement is not yet precise enough
to solve the degeneracy between different microscopic models or 
effective parameterizations of the nuclear effects. 

The main limitations on the present measurements are due to statistical uncertainties and
the flux systematic uncertainty. The increase of statistics in the extended
T2K run proposed here will strongly improve the precision.
Moreover the T2K collaboration is engaged in an effort to reduce
uncertainties on the neutrino/antineutrino flux (see Sec.\ref{sec:flux}).
To minimize the impact of flux uncertainties, ratio measurements are also
on-going. In particular, T2K is preparing the extension of the mentioned CC0$\pi$ measurement to
anti-neutrino interactions. The ratio measurement of CC0$\pi$ in neutrino/anti-neutrino
will provide powerful constraints on 2p2h modeling, as shown in~\cite{Martini:2010ex}. The measurement
of the asymmetry of neutrino and anti-neutrino rates will allow the isolation of the
axial-vector interference term in the cross-section and direct measurement of any possible
bias on the $\delta_{CP}$ measurement, due to neutrino/anti-neutrino interaction modeling.
To improve these constraints, the increase of statistics in the anti-neutrino sample
requested in this proposal will be crucial.

Similarly, the comparison of electron and muon neutrino interactions is a fundamental
input to CP violation measurements. The T2K electron neutrino differential cross-section 
measurement~\cite{Abe:2014agb} is limited by statistics in most of the bins.
With the T2K-II expected datasets of 8,000 $\nu_e$ CC
and 2,000 $\bar{\nu}_e$ CC candidates, the differences between electron and muon neutrino interactions
can be studied with good precision.

The hydrocarbon, CH, is the dominant active target in ND280. 
A major contribution to the systematic error in the oscillation analysis is due to the different 
target materials of the near- and far-detectors, where the far detector target material 
is only water. A fraction of the near detector is also 
composed of water (P$\O$D and FGD2). However, cross-section measurements on water are statistically limited. 
The CC1$\pi$ analysis~\cite{Abe:2016aoo}, which uses data up to 2014, has a 13\% error due to statistics, 
and a 36\% error from systematics. 
Not only the statistical, but also the systematic uncertainties, which are constrained from control
regions in data, would benefit by the additional data of T2K-II.

T2K is also engaged in the effort to improve the acceptance of the T2K near detector: 
in recent analyses, the reconstruction has been extended to include
backward-going tracks. The rate of backward muons or protons provides an important
input to improve neutrino-interaction modeling in kinematics regions
far from the simple CCQE configuration.

Further insight in understanding the nuclear effects will also come from
the measurement of the kinematics and the multiplicity of the outgoing nucleons
in neutrino-nucleus scattering. T2K is actively pursuing these measurements,
including the study of the topology of the energy deposited around the vertex, 
the usage of transverse kinematic imbalance~\cite{Lu:2015tcr,Lu:2015hea} and 
the measurement of event rates with two or more protons which are highly sensitive
to nuclear effects. In particular the number of expected events with at least two protons
is strongly limited by the available statistics, 
thus such a measurement will highly profit from the statistics expected in T2K-II.
New samples are also being analyzed, like neutrino interactions in the argon gas of the TPCs.
These events can provide unique information about proton multiplicity 
thanks to the very low tracking threshold (below what can be achieved
with liquid argon detectors). 
Approximately 10,600 $\nu$-Ar and 1,900 $\bar{\nu}$-Ar interactions
are expected. 
The possibility of exploiting interactions in the TPC structures (walls and cathodes)
is also under investigation.

\subsection{Non-standard Physics Studies}
The high statistics at T2K-II will enable world-leading searches for various
physics beyond the standard model.

The combination of accelerator-based long baseline measurements
with $\nu_\mu/\bar{\nu}_\mu$ beams and reactor measurements with
$\bar{\nu}_e$ flux may give redundant constraints
on ($\Delta m^2_{32}, \sin^2\theta_{23}, \delta_{CP}$).
Any inconsistency among these measurements
would indicate new physics such as unitarity violation
in the three-flavor mixing, sterile neutrinos,
non-standard interactions, or CPT violation.
For example, CP violation larger than that allowed by
the three-flavor mixing framework could result from
interference with the fourth generation sterile neutrino.
We can perform a test of the CPT theorem by comparing $\nu_{\mu}$ disappearance
and $\bar{\nu}_\mu$ disappearance. Competitive results have already been published by T2K in 2015 as shown in
Figure~\ref{fig:anuth_23}, and with the additional data already accumulated, T2K will have the world-leading sensitivity.

For non-standard neutrino-matter interactions, 
T2K-II alone will not have a high sensitivity due to the relatively short baseline.
However, comparison of the T2K-II oscillation pattern with the NO$\nu$A experiment may
show interesting results.

A comparison of $\Delta m^2_{32}$ as measured by T2K and $\Delta m^2_{ee}$ by reactors, currently at the 4\% level in both cases, will be another interesting test. With T2K-II, we expect 1\% precision on $\Delta m^2_{32}$ while reactor experiments are expected to improve their precision. Any deviation or inconsistency would imply new physics, such as non-standard neutrino-matter interactions.


At the end of T2K-II, we expect more than 500 neutral current $\pi^0$ production
samples with 97\% purity at SK. This sample can be used to measure the active neutrino flavor content and search for oscillations with a sterile neutrino in the $\Delta m^2\sim10^{-3}$~eV region.

At the near detectors, oscillations arising from mixing with sterile neutrinos with $\Delta m^2 \sim 1\; \mbox{eV}^2$ can be studied through  $\nu_\mu$ disappearance, $\nu_e$ appearance, and $\nu_e$ disappearance.
The sensitivity  of the beam $\nu_e$ disappearance analysis at T2K is already reaching
some of the allowed region of sterile oscillation parameters
by the reactor $\bar{\nu}_e$ deficit\cite{beamnuedis} and can be further improved with more data.

Sidereal time dependence of the event rate either at the near detector
or SK can be used to search for Lorentz violation.
T2K has already reported measurements
with $6.63\times10^{20}$ POT using the INGRID near detectors.
The sensitivity of this analysis will be extended with more data.
A similar study using SK could extend sensitivity by three orders of magnitude due to the longer distance.

Models with right-handed neutrinos
having $O(10^{-1}-10^2)$~GeV mass have been proposed\cite{numsm1,numsm2}.
Such heavy neutral leptons could be produced at the T2K production target
and decay in the near detector.
Reference \cite{hnlsearch} proposes to search for them at the T2K near detector and effort at T2K has now started.

Since neutrino mass likely originates from
physics at very high energy scales ($\gtrsim 10^{14}$~GeV),
new physics at these energy scales could produce effects
of comparable size to neutrino oscillation.
Redundant and precise measurements
of neutrino oscillation are equally compelling and complementary
to precision searches at colliders for 
new physics at the TeV scale.

\newpage
\section{Summary}
\label{sec:summary}
The T2K collaboration proposes to extend the run from $\onepott$ to $\twopott$
to explore CP violation in a wide range of possible true values of $\delta_{CP}$ with 99\%C.L.,
to reach $3\,\sigma$ or higher sensitivity
for the case of maximum CP violation, to precisely measure
oscillation parameters and neutrino interactions
and to search for possible new physics.
The realization of these goals requires large efforts from both J-PARC 
and the T2K collaboration.
We propose J-PARC Main Ring upgrades towards
operation at 1.3~MW following the timeline in Figure \ref{fig:POT}
with five to six months of data taking each year.
We also propose neutrino beamline upgrades to accept 1.3~MW beam and to operate
the electromagnetic horns at 320~kA.
We aim to increase the effective statistics by up to 50\%
and reduce the systematic uncertainty to 2/3 of the present one.
Following this plan, the extended T2K program would occur before the next generation long-baseline
neutrino oscillation experiments and would continue to contribute to the steady
progress of particle physics.

\section*{Acknowledgment}

We thank the J-PARC staff for superb accelerator performance. We thank the 
CERN NA61 Collaboration for providing valuable particle production data.
We acknowledge the support of MEXT, Japan; 
NSERC (Grant No. SAPPJ-2014-00031), NRC and CFI, Canada;
CEA and CNRS/IN2P3, France;
DFG, Germany; 
INFN, Italy;
National Science Centre (NCN), Poland;
RSF, RFBR and MES, Russia; 
MINECO and ERDF funds, Spain;
SNSF and SERI, Switzerland;
STFC, UK; and 
DOE, USA.
We also thank CERN for the UA1/NOMAD magnet, 
DESY for the HERA-B magnet mover system, 
NII for SINET4, 
the WestGrid and SciNet consortia in Compute Canada, 
and GridPP in the United Kingdom.
In addition, participation of individual researchers and institutions has been further 
supported by funds from ERC (FP7), H2020 Grant No. RISE-GA644294-JENNIFER, EU; 
JSPS, Japan; 
Royal Society, UK; 
and the DOE Early Career program, USA.


\pagestyle{plain}
\clearpage

\clearpage

\clearpage

\bibliography{t2k2proarxiv}

\end{document}